\documentclass[12pt,a4paper]{article}

\usepackage{slashed}

\usepackage{a4wide}
\usepackage{latexsym}
\usepackage{epsf}
\usepackage{amssymb}
\usepackage{graphicx}
\usepackage{amsmath, cite}
\usepackage{amsmath,amssymb,amsthm, slashed}
\usepackage{verbatim}
\usepackage{hyperref}
\usepackage{color}	
\usepackage{xcolor}
\usepackage{tcolorbox}
\usepackage{bbold}
\usepackage{comment}

\renewcommand{\d}{\textrm{d}}

\newcommand{\SO}{\mathop{\rm SO}}

\newtheorem{theorem}{Exercise}[section]

\newcommand{\be}{\begin{equation}}
\newcommand{\ee}{\end{equation}}
\newcommand{\beq}{\begin{equation}}
\newcommand{\eeq}{\end{equation}}
\newcommand{\ba}{\begin{eqnarray}}
\newcommand{\ea}{\end{eqnarray}}

\renewcommand{\d}{\textrm{d}}

\begin{document}
\numberwithin{equation}{section}

\begin{center}

{\LARGE Beginners lectures on flux compactifications \\ \vspace{0.3cm} and related Swampland topics}

\vspace{1.1 cm} {\large   Thomas Van Riet$^1$ and Gianluca Zoccarato$^2$}\\

\vspace{2 cm} {$^1$ Instituut voor Theoretische Fysica, K.U. Leuven,\\
Celestijnenlaan 200D B-3001 Leuven, Belgium}\\
 {$^2$ Dipartimento di Fisica and Sezione INFN, Universit\`a di Roma “Tor Vergata”,\\ via della Ricerca Scientifica 1, I-00133 Roma, Italy }


\vspace{5cm}
{\bf Abstract}
\end{center}

\begin{quotation}
These lecture notes provide a pedagogical introduction, with exercises, to the techniques used in attempts to construct vacua with stabilised moduli in string theory. The reader is only assumed to have a basic knowledge of general relativity, geometry and field theory. We emphasize physical arguments and focus on the latest developments involving the Swampland program that point to a tension for the existence of AdS vacua with small extra dimensions or dS vacua with parametric control. We include a brief summary of the current status of these thorny issues. Unlike many other reviews we make almost no use of the technicalities associated to supersymmetric geometries.   These notes are largely based on lectures given at the CERN Winter School on Supergravity, Strings and Gauge Theory and in the Tehran School on Swampland Program held in the summer of 2022.
\end{quotation}

\newpage
\tableofcontents 
\newpage

\section{Introduction}
It is not known whether string theory describes the Ultra-Violet (UV) completion of general relativity coupled to the Standard Model of particle physics, but it certainly is worth studying as a model for how UV completions can work. Nothing has happened since the invention of the theory that should temper our enthusiasm about it. However, it has become more clear that finding answers to questions in particle physics and cosmology is highly non-trivial. One reason for this is the usual decoupling principle of effective field theory and it will plague any alternative to string theory (if any). Another difficulty is our human limitation in performing string theory computations when supersymmetry is broken. 

One of the appealing features of string theory is its fundamental character; one would hope that a theory of everything does not come with arbitrary numbers in its fundamental description. The formulation of string theory does not rely on dimensionless parameters and only features a single dimensionful length scale: the string length $\ell_s$.  This means that every coupling constant, particle mass, and any other term in the effective action is a priori computable from the theory. At the same time it would be strange that the couplings and masses of the Standard Model of particle physics and cosmology are unique numbers and no other choices are consistent at a fundamental level. Indeed, no matter how constraining string theory can be, one still expects a vast landscape of possible vacua related to the possible ways extra dimensions can curl up in (meta-)stable fashions. The fluctuations of the string around those vacua will lead to a landscape of effective field theories which ``geometrizes`` couplings and masses.

This is where the Swampland paradigm, introduced first by Cumrun Vafa \cite{Vafa:2005ui}, enters the scene. The Swampland paradigm essentially asks: `which low-energy quantum field theories can not result from a UV complete quantum theory of gravity?'. Such theories are said to be in the Swampland. This question is logically identical to the old-school method of string phenomenology, which essentially asks the opposite question (``What is the Landscape of theories?''). But the negatively posed question does alter the mindset of many researchers and the focus lies on the boundaries between the Swampland and the Landscape and in particular the patterns that appear. Most of the Swampland conjectures then come in the form of inequalities on low-energy parameters that constrain EFTs. It furthermore comes with concrete suggestions of how EFTs break down when one approaches the boundary of the Landscape. This clearly differs from the seemingly endless approach to cook up interesting EFTs and try to find string theory embeddings of them, an approach which has been popular for Beyond the Standard Model Physics and cosmological inflation. 

The result of the Swampland program is a fresh impetus in string phenomenology which has enriched the subject with ideas from holography, black hole physics, mathematics, and so on. The reason black hole physics comes to the fore in this endeavour is because horizons can challenge certain held assumptions in effective field theory. So many Swampland conjectures originate from black hole gedanken experiments. 

These lecture notes do not aim to review the Swampland program since many excellent reviews exist already \cite{Brennan:2017rbf, Palti:2019pca, vanBeest:2021lhn, Grana:2021zvf, Agmon:2022thq, Cribiori:2023gcy}. Instead we focus on a specific issue in the program: \emph{what is the value (sign and size) of the vacuum energy that results from a compactification of string theory}?  The main motivation behind these notes is to convey the fact that the Swampland conjectures about the sign and size of vacuum energy, while very deep, can be motivated from basic computations understandable to graduate students. These notes outline those computations and can be seen as a reference to learn the basics of compactifications. Where we differ in existing reviews is that we omit the differential geometry needed for understanding fluctuations of supersymmetric manifolds. Instead we focus on universal aspects that are valid whether or not supersymmetry is broken and we emphasize the crucial properties of orientifold planes (negative tension objects in string theory) in this regard. For that reason we added a mini-review on orientifold planes in Appendix \ref{sec:oplaneapp} for readers familiar with the worldsheet description of perturbative string theory. The bulk of this review does not require knowledge of the worldsheet. Older pedagogical reviews on flux compactifications and moduli stabilisation are \cite{Silverstein:2004id, Denef:2007pq} and highly recommended. More in-depth and technical reviews are \cite{Grana:2005jc, Douglas:2006es,Denef:2008wq, Ibanez:2012zz, Baumann:2014nda,  Hebecker:2021egx, Tomasiello:2022dwe, Cicoli:2023opf}. 

The unavoidable  landscape character of string theory is a natural consequence of the ``extra dimensions'' predicted by the critical string. The use of quotation marks here comes from the fact that the central charge of the worldsheet does not need to be cancelled by a set of CFT fields that allow a  geometric target space interpretation, anything that cancels it will do. But whatever cancels the worldsheet conformal anomaly introduces the junk that necessarily implies many possibilities for constructing vacua that potentially have fewer non-compact dimensions than the critical dimension ($D=10$). Famous vacua are holographic backgrounds like $AdS_5\times S^5$ of which there is an infinite sequence as a consequence of the discrete infinity of possible choices for the flux through the 5-sphere (which affects the size of the 5-sphere and the curvature of the AdS part). This holographic example is not fully satisfactory for the sake of the argument we made,  because the 5-sphere fluctuations are of the same energy scale as the vacuum energy and so these vacua will not be perceived as 5-dimensional to observers. In other words, the Kaluza--Klein tower of states does not decouple because the Hubble radius of AdS $L_{AdS}$ and the average size of the extra dimensions $L_{KK}$ are of the same order:
\begin{equation}\label{scalesep}
L_{KK}/L_{AdS} \sim \mathcal{O}(1)\,,
\end{equation}
whereas a 5d EFT description would require the ratio to be significantly smaller than one. 

Whether string theory has a landscape of vacua with stabilised moduli and with decoupled massive towers (Kaluza--Klein, string excitations, $\ldots$) is currently being debated \cite{Gautason:2015tig, Lust:2019zwm, Buratti:2020kda, DeLuca:2021mcj, DeLuca:2021ojx, Demirtas:2021nlu}  and if not, we should  rethink the way string theory makes contact with low energy effective field theories (EFTs). Note how this issue connects well to the cosmological constant (cc) problem.  This problem is roughly the statement that a natural value of vacuum energy, whether positive or negative, is expected to be of the order of the cut-off scale, which would imply \eqref{scalesep} for AdS vacua from dimensional reduction. 

Another hotly debated issue is not the size of the vacuum energy but its sign. String theory famously has issues with controllable de Sitter vacua \cite{Danielsson:2018ztv}, i.e. vacua of the form $dS_d\times X_{10-d}$,  where $X$ is a compact manifold of dimension $10-d$. Clearly $d=4$ is of most importance but we will try to be more general. In this review we shall therefore not insist on de Sitter vacua that allow for lower-dimensional EFTs describing fluctuations around it. The problem of finding dS vacua is this difficult that we will be happy with anything that has de Sitter isometries or almost-de Sitter isometries (slow-roll) for the non-compact part of the string target space. 

The motivation to hunt for dS vacua should be obvious given the observation of dark energy \cite{SupernovaCosmologyProject:1998vns, SupernovaSearchTeam:1998fmf, Planck:2018vyg} or the success of inflation in explaining early universe physics. For the particular case of inflation one only requires (almost) de Sitter isometries for 60 e-folds or so but at a vastly different energy scale then what is done for late-time dark energy. An outstanding review on the status of inflation in string theory can be found in \cite{Baumann:2014nda}. In this review we discuss meta-stable dS vacua and hence not slow-roll scenarios, which are also possible in the context of dark energy as quintessence models \cite{Ratra:1987rm, Caldwell:1997ii, Copeland:2006wr}. They are famously hard to achieve in string theory as well, and we refer to \cite{Kaloper:2008qs,Cicoli:2018kdo, DavidMarsh:2018etu, Hebecker:2019csg} for a recent take on this matter. Concerning the status of dS vacua in string theory, two (complementary) reviews appeared in 2018 \cite{Danielsson:2018ztv,Cicoli:2018kdo} but much has happened since, and we  will overview some of the new insights. Our main focus is introducing the basic technical underpinnings behind the most-used attempts to construct dS vacua. A major challenge in constructing dS vacua is the art of moduli-stabilisation itself for which many excellent reviews already exist \cite{Silverstein:2004id, Grana:2005jc, Douglas:2006es, Denef:2008wq, Denef:2007pq, Ibanez:2012zz,  Hebecker:2021egx, Cicoli:2023opf}. Our emphasis will be on the specific ingredients added to moduli-stabilisation techniques that ``lift'' us to de Sitter space, if possible at all.

Readers who want to know more about the basic physics of de Sitter space first, before contemplating about its UV completion can consult \cite{Anninos:2012qw} and references therein, whereas a basic reference on the quantum gravity aspects of dS space is \cite{Witten:2001kn}.
\newpage

\section{The (conceptual) framework}
 At first sight computing vacuum energy from string theory is an awkward thing to try. String theory is a microscopic theory, by definition it should answer questions about the smallest length-scales or the highest energy densities. Dark energy relates to the Hubble length which is the highest length-scale known in Nature. So why would a late-time cosmologist need to know about strings or any other attempt for UV completion of all interactions? Similarly one would not expect we need to know about quantum gravity in order to answer biology questions. This is essentially the decoupling principle that allows progress in all of science \cite{Burgess:2020tbq}. The reason for our quest nonetheless, is the cosmological constant problem \cite{Martin:2012bt, Padilla:2015aaa, Burgess:2013ara}. This problem essentially states that the natural value to expect for the cosmological vacuum energy is of the order of the UV scale $L^{-1}_{NP}$ of New Physics (NP) beyond the Standard Model. In terms of the Hubble scale $H$:
\begin{equation}\label{ccproblem}
HL_{NP}\sim \mathcal{O}(1)\,.    
\end{equation}
The similarity between \eqref{ccproblem} and \eqref{scalesep} is noteworthy \cite{Gautason:2015tig} since the KK scale is an example of a scale at which new physics occurs and $H$ is the inverse Hubble radius. So the fact that the best controlled AdS vacua in string theory obey \eqref{scalesep} can be seen as a stringy manifestation of naturalness for AdS vacua. 

In pure quantum field theory there is at first sight not so much interesting or special about the sign of the cc. But for string theory the sign could not be more important. AdS vacua, supersymmetric or not, are often straightforward to construct and seem all abundant, but somehow de Sitter vacua are not \cite{Danielsson:2018ztv, Bena:2023sks}. Since de Sitter space cannot be supersymmetric in ghost-free theories that complete Einstein gravity \cite{Witten:2001kn} one often blames the difficulty in finding dS vacua on the SUSY-breaking.  In our opinion this is not entirely correct since many non-SUSY AdS vacua have been constructed. Although it is not expected they can be fully stable, one expects that meta-stability is possible. 

Claiming that the sign of the cosmological constant does not matter in EFT is simplifying things a bit too much. In fact there is a rather extended literature on the ``non-existence" of de Sitter space in EFT, for which a biased sample of some older works is \cite{Mottola:1984ar,Polyakov:2007mm, Polyakov:2009nq, Polyakov:2012uc}. The idea here is similar to the quantum state of a black hole which differs from its classical description. Black holes in GR have a timelike Killing vector, but this symmetry is broken at the quantum level. So if the embedding of the Schwarzschild solution in string theory can be understood one should fail to find a stringy solution with that symmetry. String theory should be smart enough to see the Hawking radiation. Similarly, some researchers expect that the de Sitter isometries are broken at the quantum level and if correct this might imply that in a theory where one has direct excess to the UV degrees of freedom one should encounter an obstacle in finding exact dS solutions. This topic is very interesting and deserves a review on its own since the number of papers is very large with authors that have all different viewpoints. \footnote{Maybe that is why a review on this topic is non-existent? }

The possible breaking of the de Sitter isometries at the quantum level it is not unrelated to the topic of the renormalization group (RG) flow of vacuum energy. If vacuum energy undergoes RG running, its value would depend on the length scales at which one measures it, which is inconsistent with assuming de Sitter isometries at all scales. For the remainder of this paper we will content ourselves with the status of de Sitter space in string theory and leave this exciting topic for what it is. However it deserves mentioning that some recent papers have made a connection between the difficulty of finding dS space in string theory and the possible RG running of vacuum energy \cite{deAlwis:2021zab, DeAlwis:2021gou} or more general the breaking of dS isometries at the quantum level \cite{Danielsson:2018qpa, Dvali:2017eba, Blumenhagen:2020doa}

We mentioned that all EFT parameters (couplings, masses, \ldots) in string theory can be computed in principle for a given vacuum. The same is true for the cosmological constant. The cosmological constant, within string theory is understood as the local minimum of a scalar potential that arises in a low-energy effective approximation around a particular vacuum, if any. We have in mind a certain compactification of the critical string over a compact manifold $X$ down to $d$ spacetime dimensions. When the effective theory thus obtained in $d$ dimensions is truncated to its scalar field sector coupled to gravity, the Lagrangian will look like:
\begin{equation}\label{Lagrangian}
e^{-1}\mathcal{L} =M_p^{d-2}\mathcal{R}_n-\tfrac{1}{2}G_{ij}(\phi)\partial\phi^i\partial\phi^j - V(\phi)\,.
\end{equation}
with $e=\sqrt{-g}$ the Vielbein determinant. 
Here $G_{ij}(\phi)$ is the metric on the scalar manifold and in general cannot be brought to a flat form $G_{ij}=\delta_{ij}$, unless its Riemann tensor vanishes. But in a specific point $\phi^*$ of the manifold we can always find field redefinitions such that this happens. Such a point can be critical point of the potential:
\begin{equation}
\partial_i V|_{\phi^*}=0\,.
\end{equation}
Finding critical points of the scalar potential involves solving a highly non-trivial system of coupled algebraic equations. Then the value of the potential at the critical point $V(\phi^*)$ determines the vacuum energy of the resulting solution since the equations of  motion read (we assume $d>2$):
\begin{align}
&R_{\mu\nu} = \frac{1}{M_p^{d-2}}\left( \frac{1}{2}G_{ij}\partial_{\mu}\phi^i\partial_{\nu}\phi^j +\frac{1}{(d-2)}g_{\mu\nu}V(\phi)\right)\,,  \label{EOM1}\\
&\partial_{\mu}\left( eg^{\mu\nu}G_{ki}\partial_{\nu}\phi^i\right)- \tfrac{1}{2}e\partial_k(G_{ij})g^{\mu\nu}\partial_{\mu}\phi^i\partial_{\nu}\phi^j - e\partial_k V =0 \,,\label{EOM2}
\end{align}
which collapse to
\begin{equation}
R_{\mu\nu} = \frac{1}{M_p^{d-2}} \frac{1}{(d-2)}g_{\mu\nu}V(\phi^*)\equiv\frac{\Lambda}{(d-2)}g_{\mu\nu}\,,    
\end{equation}
at a critical point $\phi_*$. 
\begin{theorem}
Verify the equations of motion \eqref{EOM1}, \eqref{EOM2} from the Lagrangian \eqref{Lagrangian}.
\end{theorem}
So the cosmological constant is given by the vev of the effective potential $V$ in the critical point $\phi^*$
\begin{equation}\label{LambdavsV}
    M_p^{d-2}\Lambda = V(\phi^*)\,. 
\end{equation}
The cosmological constant in our conventions has dimension of inverse length squared and relates to the Hubble radius $\ell$ via
\be
\Lambda =\pm\frac{(d-1)(d-2)}{\ell^2}\,.
\ee
For $d=4$ de Sitter space we have $\ell=H^{-1}$. 
\begin{theorem}
Inspect the mass dimensions of all objects in the Lagrangian \eqref{Lagrangian}. Does this lead to the correct mass dimensions appearing in equation \eqref{LambdavsV}?
\end{theorem}

Perturbative stability of de Sitter critical points is decided by the Hessian $H_{ij}$ of the potential in the critical point
\begin{equation}
H_{ij}=\partial_j\partial_j V|_{\phi^*}\,.
\end{equation}
\begin{theorem}
Explain why the Hessian of a scalar potential is only a bilinear form on the scalar manifold when evaluated at a critical point. 
\end{theorem}
If $H_{ij}$ is positive definite, meaning that for all vectors $v^i$ we have $v^iv^jH_{ij}>0$ then the vacuum is perturbatively stable. This means that particle masses\footnote{In our conventions the masses squared are the eigenvalues of the Hessian.} are positive and can be found from the eigenvalues of the Hessian in a field basis where $G_{ij}(\phi^*)=\delta_{ij}$, which is always possible in a point on the manifold (or along a curve if it is a geodesic).  The same applies to Minkowski vacua where $V(\phi^*)$ vanishes. On the contrary for AdS vacua negative eigenvalues  of the Hessian (``tachyons'') are allowed on the condition they are not too negative:
\be
\text{eig} (H_{ij})= m^2 \geq -\frac{(d-1)^2}{\ell^2}  \,.
\ee
This is known as the Breitenlohner--Freedman bound \cite{Breitenlohner:1982bm}.

Note that this method of thinking about vacua \emph{assumes} the validity of an EFT with a finite number of fields with masses below the cut-off scale $\Lambda_C$. Clearly such a reasoning assumes also that $\Lambda <\Lambda_C$ since otherwise the curvature radius would be smaller than the cut-off length scale. Note how this requirement is problematic when there is no scale separation \eqref{scalesep} because then 
\begin{equation}
\Lambda\sim \Lambda_C\,.
\end{equation}
In fact this will often occur for the simplest stringy set-ups and then the procedure of minimizing a lower-dimensional potential has to be considered with care. It is safer to work directly with the 10d equations of motion in this case. We will come back to this issue later. 

How do we go about and ``construct vacua" in string theory? In an ideal world we would use the equations of full-blown non-perturbative string theory. Not only do we lack such equations, it is likely that, if we would possess them, humanity would not be ready to work with them in an easy manner. Hence, as with almost all areas of physics, we become less ambitious and settle for less in life; in this case 10d supergravity with leading corrections.  The string theory corrections to 10d supergravity can be organised in two categories: first, corrections in the string coupling constant $g_s$ (either perturbative or non-perturbative). The other kind are derivative corrections which are typically controlled by the string length $\ell_s$. Since that is a dimensionful number, the corrections appear in ratios of the form $\ell_s/L$ where $L$ is a length scale of the spacetime background. Some examples include a curvature radius, cycle size, field gradient, and possibly many others.

\begin{figure}[htp]
    \centering
    \includegraphics[width=10cm]{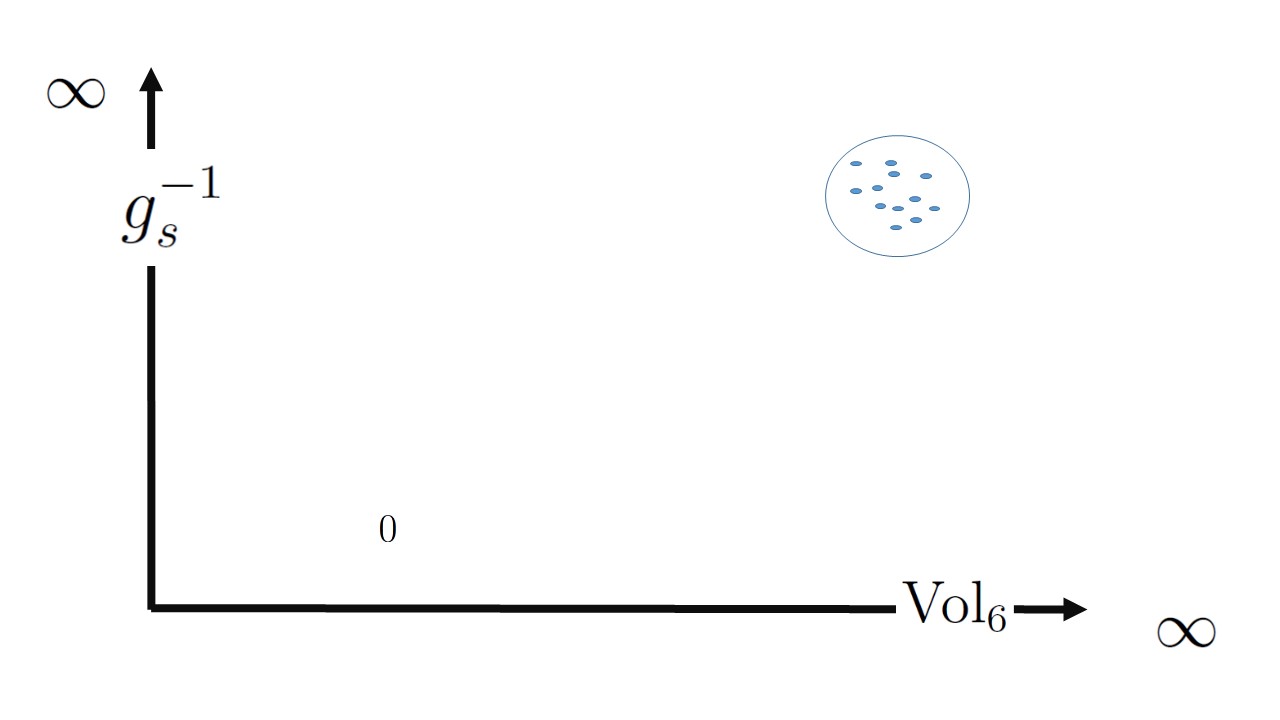}
    \caption{\small{\emph{The computable corner of string moduli space. On the horizontal axis we put the volume of the six compact dimensions in string units as an example expansion parameter suppressing derivative corrections. }}}
    \label{pic:corner}
\end{figure}

This means that one is squeezed into a corner of ``string moduli space" where computations can be done with some level of trust, see picture \ref{pic:corner}. 

Note that the vevs of the two expansion parameters shown in figure \ref{pic:corner} determine the 4D Planck mass through:
\begin{equation}
M_p^2 = g_s^{-2}\text{vol}_6 \ell_s^{-2}\,.
\end{equation}
One word of caution in interpreting the symbolic figure \ref{pic:corner}. Using string dualities we can hop around in string moduli space and what seems like an absolutely uncomputable corner in one duality frame becomes the computable corner in the other frame. So the overall situation is not as depressing as the picture would seem to suggest.

As often with perturbation theory it is unclear for which values of the expansion parameter $\epsilon=g_s$ or $\epsilon=\ell_s/L$ we can trust our expansion. Almost all perturbative expansions are asymptotic series meaning their radius of convergence vanishes and for a given value of $\epsilon$ there is a maximal order at which the expansion makes sense. This maximal order is not always easy to estimate. It has been argued that in that sense these approximations done in string theory are in a worse shape than they are in classical mechanics or quantum field theory because $\epsilon$ is a vev of a field determined by the same equations of motion one is approximating. 
In contrast, our well-known QFT expansions are different, we can decide to tune the value of the coupling constant (at a certain energy scale) to our liking and contemplate the physics for that value.  Due to the authors lack of knowledge we leave this thorny issue aside and work in the assumption that $\epsilon<0.1$ is a reasonable starting point for accepting leading corrections in string compactifications.

\begin{figure}[htp]
    \centering
    \includegraphics[width=14cm]{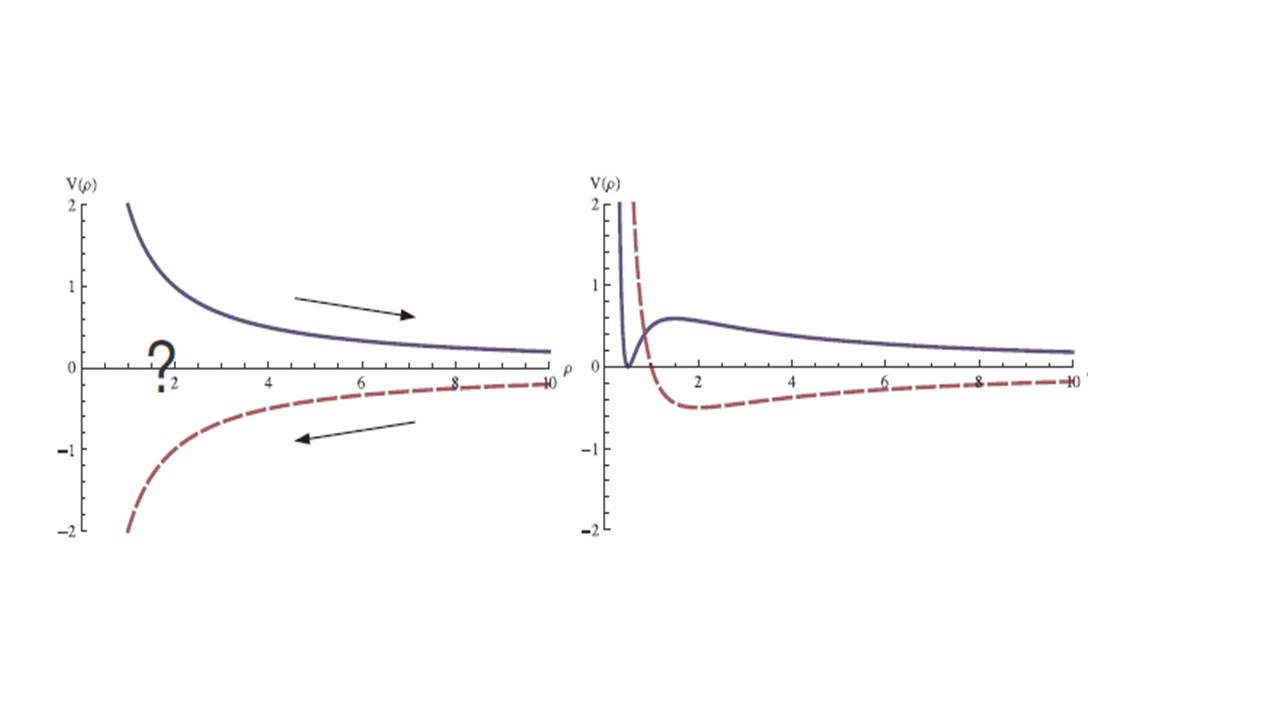}
    \caption{{\small \emph{Picture taken from \cite{Denef:2008wq}. The left figure illustrates the Dine--Seiberg problem and the right figure shows how vacua can arise at strong coupling.}}}
    \label{pic:DINESEIB}
\end{figure}

But even with this pragmatic mindset trouble is on the horizon and is known as the Dine--Seiberg problem \cite{Dine:1985he}. This problem was sloganized well in \cite{Denef:2008wq}: ``when corrections in physics can be computed they tend to be unimportant and vice versa." This slogan applies to the whole of physics and is somewhat too depressing to be correct. Still there is some truth in it. The Dine--Seiberg (DS) problem is essentially this. The idea is that on general grounds one expects a quantum effective potential to be runaway at weak coupling where corrections are small and hence if vacua exist they tend to live at strong coupling, where corrections cannot be computed as shown in figure \ref{pic:DINESEIB}

We will see in the next section that fluxes, i.e. classical vevs of form fields threading extra compact dimensions, are an ingredient to bypass the DS problem. Just consider the well-known AdS$_5\times S^5$ example: as we will see,  cranking up the flux quantum makes the parameter $\ell_s/L$ with $L$ the curvature (and length) scale of the AdS$_5$ and the $S^5$ factor (they are equal as explained before) go to zero. At the same time the string coupling is not stabilised and can take any value so we can take it to be arbitrary small. Exactly because of the high amount of supersymmetry preserved by the solution we can be confident that the string coupling is not secretly runaway, we evade the DS problem by a combination of flux and SUSY. Since we are wearing our phenomenology hats we  drop the SUSY argument made above. This can be done for vacua in which all moduli are stabilised. One of the main lessons to be learned from these lecture notes is that the flux-mechanism to evade the DS problem \emph{parametrically} can be shown to work beautifully for AdS vacua but not for dS vacua as was made explicit only recently \cite{Junghans:2018gdb, Banlaki:2018ayh, Bedroya:2019snp} and is at the core of some of the dS Swampland conjectures \cite{Ooguri:2018wrx, Hebecker:2018vxz}, all of which will be reviewed later. So dS model building is most likely going to be ``ugly business'' but somebody's gotta do this. 

It is important to be aware that corrections often organise themselves as perturbative and non-perturbative expansions near a regime of control, but this eventually breaks down away from this region because the effective field theory breaks down in a severe manner. In other words, whenever we move away from the amenable region of moduli space we expect the corrections to the tree level computation we discussed to become more and more relevant. In fact, in quantum gravity the mere validity of any effective field theory calculated in a certain region of moduli space is in jeopardy as soon as we move away from said region because this motion can upset mass-scales. For instance, having an effective field theory description in a region of moduli space implies that there are degrees of freedom that have been integrated out because their mass is larger than the EFT cutoff. However, the mass of these extra degrees of freedom will depend on the position in moduli space. This justifies our claim that motion in moduli space can potentially upset a given EFT making it necessary to integrate in or out some degrees of freedom. In fact, how and when this happens is one of the pillars of the Swampland program known as the \emph{Distance Conjecture} (DC) \cite{Ooguri:2006in}. This conjecture makes a quantitative statement about how EFTs are corrected upon moving large geodesic distances in moduli space. Unlike standard QFT approaches the DC has a very stringy feature: it immediately implies EFTs are UV completed, not by a few extra fields, but by an infinity of them, called a tower of fields. One typically has in mind that such towers comprise of fields whose masses are multiples of a certain mass scale: the DC describes how such mass scale shifts as we move in moduli space. More precisely, it claims the appearance of an infinite tower of modes that become \emph{exponentially} lighter the more we move away.  Given two points $p_1$ and $p_2$ in the moduli space in the large distance limit $\Delta \Phi \rightarrow \infty$ there exists an infinite tower of modes whose mass behaves as
\begin{align}\label{eq:distanceconj}
    M(p_1) \sim M(p_2) \, e^{-\lambda \Delta \Phi}\,.
\end{align}
Here $\Delta \Phi$ is the geodesic distance between the two points as measured with the target space metric. Moreover quite important is the constant $\lambda$: it is positive and conjectured to be $\mathcal O(1)$ in Planck units, something confirmed in all known examples. 
A word of caution about \eqref{eq:distanceconj}: one may be tempted to reverse the path in moduli space and say that going in the other direction will make modes exponentially heavier thus improving the effective field theory, however there will still be a different tower of modes that will become lighter nonetheless. As an example of the conjecture take a circle compactification of any gravitational theory: starting at any fixed value for the radius of the circle when moving towards larger and larger values we find that the infinite tower of KK-modes becomes exponentially lighter. Going at very small radius does not solve the problem and give a good EFT: in String Theory there are winding modes that become exponentially lighter in the limit of small radius.\footnote{For the reader who may be unfamiliar with some basics of String Theory, winding modes are given by strings that wrap the circle. Their mass is roughly the length of the circle times the string tension, and therefore they become heavy at large radius and light at small radius, the opposite behavior of KK-modes.}
The presence of this light infinite tower of modes is therefore how String Theory signals the presence of heavy corrections to the effective field theory when we move away from the region of large volume and strong coupling. Aside the stringy evidence for the Distance Conjecture and how it connects well with the other Swampland conjectures in a self-consistent logical framework, interesting bottom-up arguments for the correctness of the distance conjecture have appeared in \cite{Stout:2021ubb, Stout:2022phm}.

Before we close this chapter we stress one particular confusing issue. Imagine one has computed a vacuum well inside the region of parametric control given in figure \ref{pic:corner}. Let us be ignorant as to the sign of the vacuum energy, but let us assume that supersymmetry is entirely broken in the vacuum. A great question to confuse colleagues at conferences is whether $V(\phi^*)$ is the bare cc of the lower-dimensional Lagrangian or the actual physical result? The striking answer is that we are working with String Theory, so we were really working with the UV degrees of freedom and hence our computed results should be the actual physical vacuum energy, up to small corrections in the expansion parameters, which we assumed to be small. This is a striking achievement of String Theory! One can actually compute the cc without having to worry about its UV sensitivity. In other words, the loop corrections of the fields in the EFT defined by the fluctuations around the vacuum are there, but they should have been taken into account by our computations. In fact the same is true for corrections induced by all degrees of freedom even beyond the cut-off. The reason is that \emph{all} corrections have to organise themselves into stringy corrections. A recent review on the state-of-the-art in this field is \cite{Gao:2022uop}. 

In ordinary effective field theory one can only dream about this. It is well known that any particle beyond the cut-off would renormalize the CC and hence there is no control over the cosmological constant. This is the cosmological constant problem. However EFT methods are not wrong, so how do we see the EFT logic of the  problem appearing? The cosmological constant problem splits into a fine-tuning problem and a UV sensitivity. Strings can solve the second part, but the first is visible as we mentioned before: it is notoriously hard to achieve scale separation so the EFT naturalness guess:
\begin{equation}
    \Lambda\sim \Lambda_C
\end{equation}
is sensible, although in sheer conflict with observations. One way to understand the radiative instability problem of the cosmological constant in EFT, one can imagine building a vacuum in String Theory using fluxes, branes, etc. Then one decides to add extra matter multiplets by inserting extra branes. But these branes backreact on the original vacuum, shifting the positions of various moduli fields and leading to a corrected value of the vacuum energy\footnote{Those reader that cannot make sense of this sentence hopefully will after having gone through the next chapters.}. This should be the analog of the EFT computation where matter loops can have significant impact on the cosmological constant. 
\newpage

\section{Simple compactifications}

Before delving into complicated compactification scenarios we discuss the simplest basic examples; circle compactifications of pure gravity, where we introduce the notion of moduli and so-named Freund--Rubin reductions where we introduce the stabilisation of moduli.

\subsection{Circle reduction and the notion of moduli}

 Let us consider a theory with gravity in $D+1$ dimensions defined on a manifold of the form $\mathcal M_{D+1} = \mathcal M_{D} \times S^1$. We will not specify any detail about $\mathcal M_D$ and focus only on the $S^1$ compactification. Our ansatz for the $D+1$ dimensional metric is
\begin{align}
    d s^2_{D+1} = e^{2\alpha \varphi} g_{\mu \nu}\,dx^\mu dx^\nu+ e^{2 \beta \varphi} (dz+A_\mu dx^\mu)^2\,.
\end{align}
Some comments on the form of the metric:
\begin{itemize}
    \item  The coordinate on the $S^1$ is $z$ and it has periodicity $z \sim z+1$. So the radius of the circle is controlled by the factor $e^{2 \beta \varphi}$, meaning that the $D$-dimensional scalar field $\varphi$ will be the one responsible for the size of the circle;
    \item The metric on $\mathcal M_D$ is identified with $g_{\mu \nu}$ up to the warp factor depending on $\varphi$, and the $x^\mu$ are coordinates on $\mathcal{M}_D$ with indices $\mu,\,\nu$ running from $0$ to $D-1$;
    \item The off-diagonal components of the metric with one index on the circle are called $A_\mu$ and transform as a $D$-dimensional vector.
\end{itemize}
Parameters $\alpha$ and $\beta$ are arbitrary, however they can be fixed to get a $D$-dimensional effective action with a standard Einstein-Hilbert term (so-called Einstein frame). This requires $\beta = -(D-2) \alpha$, and to have canonical kinetic terms for $\varphi$ one should fix $\alpha^{-2} = 2(D-1)(D-2)$.\footnote{As the reader may have noticed these formulas start to misbehave when $D\leq 2$. In the following we shall assume $D>2$ to avoid any pathology.} With these choices, one gets the following effective action for the massless fields after reducing the action $\int \sqrt{-g_{D+1}} R_{D+1}$ 
\begin{align}\label{eq:effcirclered}
    S_D = \int_{\mathcal M_D}d^Dx \sqrt{-g_D}\left\{R_D -\frac{1}{2} (\partial\varphi)^2-\frac{1}{4}e^{2 (\beta-\alpha)\varphi}F^2\right\}\,,
\end{align}
where $F \equiv dA$.
\begin{theorem} Prove that the effective action for the massless fields after circle reduction is \eqref{eq:effcirclered}. Start with $A=0$ and then consider the more general case $A \neq 0$. To simplify note that much of the Einstein--Hilbert action $\int \sqrt{-g_{D+1}}\, R_{D+1}$ is a total derivative. 
\end{theorem}
The above exercise can be carried out entirely by relying on the following convenient formula for Weyl rescalings: consider two metrics in $D$ dimensions related by $\tilde{g}=e^{2 a\phi}g$ with $\phi$ some function and $a$ some constant, then
\begin{align}\label{eq:Weyl}
\tilde{R}_{\alpha\beta} = R_{\alpha\beta}  +  (D-2)a^2\left( \partial_{\alpha}\phi\partial_{\beta}\phi  - g_{\alpha\beta}(\partial\phi)^2\right) -(D-2)a \nabla_{\alpha}\partial_{\beta} \phi -a g_{\alpha\beta}\nabla^{\gamma}\partial_{\gamma}\phi  \,,
\end{align}
where all expressions on the right hand side are defined with respect to $g$ and not $\tilde{g}$.

Notice that \eqref{eq:effcirclered} almost achieves unification of Einstein gravity and Maxwell theory. The main issue is that the gauge coupling is actually a massless scalar which controls the size of the extra dimension. We certainly do not observe any such scalar (nor the interactions between vector and scalar in \eqref{eq:effcirclered}) so it is necessary to render it massive to justify our lack of observation of it. This general framework to give mass to scalar fields is called \emph{moduli stabilization}. In general one may use that quantum corrections will generate a potential for the scalar thus giving the action
\begin{align}
    S_D = \int_{\mathcal M_D} d^D x\sqrt{-g_D}\left\{R_D -\frac{1}{2} (\partial\varphi)^2-\frac{1}{4}e^{2 (\beta-\alpha)\varphi}F^2 - V(\varphi)\right\}\,,
\end{align}
and a potential would fix the vev of the scalar at the minimum of the potential and make it massive. However we will not try in the following to use quantum corrections but use directly classical ingredients, that is fluxes. The simplest of all flux compactifications are so-called Freund--Rubin vacua \cite{Freund:1980xh}. 

Before we do so we want to briefly discuss the extension to a compactification on a $n$-torus.  A useful reference for this is \cite{Roest:2004aqa}. One way to obtain is to simply repeat the reduction of gravity on a circle $n$ times, keeping in mind one has to also reduce the Kaluza--Klein vectors one obtains at each step. We will simply do this all at once. A useful Ansatz is
\begin{equation}\label{Ansatz1}
ds^2_{D+n} = e^{2\alpha\varphi}ds^2_D + e^{2\beta\varphi} M_{ab}(dz^a -A^a)(dz^b -A^b)\,.
\end{equation}
Let us explain notation: the $A^a$ are one-forms in $D$ dimensions $A^a= A^a_{\mu}dx^{\mu}$ which will play the role of $n$ Maxwell vectors gauging the local $U(1)^n$ symmetry of the torus. The scalar $\varphi$ measures the volume of the torus via
\begin{equation}
    \text{vol }(\mathbf T^n) = e^{n\beta\varphi}\,.
\end{equation}
The matrix $M$ is a matrix of scalar fields in $D$-dimensions that represents fluctuations of the torus at fixed volume. It is symmetric, invertible and of unit determinant. This means it describes $(n-1)(n/2 + 1)$ fields. 
\begin{theorem}
Why does $M$ describe $(n-1)(n/2 + 1)$ fields?
\end{theorem}
One can show that 
\begin{equation}\label{alphabetantorus}
\beta = -\frac{D-2}{n}\alpha\,,\qquad \alpha^2 = \frac{n}{2(D+n-2)(D-2)}\,.
\end{equation}
are required for getting $D$-dimensional Einstein frame and canonical kinetic term for $\varphi$. 
\begin{theorem}
Check this by putting $A^a=0$ and $M=\mathbb{1}$. 
\end{theorem}
\begin{theorem}
Check that $M$ in general obeys the identity $\mathrm{tr}(M^{-1}\partial M)=0$ Hint: $M$ has unit determinant.
\end{theorem}
\begin{theorem}
Check that the kinetic term for $M$ equals $\frac{1}{4}\mathrm{tr}(\partial M \partial M^{-1})$. Such a kinetic term describes the canonical metric on the coset $\mathrm{SL}(n, \mathbb{R})/\SO(n))$. Why does this coset appear?
\end{theorem}
The dimensionally reduced action is
\begin{equation}
    S_D =\int\sqrt{-g}\left(R -\frac{1}{2}(\partial\varphi)^2 +\frac{1}{4}\text{Tr}(\partial M \partial M^{-1}) -\frac{1}{4}M_{ab}e^{2(\beta-\alpha)\varphi}F^aF^b\right)\,,
\end{equation}
describing $n$-Maxwell fields coupled to several scalar fields which together form the coset $\text{GL}(n, \mathbb{R})/\SO(n))$. The reader that went through the previous exercises can reproduce all terms in the above action, aside from the Maxwell terms. The energetic reader is invited to also check these.\footnote{Warning: a lot of work is required for this.}

\subsection{A comment on Einstein frame and Planck units}
It could feel rather unnatural to, by hand, introduce a conformal factor $e^{2\alpha\varphi}$ in front of the $D$-dimensional metric in the reduction Ansatz \eqref{Ansatz1}. After all, why would the $D$-dimensional part of the higher-dimensional metric not be what an observer measures? Indeed, we were a bit quick on this (and so is the literature on the topic) so it is good to pause here and think deeper about this fact.  We refine our reduction Ansatz as follows
\begin{equation}
ds^2_{D+n} = e^{2\alpha(\varphi-\varphi_0)}ds^2_D + e^{2\beta\varphi} M_{ab}(dz^a -A^a)(dz^b -A^b)\,.
\end{equation}
Here $\varphi_0$ describes the vev of the ``radion'' scalar $\varphi$ in the $D$-dimensional vacuum around which we organise the fluctuations into an effective field theory (EFT), whose classical Lagrangian we are seeking. Note that now, in the vacuum, there is no conformal factor. It is trivial to compute the new effective action:
\begin{equation}
    S_D =\int d^D x \sqrt{-g}\left(e^{n\beta\varphi_0}R + \ldots    \right)\,,
\end{equation}
\begin{theorem}
Check this and fill in the dots. This is just a matter of rescalings. 
\end{theorem}
Note that the Einstein-Hilbert action in $D$ dimensions goes like $M_p^{D-2}R$ and so we read off that
\begin{equation}
M_p^{D-2} = e^{n\beta\varphi_0} = \text{vol}(\mathbf T^n)\,.    
\end{equation}
The reader confused by units should realise that we wrote the higher dimensional action in higher-dimensional Planck units and so was the torus volume. If we reinstate those units we have
\begin{equation}\label{planckmassKK}
(M^{(D)}_p)^{D-2} = (M^{(D+n)}_p)^{D+n-2}\, \text{vol}(\mathbf T^n)\,.    
\end{equation}
where the superscript denotes the dimension in which the Planck constant is defined and the torus volume is in $(D+n)$-dimensional Planck units such that the dimensions on both side of the equation are the same. In fact the above relation does not hold only for torus compactifications, it holds for any compactification manifold.
\begin{theorem}
Explain to yourself why that is the case.
\end{theorem}
It also entails simple physics: the larger the extra dimensions are, the weaker lower-dimensional gravity is, and that is a simple consequence of gravity ``leaking out into the extra dimensions''.

Equation \eqref{planckmassKK} can be confusing when the radion is moving far away from its vacuum and in fact keeps rolling. Then we need to adjust $\varphi_0$ since we have to give up the meaning of an EFT that describes fluctuations around a vacuum. In this case the Planck constant can run with time and one gets into issues like Brans--Dicke gravity. 

\subsection{Freund--Rubin vacua}

Consider a theory in $D+n$ dimensions whose fields consist of the metric and an $n-1$ differential form $A_{M_1 \dots M_{n-1}}$ with field strength $F_{M_1 \dots M_n}$. The action of the system is
\begin{align}
    S_{D+n} = \int d^{D+n}x \sqrt{-g}\left\{R_{D+n} -\frac{1}{2 \cdot n!} F_{M_1 \dots M_n} F^{M_1 \dots M_n}\right\}\,.
\end{align}
We are interested in considering the vacuum solutions on manifolds of the form $\mathcal M_{D+n} = \mathcal M_D \times \mathcal M_n$ where we take $\mathcal M_n$ to be compact. Our ansatz for the metric is
\begin{align}\label{eq:metricred}
    ds^2_{D+n} = ds^2_{D}+ ds^2_n = g_{\mu \nu}dx^\mu dx^\nu+ \rho\, \tilde g_{ab} dx^a dx^b\,.
\end{align}
Here coordinates $x^\mu$ are on $\mathcal M_D$ and coordinates $x^a$ are on $\mathcal M_n$, and naturally $g_{\mu \nu}$ is a metric on $\mathcal M_D$ and $\tilde g_{ab}$ is a metric on $\mathcal M_n$. We made the decision to extract an overall factor $\rho$ from the metric $\tilde g$, and $\rho$ will be a $D$-dimensional scalar field tracking down the volume of $\mathcal M_n$. This scalar is universal and does not depend on the intricacies of the geometry of $\mathcal M_n$, and moreover lots of physics can be extracted by just considering its dynamics. Given that all the information on the volume is stored in $\rho$ we need to impose the condition $\int_{\mathcal M_n} \sqrt{\tilde g} = 1$. 

Note that we could have uniformized notation with the previous subsection and wrote instead
\begin{align}\label{withvarphi}
    ds^2_{D+n} = e^{2\alpha(\varphi-\varphi_0)} ds^2_{D}+ e^{2\beta\varphi}ds^2_n \,.
\end{align}
and fix $\alpha$ and $\beta$ such that one obtains a canonical Einstein--Hilbert term and scalar kinetic term in $D$ dimensions. The reason we do not is to help the reader accessing the literature where both conventions (methods) are used and it is useful to be able to flip between both and so we leave this as an exercise further down. \footnote{Clearly we assume a vacuum here such that $\varphi=\varphi_0$ and there is no conformal factor, also note that obviously $\rho=e^{2\beta\varphi}$.}

We also need to specify the background for the $n$-form field: our ansatz is
\begin{align}
    F_n = Q\, \tilde \varepsilon_n\,.
\end{align}
Here $Q$ is a constant (the amount of flux that is threading the internal manifold $\mathcal M_n$), and $\tilde \varepsilon_n$ is  the Levi--Civita tensor for the internal manifold $\mathcal M_n$ which satisfies $\int_{\mathcal M_n} \tilde \varepsilon_n = 1$). This choice of flux automatically satisfies the equations of motion for $F_n$ and its Bianchi identity
\begin{align}
d\star F_n &= 0\,,\\
dF_n &=0\,.
\end{align}
\begin{theorem}
Verify the validity of these equations. To do so we encourage the reader to check Appendix \ref{app:forms} on $p$-forms and Hodge stars. 

\end{theorem}
Once the $p$-form equations are checked, what remains are the Einstein equations: 
\begin{align}
R_{MN} = -\frac{n-1}{2(D+n-2)} \frac{1}{n!} F_n^2 \,g_{MN} +\frac{1}{2}	\frac{1}{(n-1)!} \underbrace{F_{M L_1 \dots L_{n-1}}F_N^{\phantom{N}L_1 \dots L_{n-1}}}_{\equiv F_{MN}^2}\,.
\end{align}
We now compute the two terms on the right hand side of this equation. We have that
\begin{align}
F_n^2 = F_{M_1 \dots M_n} F^{M_1 \dots M_{n} } = Q^2 \rho^{-n}(\tilde \varepsilon _n)^2  = Q^2 \rho^{-n} n!\,.
\end{align}
 Given that all the legs of $F_n$ are in the internal manifold this means that $F^2_{M N}= 0$ if either index is chosen to be on $\mathcal M_D$, that is $F^2_{\mu N} = 0$. The remaining components are therefore $F^2_{a b}$ which we compute to be
\begin{align}
F_{ab}^2 = Q^2 \rho^{-(n-1)}\tilde \varepsilon^2_{a b} = Q^2 \rho^{-(n-1)} (n-1)! \, \tilde g_{ab} = Q^2 \rho^{-n} (n-1)! \, g_{ab}\,.
\end{align}
Summarizing we find the two equations
\begin{align}
R_{\mu \nu} &= -\frac{n-1}{2 (D+n-2)} Q^2 \rho^{-n} g_{\mu \nu}\,,\label{Einstein1}\\
R_{a b} &= -\frac{n-1}{2 (D+n-2)} Q^2 \rho^{-n} g_{a b }+ \frac{1}{2}Q^2 \rho^{-n} g_{ab} = \frac{D-1}{2(D+n-2)}\rho^{-n} Q^2 g_{ab}\,. \label{Einstein2}
\end{align}
We can rewrite these equations as
\begin{align}
R_{\mu \nu } & = -\frac{D-1}{L_{\text{AdS}}^2} g_{\mu \nu}\,,\\
R_{ab} &= \frac{n-1}{L_n^2} g_{ab}\,.
\end{align}
Notice that both $\mathcal M_D$ and $\mathcal M_n$ are Einstein spaces, with $\mathcal M_D$ having negative curvature and $\mathcal M_n$ having positive curvature. We can choose $\mathcal M_D$ to be AdS$_D$ and choose $\mathcal M_D$ to be any appropriate Einstein space $\Sigma_n$ with positive curvature: this means that our solution is AdS$_D \times \Sigma_n$. Notice that the curvature radii of the two spaces are of the same order of magnitude. There is still one aspect of these solutions to be discussed: what is the actual volume of the internal space. We chose the metric $\tilde g_{\alpha \beta}$ to have unit volume, but we still need to compute the volume as measured by the internal metric $g_{\alpha \beta}$. This is an important point as the volume of the internal manifold must be much larger than one in units of the UV scale in the higher dimension; that is
\begin{align}
\text{vol}(\mathcal M_n) \gg \ell_p^n\,.
\end{align}
This condition is necessary to ensure that we can trust our solution in the higher dimensional classical gravity theory and neglect derivative corrections. To assess which regime can be trusted let us take the case of $\mathcal M_n = S^n$, that is an $n$-dimensional sphere. With the normalized metric we have
\begin{align}
\tilde R_{a b} = \frac{n-1}{\tilde L^2} \tilde g_{ab}\,,
\end{align}
where in this case $\tilde L$ is of order 1 (up to some factors of $\pi$). Since for a constant conformal rescaling we have that $\tilde R_{a b} = R_{ab}$, and so:
\begin{align}
\frac{n-1}{\tilde L^2} \tilde g_{ab}  = \frac{n-1}{L_n^2} g_{ab} = \frac{n-1}{L_n^2} \rho\, \tilde g_{ab}\,.
\end{align}
Equating the two sides of \eqref{Einstein2} we find
\begin{align}
\rho = \frac{L^2_n}{\tilde L^2} = \frac{2(D+n-2)(n-1)}{D-1} \frac{\rho^n}{\tilde L^2 Q^n}\,,
\end{align}
and solving for $\rho$ we find
\begin{align}
\rho^{n-1} = \frac{D-1}{n-1}\frac{1}{2(D+n-2)} \tilde L^2 Q^2 \,,
\end{align}
so up to a factor which is of order 1 we find that the radius $\rho$ is proportional to (a positive power of) the charge $Q$. Therefore in a situation of large $Q$ we find also a large $\rho$ which in turn implies a small curvature: the solution is therefore trustworthy in a regime of large $Q$, that is a regime with a large amount of flux. 

\subsection{A first look at the Swampland}
One consequence of our solution is that we found that $L^2_{\text{AdS}} = L^2_{\Sigma_n}$, that is the \emph{curvature radii} are of the same order of magnitude. In the case of spheres the curvature radius can be actually identified with the radius of the sphere. So even if the internal space $\mathcal M_n$ has a finite volume we find that its volume is roughly of the same order as the 'volume' of the AdS factor.\footnote{Again, the curvature scales of the two spaces are of the same order of magnitude.} That begs the question of whether we can decouple the two volumes, or if we want to formulate this question in a more physical sense, of whether it is possible to decouple the Kaluza--Klein modes in the effective field theory in AdS. This is a good time to introduce already our first \emph{Swampland conjecture}, named the `Strong AdS distance conjecture' \cite{Lust:2019zwm} which (roughly) claims that this is not possible. We will discuss this in more depth later on.

We were a bit quick in throwing words like ``decoupling KK modes" above. We therefore briefly expand on this. Consider a massless field $\Phi$ in $D+n$ dimensions satisfying the equations of motion $\Box_{D+n} \Phi = 0$. We then expand $\Phi$ as
\begin{align}
\Phi(x_\mu,x_a)  = \phi^{(0)}(x_\mu) + \sum_{n>0} \phi^{(n)} (x_\mu) f^{(n)}(x_a)\,,
\end{align}
where $f^{(n)}(x_a)$ are eigenmodes of $\Box_n$ with eigenvalues $\lambda_n$. Then the modes $\phi^{(n)}(x_\mu)$ satisfy the equation
\begin{align}
\Box_D \phi^{(n)}(x_\mu) = \lambda_n \phi^{(n)}(x_\mu)\,,
\end{align}
so that the mass of the field is $m_n^2 = \lambda_n$ and $\phi^{(0)}(x_\mu)$ is massless. The modes $\phi^{(n)}$ are referred to as a Kaluza--Klein (KK) tower and in general couple non-trivially to other fields so the above analysis is very naive. Yet, as a rough estimate of how masses of KK modes are determined it captures the essence. 

With regard to our Freund--Rubin solution, one can wonder whether there exists a positively curved Einstein space, say with unit curvature radius, such that $\lambda_1$, the smallest KK eigenvalue, can be parametrically larger than one? Or is there a bound in the space of all Einstein manifolds with positive curvature (and unit curvature)? If there is no such bound, then we can conclude that Freund--Rubin solutions can be used for \emph{actual compactifications}, meaning that the KK sector decouples from the EFT since its mass scale is parametrically larger than the vacuum energy scale (the AdS scale). This notion of \emph{scale separation} goes against the Swampland conjecture of \cite{Lust:2019zwm}, but we just noticed that our question is purely mathematical. This is a concrete situation in which Swampland conjectures, based on physical heuristics, can be turned into formal mathematical conjectures. Indeed, it was recently conjectured by a collaboration of mathematicians and physicists that no such Einstein manifolds can be found \cite{Collins:2022nux}. This remains a conjecture and requires either proof or counter-example. Interestingly this purely mathematical conjecture can be turned into a prediction about conformal field theories, using the AdS/CFT correspondence and we will briefly explain that later in these lecture notes. But note how our simple discussion of something as basic as Freund--Rubin solutions can already bring us to the forefront of science with a very trans-disciplinary character:
\begin{equation*}
\text{Swampland} \Longleftrightarrow \text{mathematics}\Longleftrightarrow \text{conformal field theory}.
\end{equation*}

Let us contemplate that this Swampland, or mathematics, or CFT conjecture, is correct. Then there is no scale separation, meaning that we will always have that the Kaluza-Klein scale, i.e. the length scale at which we can observe extra dimensions, is of the order as the AdS Hubble scale. Then there is no standard notion of an effective lower-dimensional gravity theory, because the KK modes are simply too light to integrate them out. Then what is the meaning of a lower-dimensional action? It can only have one: \emph{that of consistent truncation}. This means that one was able to write down a lower-dimensional Lagrangian for certain fluctuations of the higher-dimensional fields, and it captures the equations of motion of those fields. When can this happen? In general such a thing is not to be expected. Let us briefly explain. Consider a compactification Ansatz of the form $\mathcal{M}_{D+n}= \mathcal{M}_{D}\times \mathcal{M}_{n}$ with $\mathcal{M}_{n}$ compact. Then take any higher dimensional field $\Phi$ and expand it in a zero mode $\Phi^{0}$, so constant along $\mathcal{M}_{n}$ and some basis of modes on $\mathcal{M}_{n}$, labeled by a discrete index $n>0$ and we write $\Phi^{(n)}$. These are our KK modes. The higher dimensional equations of motion are typically very non-linear and so we expect in general that a KK mode is sourced by a zero mode. Schematically
\begin{equation}
    \Box \Phi^{(n)} = \Phi^{(0)} +\ldots
\end{equation}
If so, we cannot, mathematically, put the KK modes to constants and let the zero modes fluctuate. But imagine now, the opposite, namely that the KK modes are very heavy compared to the zero modes and the vacuum energy. Then we can truncate the modes in a physical sense, namely in the sense of Wilsonian effective field theory: we can integrate them out. Clearly that requires scale separation, and if the Swampland conjecture against it is correct, we can only use lower-dimensional theories as consistent truncations in case the magical fine-tuning happens where the zero modes do not source the KK modes. Interestingly, this is the case for Freund--Rubin reductions with enough supersymmetry.   

A different way to look at the physical truncation to the zero modes, i.e. the construction of an EFT, is by course graining the higher-dimensional equations of motion. One can show that solving the dimensionally reduced theory of the zero-modes is equivalent to solving the  equations of motion of the higher-dimensional theory integrated over the internal space. 
\begin{theorem}
Check this for a circle compactification, using standard Fourier transformation.
\end{theorem}
Of course solving integrated equations is literally what we assume by course graining over length scales smaller than the KK scale.

Most reductions contemplated in the string phenomenology literature \emph{assume} scale separation from the get-go, get an EFT and then solve the EFT, hoping that the solution for the radion is such that it is indeed self-consistently stabilised at a value where the assumption of scale separation was correct.

\subsection{Freund--Rubin vacua redux: dimensional reduction}

We would now like to revisit the Freund--Rubin solutions discussed previously and derive them using a different technique. We will promote $\rho$ to a scalar field in $D$ dimensions and build from dimensional reduction an effective action for $\rho$ in $D$ dimensions
\begin{align}
S_D = \int d^Dx_\mu \sqrt{-g_D}\left[R_D + \frac{1}{2}(\partial \rho)^2 - V(\rho)\right]\,.
\end{align} 
The potential $V(\rho)$ controls the dynamics of $\rho$ and minimizing it one finds the vacuum at $\rho = \rho_0$ where $\partial_\rho V = 0$. When sitting at this minimum one gets that $V(\rho_0)$ acts as an effective cosmological constant $\Lambda$ in the $D$-dimensional theory, and we can identify in the case $\Lambda <0$ 
\begin{align}
\Lambda = -\frac{D-1}{L^2_{\text{AdS}}}\,.
\end{align}
The kinetic term and other terms involving derivatives of $\rho$ will not be computed for the moment as they will not be necessary. We will now derive the potential from simple dimensional reduction. Taking our dimensional reduction ansatz \eqref{eq:metricred} we find that the $(D+n)$-dimensional Ricci scalar is
\begin{align}
R_{D+n} = R_D + \rho^{-1} \tilde R_n + \dots\,,
\end{align}
where the dots include terms containing the derivative of $\rho$. Plugging this in the $(D+n)$-dimensional action and integrating over the internal space we find
\begin{align}
 S_D = \int d^D x_\mu \sqrt{-g_D} \,\rho^{\frac{n}{2}} \left[R_D + \rho^{-1}  \tilde R_n -\frac{1}{2} Q^2 \rho^{-n}  + \dots \right]\,,
\end{align}
where again the dots include terms with derivatives in $\rho$. Notice however that this action is not in Einstein frame due to the factor $\rho^{n/2}$ appearing in front of the $D$-dimensional Ricci scalar. To address this we change our dimensional reduction ansatz for the metric to
\begin{align}
ds^2 = \left(\frac{\rho}{\rho_0}\right)^\alpha ds_D^2 + \rho^2 ds_n^2\,.
\end{align}
Notice that this coincides with the old ansatz \eqref{eq:metricred} in the vacuum, that is at $\rho = \rho_0$. Plugging in the new ansatz we find the $D$-dimensional action
\begin{align}
S_D = \int d^D x_\mu \sqrt{-g_D} \,\rho^{\frac{n}{2}} \left(\frac{\rho}{\rho_0}\right)^{ \frac{D\alpha}{2}} \left[\left(\frac{\rho}{\rho_0}\right)^{-\alpha}R_D + \rho^{-1}  \tilde R_n -\frac{1}{2} Q^2 \rho^{-n}  + \dots \right]\,.
\end{align}
We can ensure that we are in Einstein frame by choosing $\alpha = -n/(D-2)$. After this choice we obtain the simple scalar potential
\begin{equation}
V(\rho) = -\rho^a \tilde R_n + \rho^b \frac{Q^2}{2} \,,    
\end{equation}
with
\begin{equation}
a = -1+ \frac{D \alpha + n}{2}\,,\qquad  b= -n + \frac{D \alpha + n }{2}\nonumber\,.    
\end{equation}
Notice that there are two components in the potential competing with each other, namely the flux and the curvature of the internal manifold. Clearly the curvature of the internal manifold has to be positive otherwise there is no minimum of the potential. Minimizing the potential we find
\begin{align}
\rho_0^{a-b} = \frac{b}{2 a} \frac{\tilde L^2 }{n (n-1)} Q^2 \,, \qquad \tilde L^2 \equiv \frac{n(n-1)} {\tilde R_n^2}\,,
\end{align}
which can be simplified to
\begin{align}
\rho_0^{n-1} = \frac{D-1}{n-1} \frac{1}{2(D+n-2)} \tilde L^2 Q^2\,,
\end{align}
which agrees with the result obtained before. When sitting at the vacuum the action in $D$-dimensions becomes
\begin{align}
S_D = \int d^D x_\mu \sqrt{-g_D} \left[ R_D -\rho_0 ^{-\frac{D n }{2(D-2)} } V(\rho_0)\right]\,.
\end{align}
\begin{theorem}
     Check that this reproduces $L^2_{\text{AdS}}$ (non-trivial).
\end{theorem}
\begin{theorem}
Redo all the above using the canonical scalar $\varphi$ defined in equation \eqref{withvarphi}. 
\end{theorem}

Let us discuss the intuition behind the fact that we get a vacuum. As mentioned before we have two competing components in the scalar potential, one coming from the curvature of the internal manifold and one from the flux. Without the presence of the flux the sphere would want to collapse to zero size to minimize the potential, however in this situation the energy density stored in the flux would increase as the sphere gets smaller.\footnote{Recall the the amount of flux cannot decrease because it is quantized: storing the same amount of flux in a smaller space clearly increases its energy density.} The vacuum sits at the point where these two components balance each other and the leftover energy density gives the cosmological constant. 

\textbf{Technical detour:} there exists an alternative Freund--Rubin solution that uses electric flux. To obtain that replace the theory so that it has a $D$-form
\begin{align}
S_{D+n} = \int d^{D+n} x \sqrt{-g_{D+n}} \left[R_{D+n} -\frac{1}{2} \frac{1}{D!} F_D^2\right]\,.
\end{align}
In this case we take the ansatz
\begin{align}
ds^2_{D+n} &= ds^2_D + \rho \,d \tilde s^2_n\,,\\
F_D &= Q \tilde \varepsilon_D\,,
\end{align}
where again $d \tilde s^2_n$ has unit volume.
\begin{theorem} Construct vacuum solution AdS$_D \times \Sigma_n$ using the equations of motion.\end{theorem}
\begin{theorem} Perform the dimensional reduction of the action and obtain the effective action for $\rho$ in $D$-dimensions. Find a problem in that $F_D^2 <0$.\end{theorem}

As a hint for the last exercise, consider this scenario: in classical mechanics one can consider the Lagrangian
\begin{align}
\mathcal L(r,\theta) = \frac{m}{2} (\dot r^2 + r^2 \dot \theta^2 )\,.
\end{align}
The equations of motion for $\theta$ imply that
\begin{align}
m r^2 \dot \theta = l\,,
\end{align}
with $l$ a constant. Then if we were to na\"ively replace this value in the Lagrangian we would get
\begin{align}
\mathcal L(r,\theta) = \frac{m}{2} (\dot r^2 +\frac{l^2}{m^2 r^2})\,,
\end{align}
which is the \emph{wrong} Lagrangian. The correct one is 
\begin{align}
\mathcal L(r,\theta) = \frac{m}{2} (\dot r^2 -\frac{l^2}{m^2 r^2})\,.
\end{align}
One can follow a practical attitude: either flip the signs by hand or work with the magnetic version (which we shall do in the following). Simply take
\begin{align}
G_n \equiv \star F_D
\end{align}
which up to a total derivative gives us back our original action.

\begin{theorem}
Now let us investigate an extension where there is an extra scalar field in the higher dimensional theory. This will prepare us for the real work (10d supergravity) discussed in the next chapter. Consider the following Lagrangian in $d=D+n$-dimensional spacetime:
\begin{equation}
    S=\int d^{D+n}x\sqrt{-g}\left[R -\frac{1}{2}(\partial\phi)^2- \frac{1}{2}{\frac{1}{p!}}e^{a\phi}F_p^2\right]\,.\label{eq:StartingAct}
\end{equation}
When compactifying this theory on a round $S^n$ it is consistent to truncate to $\phi$ and the volume scalar $\varphi$ only. The reduction Ansatz for the metric is then: 
	\begin{equation}
	d s^2 = e^{2\alpha\varphi}d s^2_{d} + e^{2\beta\varphi} d s^2_n\,,
	\end{equation}
where $d=D+n$ and $d s^2_n$ is the metric on the round $S^n$ of normalised curvature. To get $D$-dimensional Einstein frame gravity with a canonically normalised volume scalar we take as values for $\alpha$ and $\beta$ what we did in equation \eqref{alphabetantorus}:
When $n=p=D$, we can have both magnetic flux threading the $S^n$ and electric flux filling non-compact spacetime. Show that the resulting  effective potential then reads: 	
\begin{align}
 V(\varphi, \phi) =  \frac{1}{2}Q_E^2 e^{-a\phi + \left(\alpha d -\beta n\right)\varphi}  +  \frac{1}{2}Q_M^2e^{a\phi+ \left(\alpha d -\beta n\right)\varphi} - R_n e^{\left(\alpha d +n\beta-2\beta \right)\varphi}\,.	
\end{align}
Then show that whenever the product $Q_EQ_M$ is non-zero this potential allows an extremum describing an $ AdS_D\times S^D$ solution. 

\end{theorem}

\newpage

\section{Flux compactifications in String Theory}

The critical superstring theory reduces to 10-dimensional supergravity in the limit of vanishing string coupling and arbitrary small space-time curvatures and field gradients. We will be mostly concerned with the bosonic content of type IIA/IIB supergravity augmented with localised brane and orientifold sources and we therefore first introduce this. We want to emphasize that much of our terminology use originates from the string worldsheet (``Ramond, Neveu--Schwarz, D-brane, O-plane''), the reader does not need to master string perturbation theory to follow the technical details below. But we strongly advice to consult the many good books on the topic to understand where this all comes from. The physics background and intuition needed to follow the discussion below is that of general relativity and classical field theory.

The first point we stress is that supergravity is not the same thing as string theory. It arises only in a particular limit of parameters of string theory, that is at weak string coupling ($g_s \ll 1$) and a length scales much larger than the string scale $\ell_s$, see our earlier discussion around Figure \ref{pic:corner}. Supergravity has the virtue of living in a region of parameter space where we can actually perform computations in interesting backgrounds. We still need to keep in mind that eventually we need to embed supergravity solutions within string theory and this will put some constraints: symbolically one can picture the full action of string theory truncated to supergravity fields as:
\begin{align}\label{eq:stvssugra}
    \text{ST} = \text{SUGRA} + \sum_n c_n g_s^{n} + e^{-1/g_s} (\dots)  + \sum_n \left(\frac{\ell_s}{L}\right)^n d_n + e^{-L/\ell_s}(\dots) + \dots\,.
\end{align}
Here we do not presume to include all possible corrections appearing in the string theory action but highlight that perturbative as well as non-perturbative corrections in $g_s$ and $\ell_s/L$ play an important role. If nothing, equation \eqref{eq:stvssugra} serves as a cautionary tale for those venturing in string compactifications and model building to remember to keep all possible corrections under control.

While performing the computations leading to the estimation of the corrections is oftentimes prohibitive, in recent years a general approach to understanding the limitations of supergravity has been the Swampland program, and we will discuss some Swampland conjectures in the next section, especially those that arise in the special parametric regime we are interested in. The upshot is that not much String Theory knowledge will be required in the following. To go beyond the parametric regime some techniques are available, like worldsheet computations, holography, and string dualities. However, there is no guarantee that a full fledged analysis in a general regime is possible.

Compared with the previous section on Freund--Rubin vacua, compactification of String Theory modifies the situation as follows:
\begin{itemize}
\item There will be multiple $p$-forms and mutual couplings among them;
\item The scalar known as the dilaton will be present;
\item There will be sources for the $p$-forms, that is D-branes and O-planes (and more!);
\item We will obviously fix $D+n = 10$ (or $D=11$ for the case of M-theory).
\end{itemize}

\subsection{10d Supergravity}
When dealing with type II strings we sometimes work in the democratic formulation \cite{Bergshoeff:2001pv} and  double the number of $p$-forms, considering at the same time electric and magnetic forms. Let us first remind ourselves which forms are present in the case of type IIA and type IIB string theory
\begin{align}
&\text{IIA: } F_0,\, F_2,\, F_4,\, F_6,\, F_8,\, F_{10},\, H_3,\, H_7\,,\\
&\text{IIB: } F_1,\, F_3,\, F_5,\, F_7,\, F_9,\,  H_3,\, H_7\,.
\end{align}
In order to ensure that we have the right number of degrees of freedom we need to impose the condition
\begin{align}\label{eq:emduality}
F_n = (-1)^{\frac{(n-1)(n-2)}{2}} \star F_{10-n}\,,
\end{align}
where the Hodge star is done with the 10d string frame metric.
\begin{theorem}
    Prove consistency of the duality relation \eqref{eq:emduality} by verifying that
    \begin{align}
        \star^2 \gamma_p = (-1)^{p(D-p)+t} \gamma_p\,,
    \end{align}
    where $\gamma_p$ is a $p$-form in $D$-dimensions and $t$ is the number of time-like dimensions.
\end{theorem}
The 10d string frame action is
\begin{align}
S_{10} &= S_{\text{NSNS}}+ S_{\text{RR}}+ S_{source}\,.
\end{align}
Here NS stands for Neveu--Schwarz, R for Ramond, so the first two terms in the action govern the dynamics of massless fields in the NSNS and RR sectors of string theory respectively. The last term accounts for any localised source, like D-branes, orientifold planes, and more. Let us inspect the various parts of the action: the NSNS sector is given by
\begin{align}
    S_{\text{NSNS}} &= \frac{1}{2 \kappa_{10}^2} \int d^{10}x \sqrt{-g_{10}} e^{-2 \phi} \left(R-\frac{1}{2 \cdot 3!} H_{M N P} H^{M N P} +4 \partial_M \phi \partial^M \phi\right)\,.
\end{align}
The prefactor is given by
\begin{align}
    \frac{1}{2 \kappa_{10}^2 } = \frac{2 \pi}{(2 \pi \ell_s)^8}\,,
\end{align}
where $\ell_s = \sqrt{\alpha'} $ is the string length. 
The 3-form $H$ is the field strength for the B-field: $H_3=\d B_2$. 

This action can be argued to arise from assuming vanishing $\beta$ functions for the worldsheet CFT. However the Einstein-Hilbert term is not conventional because of the dilaton coupling and the physical frame is 10d Einstein frame instead. For that imagine the 10d dilaton $\phi$ has a vev $\phi_0$ and we consider fluctuations around that vev
\begin{equation}
\phi=\phi_0 + \bar{\phi}\,,
\end{equation}
where the bar indicates the dynamical part. Then write
\begin{align}
g^E_{MN} = e^{-\frac{1}{2} \bar \phi} g^s_{MN}\,.
\end{align}
where the superscripts denote Einstein and String frame. From now on we will drop these superscripts as well as the bar on the dynamical part of the dilaton. With this useful abuse of notation we arrive at:
\begin{align}\label{eq:NSNSE}
S^E_{\text{NSNS}}= \frac{1}{2\kappa^2_0 g_s^2} \int_{10}\sqrt{-g}\left(\mathcal{R}   -\frac{1}{2}(\partial\phi)^2 -e^{-\phi}\frac{1}{2}\star H_3\wedge H_3 \right)\,,
\end{align}
where $g_s=e^{\phi_0}$. As explained in Appendix \ref{app:forms} the form notation is such that $\int  \star H_3\wedge H_3 =\int \frac{1}{3!} H_{\mu\nu\rho} H^{\mu\nu\rho}\sqrt{-g}$. \begin{theorem}
Derive equation \eqref{eq:NSNSE} by relying on \eqref{eq:Weyl}.
\end{theorem}

Let us turn to the RR sector: as mentioned above this includes a set of $p$-form fields and their string frame action in the democratic formulation may be written as
\begin{align}
    S_{\text{RR}} &= -\frac{1}{2 \kappa_{10}^2 } \int d^{10}x \sqrt{-g} \,\frac{1}{4}\,\sum_{p} \star F_{p} \wedge F_{p}\,,
\end{align}
where the sum is over $p$ even for type IIA string theory and $p$ odd for type IIB string theory. It is possible to readily translate the RR action into string frame by using the following identity:
\begin{theorem}
    Verify that for a $p$-form $\gamma_p$ 
\begin{align}
\star_s \gamma_p = e^{\frac{p-5}{2} \bar \phi} \star_E \gamma_p\,.
\end{align}
\end{theorem} 

In the following we will employ a vast array of fluxes. We will consider both electric fluxes (that is fluxes with legs inside the non-compact manifold) and magnetic fluxes (that is fluxes with no legs inside the non-compact manifold). However whenever we try to compute the effective potential after dimensional reduction we will always rewrite electric fluxes as magnetic ones by just using the appropriate duality relations written before.

Given the set of $p$-forms appearing in type II string theories it is possible to write quite easily the set of Bianchi identities and equations of motion controlling their dynamics
\begin{align}
dF_n = 0\,, \qquad &n=0,1\,,\\
dF_n = H_3 \wedge F_{n-2}\,, \qquad &n>1\,.
\end{align}
This is the complete set of Bianchi identities and equations of motions because, given the duality relations \eqref{eq:emduality}, the Bianchi identity for $F_n$ gives the equations of motion for $F_{10-n}$ and vice versa. In terms of gauge potentials we can write the field strengths $F_n$ and $H_3$ as
\begin{align}
&F_1 = dC_0\,,\\
&F_2 = dC_1 + F_0 B_2\,,\\
& F_3 = dC_2 +B_2 \wedge F_1\,,\\
& \vdots\\
&H_3 = dB_2\,.
\end{align}
The equations of motion for the $H_3$ field are
\begin{align}
d(\star_s e^{-2 \phi} H_3) = -\frac{1}{2} \sum_n \star_s F_n \wedge F_{n-2}\,.
\end{align}
So far we have left out the sources and in String Theory there is a plethora of sources that can be considered: D$p$-branes, O$p$-planes, NS5-branes, KK-branes and more. Let us consider the impact of D-branes and O-planes in the equations of motion and Bianchi identities:
\begin{itemize}
\item D$p$-branes and O$p$-planes are electrically charged under $C_{p+1}$, which means that the equations of motion get modified to
\begin{align}
d(\star_s F_{p+2})  = \dots + Q_{\text{D}p/\text{O}p}\,\delta_{9-p}(\text{D}p/\text{O}p)\,,
\end{align}
where the dots include the aforementioned terms and $\delta_m$ is a delta function $m$-form with legs transverse to the world-volume of the source. The term $Q_{\text{D}p/\text{O}p}$ is included to account for the charge of the source. We will return to this point later in the lectures;
\item Using again the duality relations \eqref{eq:emduality} one sees that D$p$-branes and O$p$-planes are magnetically charged under $C_{7-p}$, which brings a modification of the Bianchi identities
\begin{align}
dF_{8-p} &= \dots +  Q_{\text{D}p/\text{O}p}\,\delta_{9-p}(\text{D}p/\text{O}p)\,,
\end{align}
where again the dots simply include the terms already present in the equations without sources. Interestingly, given that $F_5$ is self-dual, one finds that the D3-brane is a dyon;

\item F1-strings (that is the fundamental string) and NS5-branes are electric and magnetic sources respectively for $B_2$, which brings to the new Bianchi identities and equations of motion
\begin{align}
dH_3 &= Q_{\text{NS5}}\,\delta_4 (\text{NS5})\,,\\
d(\star_s e^{-2\phi} H_3 ) &= \dots + Q_{\text{F1}}\,\delta_8(\text{F1})\,;
\end{align}
\item All extended objects carry some energy density and therefore modify the energy-momentum tensor. This adds some source terms to the Einstein equations. Moreover they couple to the dilaton thus introducing some source terms in the dilaton equations of motion as well.

\end{itemize}
Finally, a brief comment: when one builds flux vacua we request that all extended objects fill all 4d directions in order to preserve 4d Poincar\'e symmetry. If one nonetheless adds sources which do not fill 4d, then one is looking at excitations around the vacuum. For example a D3 brane wrapped around 3 compact directions corresponds to a particle state in the vacuum. Many D3 branes wrapped around the same 3 compact dimensions could become a black hole, and so on. 

\paragraph{O-planes: a small practical guide for supergravity.}

O-planes are common in string compactifications and in  Appendix \ref{sec:oplaneapp} we have provided a quick summary of what they are for readers familiar with the basics of String Theory. In the bulk of the text we provide the essential tools to be able to perform computations in backgrounds that include these sources. The most crucial property is that they correspond to objects of negative tension, without destabilising the theory. This is not something we are used to in ordinary quantum field theory.  

Without going into too many details about the worldsheet point of view we can say that O-planes are located at fixed points of  $\mathbb Z_2$ involutions $\sigma$. This involution  involves a change of world-sheet orientation and a space involution $\mathcal R$ and it is the set of fixed points under $\mathcal R$ that determine the location of the orientifold planes.  Once orientifold planes are present one needs to be careful and consider their impact on various supergravity fields, and one finds the following transformation laws
\begin{align}
&\sigma(H_3)  = - H_3\,,\\
&\sigma(F_i) = \alpha (F_i)\,, \qquad \text{for O2/O3/O6/O7-planes}\,,\\
&\sigma(F_i) = -\alpha (F_i)\,, \qquad \text{for O0/O1/O4/O5/O8/O9-planes}\,.
\end{align}
Here $\alpha$ is the operator that reverses all indices, so for instance
\begin{align}
&\alpha(F_{m n}) = F_{nm}\,,\\
&\alpha(F_{mnp}) = F_{pnm}\,.
\end{align}
After rearranging all the indices one finds that
\begin{align}
\alpha(F_n) = (-1)^{\lfloor \frac{n}{2}\rfloor} F_n\,,
\end{align}
where $\lfloor \cdot \rfloor$ is the floor function. When it comes to the dilaton and metric they are simply invariant under the orientifold involution $\sigma$. This is roughly all the information needed for flux compactifications concerning orientifold planes. Basically in addition to adding source terms to the equations of motions and Bianchi identities orientifold planes also restrict the fluxes that are available as they need to obey the aforementioned transformation rules. Let us discuss in a simple example orientifold planes. 

\textbf{Example:} take a $\mathbf{T}^3$ compactification. Call the coordinates on the torus $\theta_a$ with  $a=1,2,3$, and we choose them to take values in the interval $[0,1]$. The general internal metric is
\begin{align}
ds^2_{\mathbf{T^3}} &= \mathbb M_{ab} d\theta^a d\theta^b\,.
\end{align}
Consider an orientifold involution with $\mathcal R$ acting as $\theta_m \rightarrow - \theta_m$. There are fixed points whenever any of the coordinates $\theta_m$ is equal to $0$ or $1/2$, which means that in total there are $8$ fixed points. This solution with O6-planes has the 10d metric
\begin{align}
ds_{10}^2 = ds_7^2 + \mathbb M_{ab} d\theta^a d\theta^b\,,
\end{align}
and notice that because of the involution there are no massless KK-vectors, that is in
\begin{align}
d\theta^a +A_\mu^a(x^\mu) dx^\mu\,,
\end{align}
the field $A_\mu^m(x^\mu)$ is removed because it does not transform appropriately under $\sigma$.

In the following we will not need much more information about how orientifold planes work: just remember that the location of orientifold planes is determined by an involution (and therefore by topology) and that their charge is negative. For those interested in more details about the worldsheet point of view we refer to Appendix \ref{sec:oplaneapp} and references therein.

A crucial difference with adding D-branes into a manifold is that D-branes can be put on top of each other and we can contemplate adding $N$ D-branes with $N$ a number we decide. That is not true for O planes. The number of O planes (and thus their charge) is fixed by the number of fixed points under the spacetime involution, and therefore by the topology of the internal manifold. 

\subsection{10d equations of motion versus dimensional reduction}\label{sec:10d}

In this section we discuss the two approaches to find solutions discussed previously in the context of Freund--Rubin solutions;using the equations of motion of the 10d system and performing a dimensional reduction. We will employ them directly in the context of type II string theory solutions. For additional discussion see for instance the appendix of \cite{Danielsson:2009ff}. 

Our starting point is the 10d bosonic action in Einstein frame. Up to Chern--Simons terms the action is
\begin{align}
S_{10d} \sim \int d^{10}x \sqrt{-g} \left[R-\frac{1}{2} (\partial \phi)^2 -\frac{1}{2\cdot 3!} e^{-\phi} H_3^2 -\frac{1}{2}\sum_n e^{a_n\phi }\frac{F^2_n}{n!}\right]+S_{loc}\,.
\end{align}
Here $S_{loc}$ includes contributions from localized sources and $a_n = \frac{5-n}{2}$. We will choose not to work in the democratic formulation in the following. As a word of caution, notice that in type IIB string theory we have that $\star F_5= F_5$ on shell, which means that $F_5^2 = 0$. 
The action of localized sources takes the general form
\begin{align}\label{eq:Sloc}
S_{loc} = -T_p \int_{\Sigma_{p+1}} d^{p+1} x \sqrt{-\text{det}(g+\dots)} + \mu_p\int C_{p+1}\,.
\end{align}
The term $T_p$ is called the \emph{tension} and for D-branes it is positive while for O-planes it is negative\footnote{There also exist positive tension orientifolds but we do not discuss them here.}. The term $\mu_p$ is the charge of the source, and in particular in order to have a source that satisfies the BPS-bound we need to have
\begin{align}
T_p = \pm e^{\frac{p-3}{4}\phi} \mu_p\,.
\end{align}
The physics about this equation is simple: when two D-branes or O planes of the same charge are put next to each other in flat space, the gravitational attraction exactly cancels the electromagnetic repulsion. Similarly if a D$p$-brane is placed next to a O$p$-plane then the gravitational repulsion exactly cancels the electromagnetic attraction, since in our convention a O$p$-plane has opposite charge and tension to its D$p$-cousin.
\begin{theorem}
What happens when we put a brane next to an anti-brane? Or a D$p$-brane next to an anti-O$p$-plane? 
\end{theorem}
Note that O$p$-planes have fixed positions in space and cannot move; they are non-dynamical objects from the viewpoint of perturbative string theory. This is one reason they do not induce instabilities.
There are additional terms in the source action that we did not mention but the structure written in \eqref{eq:Sloc} is sufficient for our purposes. 

Let us start discussing the equations of motion when $\phi$ is constant. We have that
\begin{align}\label{eq:boxphi}
\Box \phi = 0 = -\frac{1}{2} e^{-\phi} \frac{H_3^2}{3!} + \sum_n \frac{a_n}{2} e^{a_n \phi} \frac{F_n^2}{n!} \pm \frac{p-3}{4}e^{\frac{p-3}{4} \phi} |\mu_p| \delta_{p+1}(\Sigma)\,.
\end{align}
\begin{theorem} Verify equation \eqref{eq:boxphi} by varying the action with respect to $\phi$. \end{theorem}
Now take a compactification scenario where $\mathcal{M}_{10} =\mathcal{M}_4\times \mathcal{M}_6$. 
The integrated version of the above equation of motion is:
\begin{align}\label{eq:dilinteg}
\int_{\mathcal M_6} d^6x \sqrt{g_6}\left[ -\frac{1}{2} e^{-\phi} \frac{H^2_3}{3!} + \sum_n \frac{a_n}{2} e^{a_n \phi} \frac{F_n^2}{n!} \pm \frac{p-3}{4}e^{\frac{p-3}{4} \phi} |\mu_p| \delta_{p+1}(\Sigma)\right]=0\,.
\end{align}
This equation should be reproduced by the lower-dimensional EFT. Similarly we can consider the internal Einstein equations with constant $\phi$ (assuming a vacuum solution):
\begin{align}
R_{ab} =& \sum_n \left[ -\frac{n-1}{16 n!} g_{ab} e^{a_n \phi} F_n^2 +\frac{1}{2(n-1)!} e^{a_n \phi} (F^2_n)_{ab}\right]+\\
& -\frac{1}{48} g_{ab} e^{- \phi} H_3^2 +\frac{1}{4} e^{- \phi} (H^2_3)_{ab}+\frac{1}{2}\left(T^{loc}_{ab} -\frac{1}{8} T^{loc} g_{ab}\right)\,.\nonumber 
\end{align}
Here $T^{loc}_{ij}$ is the energy-momentum tensor produced by the localized sources, and $T^{loc}$ its trace. The tensor reads:
\begin{align}
\text{4d: } &T_{\mu \nu}^{loc} = - T_p g_{\mu \nu} \delta(\Sigma)\,,\\
\text{6d: }&T_{ij}^{loc} = - T_p \Pi_{ij} \delta(\Sigma)\,,
\end{align}
where $\Pi_{ij}$ is the projector on $\Sigma$.
\begin{theorem}
Derive at least the flux terms in this trace-reversed Einstein equation. 
\end{theorem}

Now we will integrate over $\mathcal M_6$ and trace. In particular we need to use that
\begin{align}
\int_{\mathcal M_6} \delta(\Sigma) = 1\,,\\
\int_{\mathcal M_6} \Pi_{ij}\delta(\Sigma) = \frac{p-3}{6}g_{ij} \,,
\end{align}
where the last equation is valid for D$p$-branes and O$p$-planes in the simplest cases. The resulting equation, which is nothing but the 6d trace-reversed and integrated Einstein equation, is
\begin{align}\label{eq:einstint}
\int_{\mathcal M_6} d^6x \sqrt{g_6} \left[R_6 - \sum_n \frac{1}{8(n-1)!}e^{a_n\phi} F_n^2 -\frac{1}{16} e^{-\phi} H_3^2 +\frac{1}{8} T^{loc}\right] = 0\,.
\end{align}
\begin{theorem}
Check equation \eqref{eq:einstint}.
\end{theorem}

We can finally consider the 4d Einstein equations
\begin{align}\label{eq:einst4d}
G_{\mu \nu}^{(10)} &= \sum_n \frac{1}{2 n! }e^{a_n \phi} \left[n (F_n^2)_{\mu \nu}-\frac{1}{2}g_{\mu \nu} F_n^2 \right] +\frac{1}{12 }e^{- \phi} \left[3(H_3^2)_{\mu \nu}-\frac{1}{2}g_{\mu \nu} H_3^2 \right] +\frac{1}{2} T_{\mu \nu}^{loc}\,.
\end{align}
\begin{theorem}
Check  equation \eqref{eq:einst4d}
\end{theorem}
In the previous equation note that the terms $(F_n^2)_{\mu \nu}$ and $(H_3^2)_{\mu \nu}$ are actually zero because we are considering only magnetic fluxes. We will take the trace of and integrate the previous equation over 6d and use the relation
\begin{align}
R_4 = 2 \Lambda = 2 V_{on-shell}\,.
\end{align}
The scalar potential $V$ is the sum of various energy density terms
\begin{align}
V = V_R + V_H + \sum_n V_n +V_{source}\,,
\end{align}
where
\begin{align}
V_n &= \int_{\mathcal M_6} d^6x \sqrt{g_6} \frac{1}{2 n!} e^{a_n \phi} F_n^2\,, \qquad V_H= \int_{\mathcal M_6} d^6x \sqrt{g_6} \frac{1}{2 n!} e^{- \phi} H_3^2\,,\\
V_R &= -\int_{\mathcal M_6} d^6x \sqrt{g_6} R_6\,,\qquad \quad V_{source} = T_p\int_{\mathcal M_6} d^6x \sqrt{g_6} \, \delta(\Sigma)\,.
\end{align}
We will later show that 
\begin{align}
&\int_{M_6}d^6x \sqrt{g_6} \Box \phi = \dots \qquad \Longleftrightarrow \qquad \partial_\phi V = 0\,,\\
&\int_{M_6}d^6x \sqrt{g_6} R_6  = \dots \qquad \Longleftrightarrow \qquad \partial_{\text{volume}} V = 0\,.
\end{align}
In other words we have the following match
\begin{itemize}
\item The 4d trace and integrated Einstein equations \eqref{eq:einst4d} give the definition of the 4d scalar potential;
\item The integrated equation of motion of the dilaton \eqref{eq:dilinteg} gives the minimization of the potential with respect to the dilaton;
\item The integrated and traced 6d Einstein equations \eqref{eq:einstint} give the minimization of the potential with respect to the volume.
\end{itemize}
Therefore, regardless of the various complications of a particular compactification, the dilaton and the volume are universal scalars that allow us to extract some model independent information. The full 4d effective action will in general take the form \eqref{Lagrangian}.  The equations that determine the vacuum are then $\frac{\partial V}{\partial \phi^i} = 0$
and in the cases of de Sitter and  metastability requires one to compute the Hessian. 

One would expect that $\partial_{\phi_i} V = 0$ is equivalent to solving the full set of 10 equations of motion. However as we already anticipated the equivalence is restricted to only the integrated equations of motion so it is not one to one. We are thus coarse-graining and lose some information stored in the 10d equations of motion.  
%
%
%
%

We will now start analyzing the 4d scalar potential by extracting its dependence on volume and dilaton.

\subsection{Volume and dilaton dependence of 4d scalar potential}

We follow the treatment first pioneered in \cite{Hertzberg:2007wc, Silverstein:2007ac} and later more extensively in \cite{Danielsson:2009ff, Wrase:2010ew, VanRiet:2011yc}. 

We start by writing the 10d string frame metric as
\begin{align}\label{eq:10dmetric}
ds^2_{10} = \tau^{-2} ds_4^2 +\rho\, ds^2_6\,.
\end{align}
We take the metric $ds_4^2$ to be in Einstein frame and $ds^2_6$ has unit volume. With this choice the total volume of $\mathcal M_6$ in 10d string frame is
\begin{align}
\text{vol}(\mathcal M_6) = \rho^3\,.
\end{align}
If we go to 4d Einstein frame this fixes
\begin{align}\label{eq:dilatonscale}
\tau = \rho^{3/2} e^{-\phi}\,.
\end{align}
\begin{theorem} Prove that for a compactification on a $D$-dimensional manifold one has that
\begin{align}\label{tau}
    \tau^{8-D} = \rho^{D/2}e^{- 2\phi}\,.
\end{align}
which generalizes equation \eqref{eq:dilatonscale}.
\end{theorem}

Alternatively we can write the 10d metric as
\begin{align}
ds_{10}^2 = \tau^{-2} \tau_0^{2} ds_4^2 + \rho ds_6^2\,,
\end{align}
where $\tau_0$ is the vev of $\tau$. Then the units in 4d are the same as 10d Planck units and $M_P^{(4)} = \tau_0$.

It is possible to show the scaling behavior of the various terms in the potential with respect to $\rho$ and $\tau$ and the result is
\begin{align}
\label{eq:VRscale} V_R &= U_R \rho^{-1} \tau^{-2}\,,\\
\label{eq:VHscale}V_H &= U_H \rho^{-3} \tau^{-2}\,,\\
V_n^{RR} &= U_n \tau^{-4} \rho^{3-n}\,,\\
\label{eq:Vsource}V_{source} &= U_s \tau^{-3} \rho^{\frac{p-6}{2}}\,.
\end{align}
Here $U_R$, $U_H$, $U_n$, and $U_s$ do not depend on $\tau$ and $\rho$.
A remarkable aspect is that the scalings hold for any model, thus allowing us to extract model independent information from this analysis of dilaton and volume. 

\paragraph{Example:} Let us consider in detail the case of $V_R$. In 10d string frame one has
\begin{align}
    S \sim \int d^{10}x e^{-2\phi} \sqrt{-g_{10}}\,R_{10}\,.
\end{align}
We can decompose the 10d Ricci scalar as
\begin{align}
    R_{10} = R_4 + R_6 + \dots\,,
\end{align}
where the dots include terms that are unimportant in the determination of the scalings. Then one gets the potential
\begin{align}
    V_R = \textcolor{red}{-} \int_{\mathcal M_6} d^{6}x \sqrt{-g_4}\sqrt{g_6} e^{-2\phi }R_6\,.
\end{align}
Pay attention to the sign in $V_R$.
By using the form of the 10d metric \eqref{eq:10dmetric} as well as the relation \eqref{eq:dilatonscale} it is possible to extract the scalings of the various components appearing in $V_R$
\begin{align}
    R_6 &\sim g_6^{-1} \sim \rho^{-1}\,,\\
    \sqrt{g_6} &\sim \rho^3\,,\\
    \sqrt{-g_4} & \sim \tau^{-4}\,,\\
    e^{-2\phi} &= \tau^2 \rho^{-3}\,.
\end{align}
Using these relations we find that
\begin{align}
    V_R \sim \rho^{-1}\rho^{3}\tau^{-4 }\tau^2 \rho^{-3} = \rho^{-1}\tau^{-2}\,.
\end{align}
Following a similar strategy it is possible to obtain the other densities. 
\begin{theorem}
    Prove the scaling properties of the various components of the scalar potential in equations \eqref{eq:VRscale}-\eqref{eq:Vsource}.
\end{theorem} 
One important observation is that the potential discussed so far, that is 
\begin{align}
    V = V_R + V_H + V_{source}+ \sum_n V_n\,,
\end{align}
is merely the \emph{tree-level potential}. There will be a plethora of corrections, both $\alpha'$ and $g_s$ corrections as well as non-perturbative corrections. It is therefore of utmost importance in every solution found to check whether such corrections are sufficiently suppressed. 

We are now ready to prove: 

\begin{theorem} Check that the minimization conditions of the scalar potential with respect to $\rho$ and $\tau$ give our previous integrated 10d equations of motion \eqref{eq:dilinteg} and \eqref{eq:einstint}. \end{theorem}

The simple analysis of the scalar potential we found above already allows us to discuss two important aspects of the Swampland program, discussed in the next section:
\begin{itemize}
    \item [i.] dS no-go theorems;
    \item [ii.] no-go (or potentially go) theorems for scale separated AdS solutions.
\end{itemize}
Surprisingly much of these conjectures can be understood based purely on the scaling of the energy densities with respect to the 10d dilaton and volume scalar, and thus $\rho$ and $\tau$. 

But before going into that let us consider an important warm-up example: compactifications with fluxes down to classical Minkowski vacua. Since Minkowski space has vanishing vacuum energy, it is on the border between dS space and scale-separated AdS space. At first sight it is difficult, if not impossible, to find Minkowski vacua with fluxes since it would require an exact cancellation between positive flux energies and negative energies, either from O-planes or internal curvature. We have seen for Freund-Rubin vacua that such a cancellation does not occur, but interestingly the equations of motion of 10d supergravity with fluxes and O-planes can sometimes naturally enforce such a classical cancellation as we will now demonstrate. The basic references that uncovered this mechanism first are \cite{Giddings:2001yu,Dasgupta:1999ss}. 

\subsection{Minkowski vacua from IIB with fluxes and O3/D3-sources}

We consider the case of type IIB compactifications with $F_3$, $H_3$ fluxes as well as O3-planes and D3-branes. This means the O3/D3 sources fill 4d spacetime and the fluxes are threading internal 3d submanifolds. Note that space filling O3 planes are consistent with the parity of 3-form fluxes inside the internal dimensions.
\begin{theorem}
Consider a six-torus with O3 sources at fixed positions by using proper involutions. Now check that the 3-form fluxes have the correct parity to be present.
\end{theorem}
We take the case of a Ricci-flat metric $g_6$, such as for instance a Calabi--Yau metric,  and therefore the potential component $V_R$ is absent. The scalar potential becomes
\begin{align}
    V = \underbrace{U_H}_{>0} \rho^{-3} \tau^{-2} + \underbrace{U_3}_{>0} \tau^{-4} + \underbrace{U_{source}}_? \tau^{-3}\rho^{-3/2}\,.
\end{align}
Note that given that they correspond to magnetic flux energy densities both $U_H$ and $U_3$ are positive, however it is not possible to say anything at the moment about $U_s$ given that its sign will depend on the relative number of O3-planes and D3-branes.\footnote{Remember that O-planes contribute negative energy density.} Let us minimize the potential
\begin{align}\label{eq:rhopartrho}
    \rho\, \partial_\rho V = 0\,, \qquad \Longrightarrow \qquad -3V_H -3/2 V_{source} = 0\,.
\end{align}
Notice that a solution is possible only if $V_{source}< 0 $, that is in a situation where the orientifold energy overcomes the D-brane energy. The other minimization condition is
\begin{align}
    \tau\, \partial_\tau V = 0 \,, \qquad \Longrightarrow \qquad -2V_H -4V_3-3V_{source} = 0\,.
\end{align}
Replacing in this equation $V_{source}$ using \eqref{eq:rhopartrho} one finds that $V_3 = V_H$. Combining all these conditions we find that at the minimum
\begin{align}
    V = V_3 +V_H+ V_{source} = V_H + V_H -2 V_H = 0\,.
\end{align}
So only Minkowski solutions are possible in this scenario: there is indeed a perfect cancellation between the positive energy sourced by the fluxes and the negative energy sourced by the orientifold planes. How is this balance achieved? Following \cite{Giddings:2001yu, Blaback:2010sj} we claim that the potential is actually a perfect square
\begin{align}
    V = (\rho^{-3/2}\tau^{-1} U_H -U_3 \tau^{-2})^2\,,
\end{align}
which happens when $U_{source} = -2 U_3 \,U_H$. To justify this magic let us go back to the 10d picture. The scalar potential computed from the 10d string frame, is
\begin{align}
    V = \int_{\mathcal M_6 } \frac{1}{2} \star_6 F_3 \wedge F_3 + \frac{1}{2}e^{-2 \phi} \star_6 H_3 \wedge H_3 - |\mu_3| e^{-\phi} \delta(x)\,.
\end{align}
Notice that for a three-form $\alpha$ in a six-dimensional space we have that $\star_6\alpha \wedge \alpha = \sqrt{g_6}\,\alpha^2/3!$, and $\mu_3$ is the net negative tension of an O3/D3 mixture. It is important to use that there is a tadpole condition: by taking the Bianchi identity for the type IIB five-form we find that
\begin{align}
    dF_5 = H_3 \wedge F_3 + Q_3 \,\delta(x)\,.
\end{align}
Requiring that the right-hand side of the equation integrates to zero on $\mathcal M_6$ one finds that
\begin{align}
    \int_{\mathcal M_6} H_3 \wedge F_3 = -Q_3 = \pm |\mu_3|\,.
\end{align}
In the last passage we equated the charge and tension of the sources because they are BPS objects. Therefore using this relation we can rewrite the scalar potential as 
\begin{align}
    V &= \int_{\mathcal M_6 } \frac{1}{2} \star_6 F_3 \wedge F_3 + \frac{1}{2}e^{-2 \phi} \star_6 H_3 \wedge H_3 \mp e^{-\phi} H_3 \wedge F_3 =\\
    &= \int_{\mathcal M_6 }\frac{1}{2} \star_6 (\star_6 F_3 \mp e^{-\phi} H_3) \wedge (\star_6 F_3 \mp e^{-\phi} H_3)\,,
\end{align}
which is a perfect square. The minimum sits at 
\begin{align}\label{ISD}
    \star_6 F_3 = \mp e^{-\phi} H_3\,,
\end{align}
which means that the fluxes are (anti) imaginary self-dual (oftentimes shortened to (A)ISD fluxes). Indeed write $G_3 = F_3 -i e^{-\phi}H_3$, then the previous condition can be rewritten as $\star_6G_3 =\pm i G_3$. These solutions are called \emph{no-scale} Minkowski solutions since there are still some flat directions. Indeed writing the potential in terms of $\rho$ and $\tau$ one can see that
\begin{align}
    V = \rho^{-3/2} \tau^{-3}\left[U_H \tau^{1/2}\rho^{-3/4}-U_3 \frac{1}{\tau^{1/2}\rho^{-3/4}}\right]^2\,.
\end{align}
Given that it appears as an overall prefactor the combination $\rho^{-3/2} \tau^{-3}$ is not fixed and therefore free. The 10d viewpoint on the presence of a flat direction is as follows: the condition \eqref{ISD}
is unaffected by a rescaling of the volume of $\mathcal M_6$, thus implying that the volume is a flat direction. To check this recall that the Hodge star in 6d acting on a 3-form involves the following combination of the Levi-Civita tensor
\begin{align}
    {\epsilon_{abc}}^{def} \sim \epsilon_{abcdef} (g^{-1})^3 \sim \sqrt{g_6} (g^{-1})^3\,,
\end{align}
and it is rather easy to check that the last combination is invariant under a rescaling of the internal volume.

The same reasoning outlined so far holds for a case with a mixture of O3-planes and D3-branes given that
\begin{align}
    \text{sgn } Q_{\text{D3}} &= - \text{sgn } Q_{\text{O3}}\,,\\
    \text{sgn } T_{\text{D3}} &= - \text{sgn } T_{\text{O3}}\,,
\end{align}
so a mixture of D3-branes and O3-planes is still BPS since $|T_{mix}| = |Q_{mix}|$. However some care needs to be taken: an excessive amount of D3-branes such that $T_{mix}>0$ would destroy the vacuum. 

\begin{theorem}
Now that we achieved no-scale Minkowski vacua in $D=4$ repeat this exercise for Minkowski vacua in $D=3, 5, 6, 7$ where space filling Op planes have a charge canceled by a combination of $H_3$ and $F_{6-p}$ flux. The answers can be found in \cite{Blaback:2010sj}. 
\end{theorem}

\begin{theorem}
Consider a 6-torus with angles $\theta^a$ and an O3 involution that acts as $\theta^a\rightarrow -\theta^a$. How many O3 planes do you count? Now consider the simple $F_3$ flux $F_3 = f d\theta^1\wedge d\theta^2\wedge d\theta^3$. What is the corresponding $H_3$ flux and at what value is the dilaton stabilised? 
\end{theorem}
\subsection{A brief look at uplifting and anti-branes}
An even more interesting case is the combination O3/$\overline{\text{D3}}$. One can check that
\begin{align}\label{upliftGKP}
    V = V_{GKP} + 2 T_{\overline{\text{D3}}}\,.
\end{align}
With $V_{GKP}$ we denote the previously derived scalar potential for the no-scale Minkowski vacua. 
The last term is positive and naturally brings about the idea of an $\overline{\text{D3}}$-uplift \cite{Kachru:2003aw}.
\begin{theorem}
Check this equation. 
\end{theorem}
\begin{theorem}\label{th:upliftGKP}
Check that the potential \eqref{upliftGKP} has no vacuum since the volume scalar is runaway.
\end{theorem}

Instead of just computing the factor of two appearing in front of the tension, it is instructive to understand it on physical grounds. The extra energy compared with the no-scale vacuum is due to the fact that in addition to the extra tension coming from the $\overline{\text{D3}}$-brane it is necessary to add some flux to cancel the tadpole, and the extra flux and the tension give the same contribution and combine. To see this one can think of adding to an existing solution a pair of a  $\overline{\text{D3}}$-brane and D3-brane. The total pair has no charge and will not upset the tadpole and thus the fluxes remain the same. But the tensions add up giving the factor of two. This argument was first presented in \cite{Maldacena:2001pb}.  

\begin{figure}[htp]
    \centering
    \includegraphics[width=14cm]{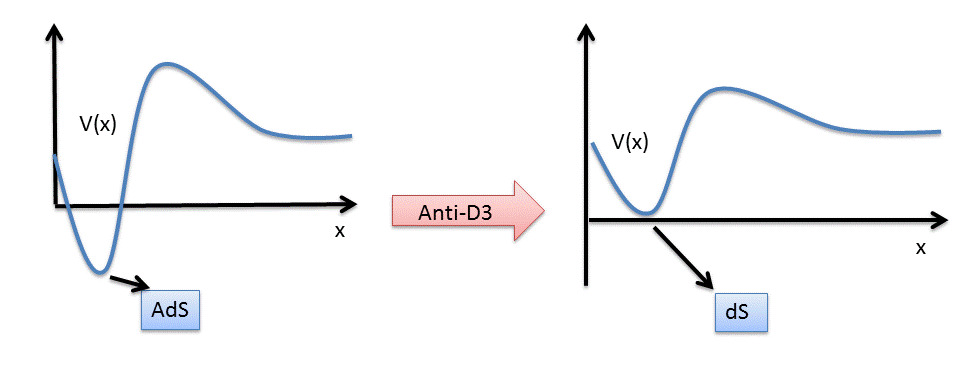}
    \caption{Uplifting AdS to dS using anti-D3 branes at a warped throat. }
    \label{fig:uplift}
\end{figure}

Some word of caution regarding the $\overline{\text{D3}}$-uplift: it is usually required that the $\overline{\text{D3}}$-brane sits at the bottom of a warped throat, that is a region of the internal manifold with high warping. To define warping let us slightly modify the 10d metric to
\begin{align}
    ds_{10} = e^{2A} ds_4^2 + ds_6^2\,.
\end{align}
Warping is accounted for by the function $A$, a function of the \emph{internal} coordinates. In a highly warped region one has $-A\gg1$ implying that $e^{2A} \rightarrow 0$. The implication for the $\overline{\text{D3}}$-uplift scenario is that in a highly warped region a redshift of the $\overline{\text{D3}}$-brane tension occurs
\begin{align}
    \mu_{\overline{\text{D3}}} \int_{\overline{\text{D3}}} d^4x \sqrt{-g_4}  \sim \mu_{\overline{\text{D3}}}\, e^{4A}\,.
\end{align}
This is a desirable feature since it was understood that taking into account the backreaction of the O3/D3 planes properly, tends to create such warping and it can even lead to exponential hierarchies, in the right geometry \cite{Giddings:2001yu}. Hence the uplift energy can be very substringy and one can hope that it creates an almost tunably small SUSY-breaking energy. This is the idea behind the famous KKLT scenario \cite{Kachru:2003aw}: quantum corrections to the no-scale Minkowski vacua create a SUSY AdS vacuum, which is then uplifted to a meta-stable dS on the condition there is a tunable small uplift, as depicted in Figure \ref{fig:uplift}

Note that the quantum corrections to the no-scale Minkowski vacua are required since otherwise the uplift leads to a runaway of the volume scalar instead a of stabilisation into a vacuum, as shown in exercise \ref{th:upliftGKP}.

The reader could worry that this seems very ad hoc; why would an anti-D3 sit exactly at the region of high warping, known as the ``tip of the throat''? 
\begin{theorem}
Explain why anti-D3 branes will be drawn towards the tip of the throat dynamically. 
\end{theorem}

\subsection{A brief look at 4d supergravity}
Finally, we discuss the 4d supergravity viewpoint on these solutions. 4d supergravity only arises as an effective field theory if SUSY is broken well below the KK scale. This means the compactification manifold by itself cannot break SUSY entirely and for minimal supersymmetry this requires it to be (conformal) Calabi-Yau with the flux and source ingredients we have used (3-form fluxes and D3/O3) \cite{Grana:2005jc,Tomasiello:2022dwe}.   For such manifolds  we have two kind of moduli that descend from the metric and fluxes
\begin{itemize}
    \item Complex structure moduli (that is moduli that control deformations of the complex three form $\Omega_3 \sim dz_1 \wedge dz_2 \wedge dz_3$);
    \item K\"ahler moduli (that is moduli that control deformations of the two form $J \sim dz_1 \wedge d\bar z_1+dz_2 \wedge d\bar z_2+dz_3 \wedge d\bar z_3$).
\end{itemize}
Introduction of fluxes induces in the 4d supergravity action the Gukov--Vafa--Witten superpotential
\begin{align}
    W_{GVW} = \int_{\mathcal M_6} \Omega_3 \wedge G_3 \,.
\end{align}
This superpotential depends solely on the complex structure moduli (via $\Omega_3$) and the dilaton (via $G_3$). Given any superpotential $W$ the 4d scalar potential can be computed via the canonical supergravity formula
\begin{align}
    V = e^K \left(K^{i \bar \jmath} D_i W D_{\bar \jmath} \overline W -3 |W|^2\right)\,.
\end{align}
Here $K$ is the K\"ahler potential and $K^{i \bar \jmath}$ is the inverse K\"ahler metric $K_{i \bar \jmath}=\partial_i\partial_{\bar\jmath}K$. Moreover the covariant derivative of the superpotential is defined as
\begin{align}
    D_i W = \partial_i W - W\partial_i K \,.
\end{align}
In our scenario let us differentiate between K\"ahler and complex structure moduli, calling $\phi_\alpha$ the K\"ahler moduli and $\phi_m$ the complex structure moduli. Given the structure of the K\"ahler potential one can show (we will not do that here) 
\begin{align}
    K^{\alpha \bar \beta } \partial_\alpha K \partial_{\bar \beta }K = 3\,,
\end{align}
which, together with the fact that the superpotential does not depend on the K\"ahler moduli, leads to a highly simplified form of the scalar potential
\begin{align}
    V = e^K \left(K^{m \bar n}D_m W D_{\bar n} \overline W\right)\,,\label{noscale}
\end{align}
which is a sum of squares matching therefore our previous discussion.
\begin{theorem}
Verify equation \eqref{noscale}
\end{theorem}

Given that all K\"ahler moduli enter in the scalar potential only via the $e^K$ factor we have that these scalars are not stabilized at all: this extends our previous result about the overall volume being a flat direction to the whole set of K\"ahler moduli. In the literature there have been attempts to stabilize them by exploiting some quantum corrections present in the compactification, most notably in the KKLT \cite{Kachru:2003aw} and LVS \cite{Balasubramanian:2005zx} scenarios. 

Let us briefly recall how the KKLT scenario of \cite{Kachru:2003aw} works: one considers a 4d EFT with superpotential
\begin{align}
    W = W_0 + A e^{i a \rho}\,.
\end{align}
Here $W_0$ is a constant that descends from the fluxes stabilizing complex structure and dilaton, $A$ and $a$ are constants and $\rho$ is a K\"ahler modulus. This simple superpotential works for cases in which the Calabi--Yau manifold has a single K\"ahler modulus. The imaginary part of $\rho$ is related to the internal volume of the Calabi--Yau. The exponential term in $\rho$ is produced by instantons in the internal space or gaugino condensation on D7-branes (\emph{aka} fractional instantons). If a solution to the F-term equation $D_\rho W = 0$ exists, we obtain a supersymmetric AdS vacuum. However in order to find such a solution one needs to balance the classical contribution from fluxes in $W_0$ against the quantum corrections. This obviously requires a lot of fine tuning in $W_0$ to make sure that it is sufficiently small which is one of the critiques leveled against this paradigm to build  Anti-de Sitter solutions. 

In this scenario one has that
\begin{align}
    L^2_{\text{AdS}} \approx \frac{1}{e^K |W|^2} \approx \frac{e^{2a t_0} t_0}{a^2 |A|^2}\,,\label{AdSKKLT}
\end{align}
where $t_0$ is a solution to
\begin{align}
    3 W_0 + A e^{-a t_0}(3+2a t_0) = 0\,.
\end{align}
\begin{theorem}
Derive this using the supergravity F-term equation $D_\rho W = 0$  . 
\end{theorem}
Worries about this procedure are explained in the next section.

\newpage
\section{Applications to the Swampland program}
\subsection{A brief look at the tadpole conjecture}

We ended the last section by explaining how quantum corrections could stabilise the K\"ahler moduli, which are left unstabilized at the classical level in IIB string theory with 3-form fluxes. If we forget for a moment about K\"ahler moduli we can ask ourselves the even more basic question of whether we can actually stabilize all complex structure moduli. One important statement regarding this is the swampland tadpole conjecture \cite{Bena:2020xrh}. Let us try to roughly summarize the conjecture: for a case where we have a large number $n_{mod}$ of moduli we will need more fluxes to stabilize them at a \emph{generic} point in moduli space. But unfortunately fluxes contribute to the tadpole
\begin{align}
    \int H_3 \wedge F_3 \equiv Q_{flux} = Q_{\text{O3}}-Q_{\text{D3}} \leq Q_{\text{O3}}\,.
\end{align}
The total O3-plane charge available is actually fixed by the topology of the compactification and therefore some care has to be taken not to overshoot. The tadpole conjecture states that the flux charge $Q_{flux}$ needed to stabilize $n_{mod} \gg 1 $ at a generic point in moduli space  satisfies the condition
\begin{align}
    Q_{flux} > \frac{1}{3}  n_{mod}\,,
\end{align}
yet the maximal O3-plane charge coming from topology seems to satisfy
\begin{align}
    Q_{\text{O3}}^{max} < \frac{1}{3} n_{mod}\,.
\end{align}
This clearly indicates a tension about the prospect of being able to stabilize all complex structure moduli. There have been several tests of the conjecture since its first appearance, see for instance \cite{Bena:2021wyr,Marchesano:2021gyv,Plauschinn:2021hkp,Lust:2021xds,Grimm:2021ckh,Grana:2022dfw,Tsagkaris:2022apo,Coudarchet:2023mmm}, see also \cite{Lust:2022mhk} for a stronger version of the conjecture.

\subsection{de Sitter}

Let us focus on possible dS solutions using fluxes and localized sources. Since IIB was giving us Minkowski solutions, at least with the easiest available ingredients, let us move to IIA and revisit the setup considered in \cite{Hertzberg:2007wc}; compactifications with $F_0$, $F_2$, $F_4$, $F_6$, and $H_3$ fluxes as well as O6-planes and D6-branes. The internal space is taken to be a Calabi--Yau threefold meaning that the internal curvature is taken to be zero. Minimization of the potential with respect to $\tau$ and $\rho$ gives
\begin{align}
    \label{eq:Vrho}&\rho \partial_\rho V = 0\,, \qquad \rightarrow \qquad  -3 V_H+ 3 V_0+ V_2 - V_4 -3V_6 = 0\,,\\
    \label{eq:Vtau}&\tau \partial_\tau V = 0\,, \qquad \rightarrow \qquad  -2 V_H- 3 V_{source}-4(V_0 +V_2 + V_4 +V_6) = 0\,.
\end{align}
\begin{theorem}
Prove equations \eqref{eq:Vrho} and \eqref{eq:Vtau}.
\end{theorem}
Note that the last equation admits a solution only if $V_{source}<0$ given that $V_H$ and all the $V_n$ are positive. Therefore similarly to the case of type IIB compactifications described previously we have that the charge of orientifold planes must be larger than the charge of D-branes. Using the minimization conditions one can write the potential at the minimum
\begin{align}
    V = -\frac{2}{9}V_2 - \frac{8}{9} V_4 -\frac{2}{3} V_6 <0\,,
\end{align}
finding therefore only AdS vacua or Minkowski in the particular case in which $F_2 = F_4 = F_6 =0$. Such Minkowski solutions indeed have been found in for instance \cite{Grana:2006kf} whereas the AdS solutions were for instance constructed in \cite{DeWolfe:2005uu} and we will come back to the AdS solutions later in these lectures. 

We also need to consider tadpoles, and in this scenario the relevant Bianchi identity is the one for $F_2$
\begin{align}\label{eq:F2Bianchi}
d F_2 = F_0 H_3 + Q_6 \delta(\text{D6}/\text{O6})\,.
\end{align}
\begin{theorem}
    Check that \eqref{eq:F2Bianchi} is the only Bianchi identity that gives a non-trivial tadpole condition.
\end{theorem}
If we integrate this over a 3-cycle perpendicular to the source we find
\begin{align}
    F_0 \int_{\Sigma_3} H_3 = - Q_6\,.
\end{align}
The right hand side is fixed by topology and cannot therefore be too large. Following a similar strategy one can show that
\begin{itemize}
    \item [i.] If there is no negative tension then the internal curvature has to be positive, that is $R_6 >0$. This in turn implies that $V_R <0$; This is part of a general result that negative internal curvature at tree-level is only possible with negative tension objects \cite{Douglas:2010rt};
    \item [ii.] de Sitter solutions in this scenario are possible only if $R_6<0$ (which implies $V_R>0$), $V_{source}<0$, and $F_0 \neq 0$.
\end{itemize}

\begin{theorem} Prove these two statements. \footnote{In case you do not find it, you can consult \cite{Haque:2008jz}, but keep in mind that the claimed dS solutions in that paper are flawed.}\end{theorem}
\begin{theorem}
Try to find no-go theorems in IIB compactified to 4d with O3-planes and all possible fluxes. 
\end{theorem}
This game is rather straightforward and one can obtain an interesting list of no-go theorems that way, see for example \cite{Wrase:2010ew, VanRiet:2011yc}.  The fact that such an analysis shows that orientifolds are always needed for de Sitter, despite delivering negative energy was well-known \cite{Maldacena:2000mw, deWit:1986mwo} and is usually referred to as the ``Maldacena--Nu\~{n}ez no-go''. Furthermore note that no-go theorems derived from solving the equations $\partial_{\tau}V=\partial_{\rho}V=0$ with the condition $V>0$ inform us what is minimally needed to find solutions. So no-go theorems are useful as a tool to learn what is needed to achieve the goals. In that regard one can also look at the $2\times 2$ Hessian $\partial_i\partial_j V$ and demand is does not have negative eigenvalues for positive $V$ critical points in the $\rho, \tau$ subspace. This by itself restricts strongly the further possibilities \cite{Shiu:2011zt, VanRiet:2011yc}.
\begin{theorem}
Since the Hessian in the 2d subspace spanned by $\rho$ and $\tau$ is only a submatrix of the full Hessian that includes all scalar field directions, why can one still infer that negative eigenvalues of the submatrix imply negative eigenvalues of the full matrix?
\end{theorem}

The previous statement about dS solutions in type IIA can be made more precise by saying that whenever $V>0$ then \cite{Hertzberg:2007wc}
\begin{align}\label{eq:slowrollIIA}
    \varepsilon \geq \frac{27}{13}\,.
\end{align}
Here $\varepsilon$ is the multi-field slow roll parameter
\begin{align}
    \varepsilon = \frac{G^{ij} \partial_i V \partial_j V}{V^2}\,.
\end{align}
The implication of \eqref{eq:slowrollIIA} is momentous: even slow-roll inflation is impossible. However, one needs to treat this inequality with caution since it does not exclude solutions with a brief period of cosmic acceleration as pointed out in for instance \cite{Emparan:2003gg}. In fact string theory derived potentials will always allow solutions with some mild cosmic acceleration and the argument is trivial. Consider just a reduction over a flat space with flux. This means that the potential energy is positive $V \sim \int_6 F^2$, yet the volume scalar is runaway. When canonically normalised such a potential is of the exponential form in the volume scalar $\varphi$
\begin{equation}
    V\sim \Lambda\, e^{c \varphi}\,.
\end{equation}
with $\Lambda>0$. Now imagine $c>0$ (in case $c<0$ we consider the field redefinition that flips the sign of the scalar)). We can consider a cosmology with an initial condition for which the scalar velocity points upward the potential. Then after a while the scalar has to turn and move backward again towards negative $\varphi$. At the moment of return, there must unavoidably be a period of cosmic acceleration. Yet such models are not useful for cosmology since the number of e-folds tends to be very small, and more importantly a runaway volume scalar (or dilaton) would be problematic for the validity of the lower-dimensional effective field theory. Computing the number of e-folds for such cosmologies in full fledged models is not easy and was carried out in for instance \cite{Marconnet:2022fmx, Blaback:2013fca}. 

The statement
\begin{align}
    \varepsilon \gtrsim \mathcal O(1)\,,
\end{align}
is true in many classical flux compactifications. It is possible to evade this if some very specific requirement are met: in the case of type IIA supergravity with classical ingredients one needs all the ingredients to make a dS solution, that is negative curvature, orientifold planes, and Romans mass. In the literature there are further results about the presence of de Sitter solutions in classical flux vacua which started from the original works \cite{Silverstein:2007ac, Flauger:2008ad, Caviezel:2008tf, Danielsson:2009ff} and most further developments were reviewed in \cite{Danielsson:2011au, Andriot:2022yyj, Andriot:2019wrs}. We can summarize the bare essentials of all this work in two statements
\begin{itemize}
    \item There are no de Sitter critical points at small $g_s$ and large volume;
    \item There are a few de Sitter critical points at $g_s \sim \mathcal O(1)$ and $\text{vol} \sim \mathcal O(100)$, but all of these come with tachyons.
\end{itemize}
The status of de Sitter solutions in classical flux compactifications is therefore unclear. Even if this seems discouraging here are some further remarks
\begin{itemize}
    \item [i.] There is some understanding of the tachyons, see for instance \cite{Danielsson:2012et,Andriot:2021rdy};
    \item [ii.] There are some arguments that suggest why large volume and small $g_s$ are unlikely, see for instance \cite{Junghans:2018gdb, Banlaki:2018ayh}. 
\end{itemize}
Let us sketch a proof for the second statement. We know that a de Sitter solution in type IIA with O6-planes requires $R_6 <0$ and $F_0 \neq 0$. Notice that $F_0$ and $H_3$ are bounded by tadpoles and therefore cannot be made too large. On the other hand $F_4$ is not bound by tadpoles. Typically to get small $g_s$ and large volume it is necessary to turn on the fluxes to large values, so calling for instance $N = \int F_4$ it is necessary to have $N \rightarrow \infty$. Hence in this situation we have that
\begin{align}
    V_4 \sim N^2 \tau^{-4}\rho^{-1}\,,
\end{align}
and we moreover have that
\begin{align}
    &V_R \sim U_R \rho^{-1}\tau^{-2}\,,\\
    &V_{\text{O6}} \sim T_6 \tau^{-3}\,.
\end{align}
If we want to be able to find a minimum, that is a solution of $\partial V = 0$, in the limit $N \rightarrow \infty$, all the various terms have to balance against each other. Parametrizing $\rho$'s and $\tau$'s scalings with $N$ as
\begin{align}
    \rho \sim N^a\,, \qquad \tau \sim N^b\,,
\end{align}
we find that
\begin{align}
    V_4 \sim N^{2-a-4b}\,, \qquad V_R \sim N^{-a-2b}\,, \qquad V_{\text{O6}} \sim N^{-3b}\,.
\end{align}
In order to have a dS vacuum we need all terms to balance each other, and in particular we need that the O6-plane and curvature contributions scale in the same fashion. Requiring this imposes that $a=b=1$, however in this condition we have that the string coupling would scale as
\begin{align}
    g_s = \tau^{-1}\rho^{\frac{3}{2}} \sim N^{\frac{1}{2}}\,.
\end{align}
So in the large $N$ limit the string coupling becomes infinite.

All these investigations on the nature of de Sitter space inspired the dS-swampland criterion \cite{Ooguri:2018wrx}.\footnote{The condition on the second derivative, which is the reason behind the word `refinement', was first suggested in \cite{Andriot:2018wzk} and \cite{Garg:2018reu} based on the tachyons present in the classical dS solutions known at the time. But strictly speaking this cannot be argued since none of the unstable classical dS solutions known are self-consistent, in the sense that they live in a regime of small coupling and large volume.} The statement is as follows: 
\begin{tcolorbox}
In asymptotic regime of field space we must have either
\begin{equation}
    |\nabla V| \geq c\, V\,, \label{dSconj1}
\end{equation}
    
or
\begin{equation}
        \min( \nabla_i \nabla_j V) \leq -c'\, V \label{dSconj2}
\end{equation}
where $c$ and $c'$ are positive and of order 1.
\end{tcolorbox}
Two brief comments on this statement: first, it is phrased in units where $M_P = 1$. Second, an appropriate field basis is chosen such that at the minimum $\phi_0$ the moduli space metric is $G_{ij}(\phi_0)=\delta_{ij}$. This statement is compatible with all the studies in the literature we mentioned before. Asymptotic regime means classical dS space, however all the examples we cited before lack any sort of parametric control (and they have tachyons). There is also a famous case of a scalar with an extremal point with positive energy: the Higgs field \cite{Denef:2018etk}. However, at the origin of the Higgs potential, even if $V>0$ and the first derivative is zero, the potential is tachyonic and the negative squared mass is of order 1.\footnote{In fact there is no clear understanding of the Higgs potential near the origin, but there is a Swampland conjecture, called Festina Lente \cite{Montero:2019ekk} (see also \cite{Guidetti:2022xct}), in spirit similar to Weak Gravity, which forbids the origin to have a locally stable minimum \cite{Montero:2021otb}, see also \cite{Lee:2021cor}.} The arguments behind the conjecture involve the following ingredients
\begin{itemize}
    \item Distance conjecture, which we briefly explained around equation \eqref{eq:distanceconj}.
    \item The fact that the dS-entropy scales as $S_{\text{dS}} \sim H^{-2}$.
\end{itemize}
Here we will present a shorter alternative argument that appeared in \cite{Hebecker:2018vxz}. The starting point is the species bound \cite{Veneziano:2001ah,Arkani-Hamed:2005zuc,Dvali:2007hz,Dvali:2007wp}, which states that the cutoff $\Lambda$ of a theory of quantum gravity with $N$ species is $\Lambda \sim M_P/\sqrt{N}$. The distance conjecture claims that when covering large distances in scalar field space an equally spaced tower of modes should appear with masses of order
\begin{align}
    m \sim M_P e^{-b \phi}\,.
\end{align}
Therefore the number of particles below a cutoff $\Lambda_c$ is estimated to be $N \sim \Lambda_c/m$. Eliminating $N$ in these two relations one gets
\begin{align}
    \Lambda \sim (m M_P^2)^{\frac{1}{3}} \sim M_P e^{-\frac{b\phi}{3}}\,.
\end{align}
In a dS solution the potential can be written as $V \approx H^2 M_P^2 $ where $H$ is the Hubble constant. In general we wish to be in a regime where $H \leq \Lambda_c$, which combined with the previous estimate of the cutoff gives
\begin{align}
    M_P^4 e^{-\frac{2b \phi}{3}} \geq V(\phi)\,,
\end{align}
which implies that the potential falls off exponentially. This is indeed compatible with all known examples in flux compactifications. We stress however that this does \emph{not} rule out de Sitter solutions. The conjecture rules out de Sitter solutions only in the parametric regime, but not otherwise.

A related swampland conjecture about the presence of de Sitter is the \emph{Transplanckian Censorship Conjecture} (TCC) \cite{Bedroya:2019snp, Bedroya:2019tba}. The TCC assumes that in a consistent quantum theory of gravity sub-Planckian quantum fluctuations should remain quantum and never become larger than the Hubble horizon which then freeze in an expanding universe. For scalar-driven cosmologies it leads to the refined dS Swampland conjecture in the asymptotic regime of moduli space and fixes the unknown constants $c, c'$ in the Refined de Sitter Conjecture \eqref{dSconj1} \eqref{dSconj2}. Away from that regime the TCC forbids long-lived meta-stable dS spaces, but it does allow sufficiently short-lived ones. When applied to inflationary models it is a significant constraint (in line with previous strong Swampland constraints \cite{Rudelius:2019cfh}). Interestingly the coefficients $c, c'$ predicted by the TCC behave well under dimensional reduction \cite{Rudelius:2021oaz} and show up in explicit 10d supergravity constraints \cite{Andriot:2022xjh}. For some criticism regarding the validity of the TCC we refer to \cite{Burgess:2020nec}.

\subsection{Arguments against stringy de Sitter?}
It appears that the combination of heuristic arguments (of the type described in \cite{Ooguri:2018wrx, Hebecker:2018vxz}) and the absence of classical dS solutions, despite much effort in this direction, gives some confidence for the correctness of the de Sitter Swampland conjecture. It would not forbid meta-stable de Sitter, it only forbids it in regimes of parametric control. Away from that regime, the TCC conjecture would then significantly constrain the lifetime of meta-stable dS vacua, if any at all.

If so, we are in a sticky situation, similar to QCD problems, where we lack simple computational tools. Of course, scenarios like KKLT or the LVS construction do suggest these meta-stable dS vacua in the non-parametric regime, yet we must add that these scenarios received quite some criticism over the years. Below we highlight some of the most important recent criticisms and provide pointers to the literature.\footnote{We refer to the review \cite{Danielsson:2018ztv} for problems related to meta-stable dS vacua discussed before 2018.} 

\paragraph{Small $W_0$?:} KKLT requires a value $W_0$ that can be tuned to be arbitrarily small (see the discussion above equation \eqref{AdSKKLT}). This can be in tension with the tadpole conjectures since small $W_0$ is argued to arise from large flux numbers canceling against each other in the equation that determines $W_0$. 
Note that this constitutes a criticism of the KKLT scenario, before even the uplift to de Sitter space is addressed. Although it must be said there are very non-trivial recent breakthroughs in obtaining small $W_0$ \cite{Demirtas:2019sip, Demirtas:2021nlu, Demirtas:2021ote} (see also \cite{Alvarez-Garcia:2020pxd}).  Note that these references compute also with great precision the non-perturbative superpotential needed for the KKLT K\"ahler moduli stabilization.\footnote{See also \cite{Kim:2023sfs} for a discussion of string loop corrections in these solutions.}

More general, there exist worries about the KKLT effective field theory approach, which could potentially be organised as an expansion in $W_0$-corrections, see for instance the discussion between \cite{Sethi:2017phn} and \cite{Kachru:2018aqn}.  Especially the recent work \cite{Lust:2022lfc} seems to point towards a contradiction with holography, which, if correct, would deem the KKLT AdS vacuum as a fake solution.

\begin{figure}[htp]
    \centering
    \includegraphics[width=12cm]{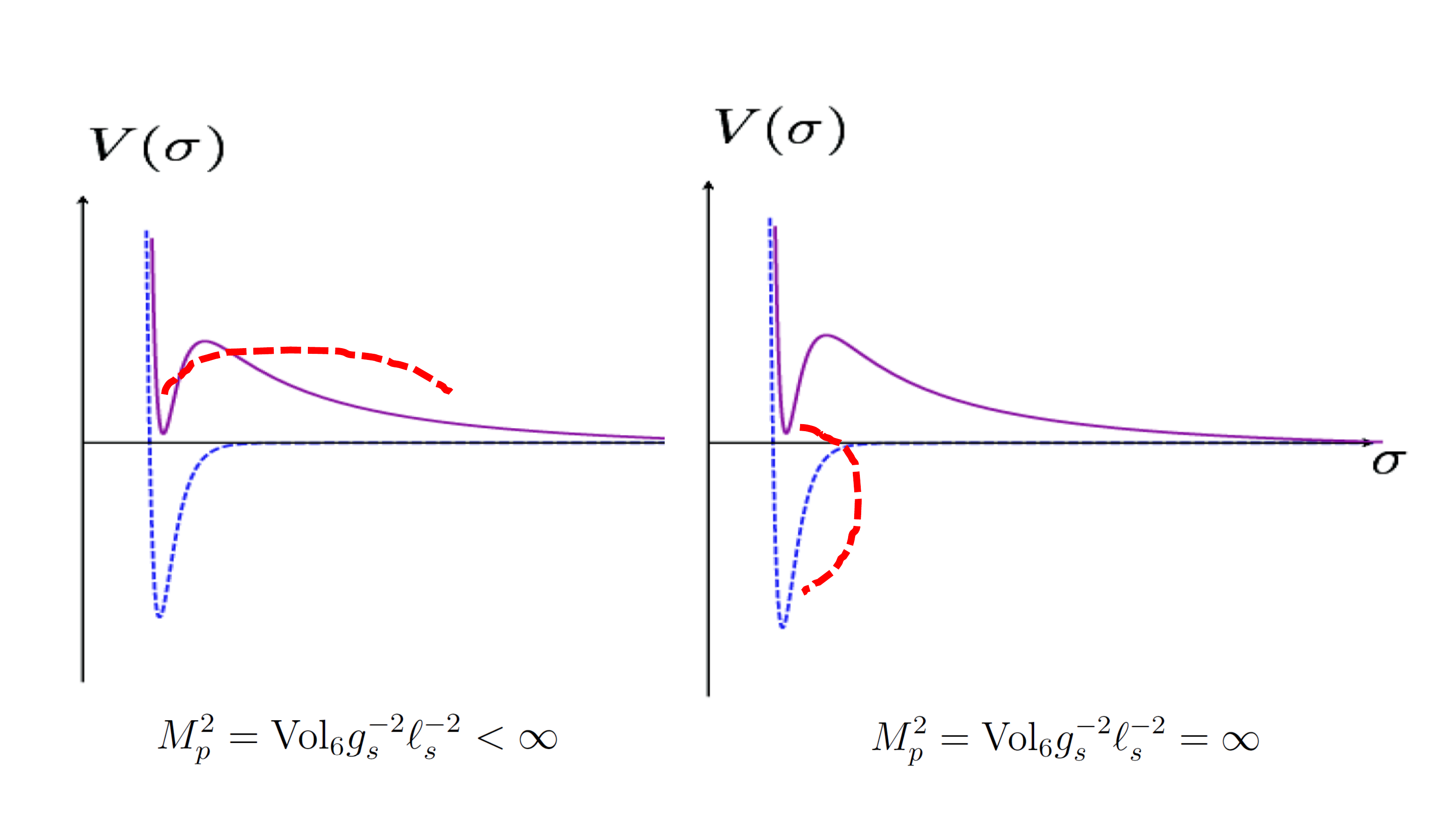}
    \caption{{\small \emph{Two natural ways an uplifted vacuum can decay. The decay process is represented by the red dotted line in both pictures. On the left a `closed string decay' where the volume modulus tunnels out its vacuum and causes decompactification. This can only happen at finite Planck mass. On the right an open string decay in which the anti-D3 brane decays against background fluxes. This can even happen in a non-compactification, so at infinite Planck mass.} }}
    \label{pic:decay}
\end{figure}

\paragraph{Singular fluxes destabilise anti-branes?:} 
The validity of the $\overline{\text{D3}}$-brane uplift in KKLT and LVS has been debated since its invention.
We start with one of the most studied criticisms regarding anti-brane uplifts and that is the stability of the anti-D3 brane with respect to annihilation against its surrounding fluxes. This was studied first in \cite{Kachru:2002gs} in the non-compact, holographic setup of anti-D3 branes at the tip of the so-called Klebanov--Strassler geometry \cite{Klebanov:2000hb}. This geometry is an explicit \emph{local} model of how regions with large warping can be generated inside flux compactifications \cite{Giddings:2001yu}. The rough picture of so-called brane-flux annihilation \cite{Kachru:2002gs} is the following. If one looks at the $F_5$ Bianchi identity
\begin{equation}
    dF_5 = H_3\wedge F_3 + 
    \emph{local sources}\,,
\end{equation}
where `local sources' denote the 6-form source terms for background orientifold planes but also the anti-D3 branes. The anti-D3 branes are not mutually SUSY with the 3-form fluxes and carry opposite orientation. In other words; the background fluxes generate D3 charges instead of anti-D3 charges. What Kachru, Pearson and Verlinde \cite{Kachru:2002gs} showed is that string theory allows a mechanism in which these background fluxes materialise into actual D3 branes that annihilate the anti-D3 branes. In general one expects this to be a non-perturbative instability which means that uplifts can still generate meta-stable, long-lived vacua. Pictorially, brane-flux decay represents an open-string instability which makes a dS vacuum obtained via uplifting decay following the dotted red line in the right picture of Figure \ref{pic:decay}.

The condition for meta-stability was derived in \cite{Kachru:2002gs} and, \emph{within the probe approximation}, requires that the flux density is low enough. However the first explicit computations of anti-D3 backreaction implied that the fluxes became singular near the anti-D3 branes \cite{Bena:2009xk, Gautason:2013zw, Blaback:2014tfa}! In fact the singularity is indeed such that the anti-D3 branes attract flux of opposite orientation leading to a direct brane-flux annihilation, which would make the singular fluxes smooth by a time-dependent process \cite{Blaback:2012nf}. In the next section we describe the arguments against this perturbative decay process idea. 

\begin{figure}[htp]
    \centering
    \includegraphics[width=12cm]{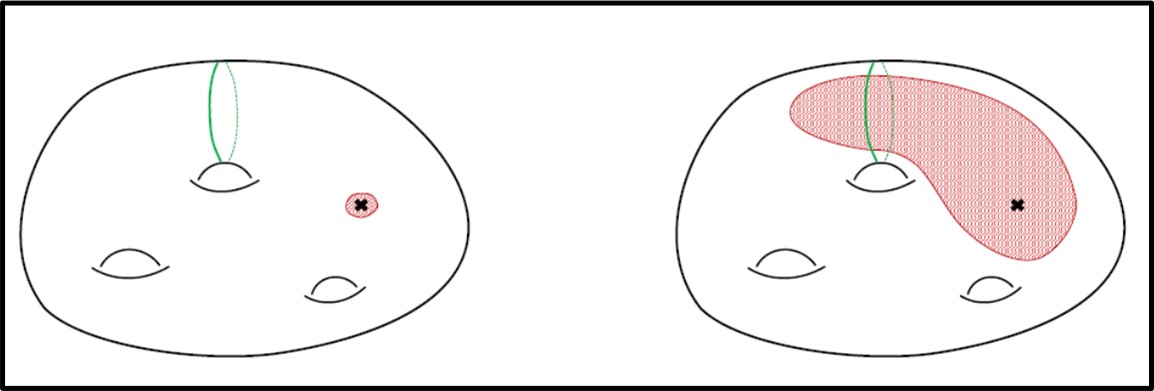}
    \caption{{\small \emph{Picture taken from \cite{Gao:2020xqh} representing the singular bulk problem. The green curve represents a stack of D7 branes wrapping a 4-cycle which provide the non-perturbative stabilisation of the volume modulus. The cross represents local orientifolds needed in the background GKP solution. The right picture shows the large region near the orientifolds that develops an unphysical singularity when one insists on gluing a local throat geometry that can successfully accommodate an anti-brane uplift. }}}
    \label{pic:singularbulk}
\end{figure}

\paragraph{Singular anti-brane uplift in compact geometries?}
Whereas the previous discussion about singularities can be discussed at infinite volume, so in local geometries (like the non-compact Klebanov--Strassler throat), there is yet another striking singularity that is specific to the details of compact geometries; i.e. the problem arises when one wants to glue a local throat geometry in a \emph{compact} Calabi-Yau and generate a long enough throat for a successful uplift (hence with fine tuned large warping) and such that the fluxes do not annihilate the anti-D3 brane directly. The problem was first sketched in \cite{Carta:2019rhx} where it was argued that a large enough throat with a meta-stable anti-brane would have a volume larger than the volume predicted by the moduli-stabilisation process. This has been called the `throat fitting problem'. It was then realised that the problem should be more subtle \cite{Gao:2020xqh} since in a GKP background at finite volume any size throat can be glued. Instead it was uncovered that this problem of throat fitting really becomes a `singular bulk problem' in the sense that unphysical singularities are developed in the bulk once one requires a throat geometry that allows a well-defined de Sitter uplift as depicted in Figure \ref{pic:singularbulk}. In other words, reference \cite{Gao:2020xqh} suggests that the geometry will be out of control because the warping needed to redshift the anti-brane tension is not consistent with the background fluxes used to stabilise the moduli. 

 \paragraph{No uplift in Large Volume Scenario?:} It appears that uplift scenarios in the LVS context \cite{Balasubramanian:2005zx} are out of control due to derivative corrections\cite{Junghans:2022exo} and this tension is related to tension of solving the tadpole problem \cite{Gao:2022fdi}. 
 \paragraph{Collapsing throat geometries?:}
 Another potential issue with the uplift scenario was pointed out in \cite{Bena:2018fqc} (see also \cite{Randall:2019ent, Dudas:2019pls, Bena:2019sxm}); The tip of a throat region is a local 3-cycle whose size is controlled by a modulus known as the conifold modulus, which is warped down in mass and so rather light. Hence, fluctuations of the 3-cycle cost little energy and one should avoid that this cycle collapses to zero size when anti-D3 branes are added. Such a collapse would lead to direct brane-flux annihilation \cite{Scalisi:2020jal} or other instabilities \cite{Blumenhagen:2017vsi}. 
 
\begin{figure}[htp]
    \centering
    \includegraphics[width=10cm]{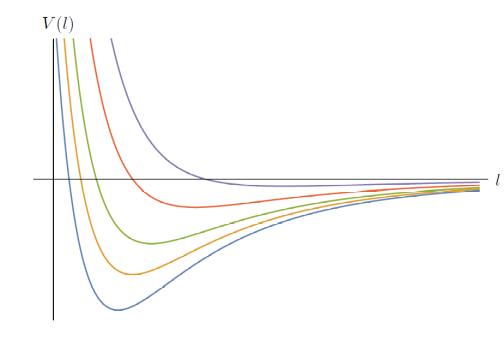} 
    \caption{\small{\emph{Picture taken from \cite{Moritz:2017xto}. Flattening effects of the anti-brane uplift: instead of uplifting, flattening effects can cause a runaway exactly when the uplift should lead to a dS vacuum instead of an AdS vacuum. Here $\ell$ can represent a general size modulus of compact dimensions and the blue lowest curve is the potential without anti-brane uplifts. The subsequent lines are effective potentials with an ever growing number of anti-branes. Exactly when the critical point would have positive energy there is a runaway. }} }
    \label{pic:flattening}
\end{figure}

\paragraph{No dS uplift due to flatting effects?:}
In an attempt to understand how to translate the KKLT gaugino condensation, obtained within 4d EFT, to a 10d picture\footnote{See also \cite{Koerber:2007xk,Grana:2020hyu, Grana:2022nyp}} a so-called \emph{flattening effect} was discovered in \cite{Moritz:2017xto} for the anti-brane uplift. This flattening is a 4d EFT interpretation of something seen in a 10d language and is depicted in Figure \ref{pic:flattening}. To make a long story short, the upshot is that anti-D3 branes did not seem to contribute in an uplifting manner from the viewpoint of 10d Einstein equations, rather to a runaway behavior, exactly when the anti-brane uplift energy is high enough to lift AdS to dS.  This is then interpreted as the flattening of the effective potential due to backreaction effects beyond the probe approximation used in the usual uplift argument. 

\paragraph{Goldstino induced tachyons from uplifting?:} Finally there are issues with anti-brane uplifts visible at low energies. Below the SUSY breaking scale one can write down an effective field theory which is of the Volkov--Akulov type for the goldstino and they seem to feature tachyonic instabilities \cite{DallAgata:2022abm, Farakos:2022jcl}. To date this has not been addressed or interpreted as harmless.

\paragraph{General dS no-go results in the quantum regime: } Outside of the concrete context of KKLT or LVS some interesting no-go theorems for the non-parametric regimes (i.e. non-classical regimes) have also been found and contribute to the tension for the existence of well-controlled dS vacua in String Theory. In particular there are some striking dS no-go results deep in the non-classical regime of string theory \cite{Gautason:2013zw, Kutasov:2015eba, Green:2011cn, Quigley:2015jia, Plauschinn:2020ram, Grimm:2019ixq}.\footnote{Some of the non-classical dS constructions rely on non-geometric fluxes which is a concept that is nicely reviewed in \cite{Plauschinn:2018wbo}.} 

\paragraph{Quintessence to the rescue?:} It has been suggested that the Swampland tensions for de Sitter naturally points to quintessence scenarios in order to explain the nature of the observed dark energy \cite{Agrawal:2018own}. However, no-go results for quintessence scenarios also exist, see \cite{Cicoli:2021fsd, Cicoli:2021skd, Shiu:2023nph} and a general tension for obtaining controlled quintessence in String Theory was pointed in \cite{Hebecker:2019csg}. A review on quintessence attempts from string theory can be found in \cite{Cicoli:2018kdo}.
\bigskip

The claim that there is evidence for a landscape of de Sitter vacua in string theory can only be justified when at least all of these criticisms are carefully addressed. In the next section we explain how there are also reasons to be more optimistic about dS scenarios because there exist new ideas and some of the criticism above has been refuted, especially the criticism regarding singular fluxes near anti-branes.      

\subsection{New arguments in favor of stringy de Sitter? }
De Sitter model building in String Theory is an activity that started roughly in the beginning of 2000 and never stopped since. It is fair to say that the creation of models, or better said, suggestions for  constructions, outpaces the scrutiny. In other words, the amount of energy spend on suggesting how de Sitter can be constructed outweighs the energy spend on cleaning up the details who potentially render the suggestions inconsistent. Luckily with the advent of the Swampland Program much more activity exists now in scrutinizing existing de Sitter scenarios. Nonetheless it is important to come up with new ideas as well. In what follows we present both the results of this scrutiny, which partially refutes the previous criticism, and we present brief pointers to the literature on new suggestions for dS model building since 2018, whereas a classification of de Sitter attempts before 2018 has already been reviewed in \cite{Danielsson:2018ztv}, see also \cite{Cicoli:2018kdo}.

\paragraph{No singular fluxes near anti-D3 branes in dS uplifts.} 
The first reference to argue that the singular flux-clumping around anti-D3 branes could be absent is \cite{Michel:2014lva} (see also \cite{Polchinski:2015bea}). The assumption in that reference was that within supergravity the singularity seems unavoidable, but that stringy corrections resolve it. This was argued on the basis of \emph{brane effective field theory}. The resulting corrected flux clumping would then not be large enough to trigger direct brane-flux decay. This argument is very sensible but some details have been criticised in \cite{Bena:2014jaa, Bena:2016fqp}. Nonetheless references \cite{Cohen-Maldonado:2015ssa, Cohen-Maldonado:2016cjh} revised the supergravity arguments that lead to the singular fluxes and found that exactly the ingredients used to derive brane-flux decay provide the required mechanism to avoid the singularity \emph{within supergravity}. 

The picture is not too complicated to understand intuitively; according to Kachru, Pearson and Verlinde \cite{Kachru:2002gs} the details of brane-flux decay proceed via the anti-D3 brane ``puffing'' into a spherical 5 brane that behaves effectively as the anti-D3 brane from some distance. The only effect is that the anti-D3 brane is smeared out a bit and stops being pointlike inside the extra dimensions. Hence the manner in which that anti-brane attracts flux becomes diluted.  Using fully backreacted supergravity equations of motion  references \cite{Cohen-Maldonado:2015ssa, Cohen-Maldonado:2016cjh} found that this ``polarisation'' into a 5-brane seems a perfect way around the singularity theorems derived in \cite{Gautason:2013zw, Blaback:2014tfa}. 

\begin{figure}[htp]
    \centering
    \includegraphics[width=10cm]{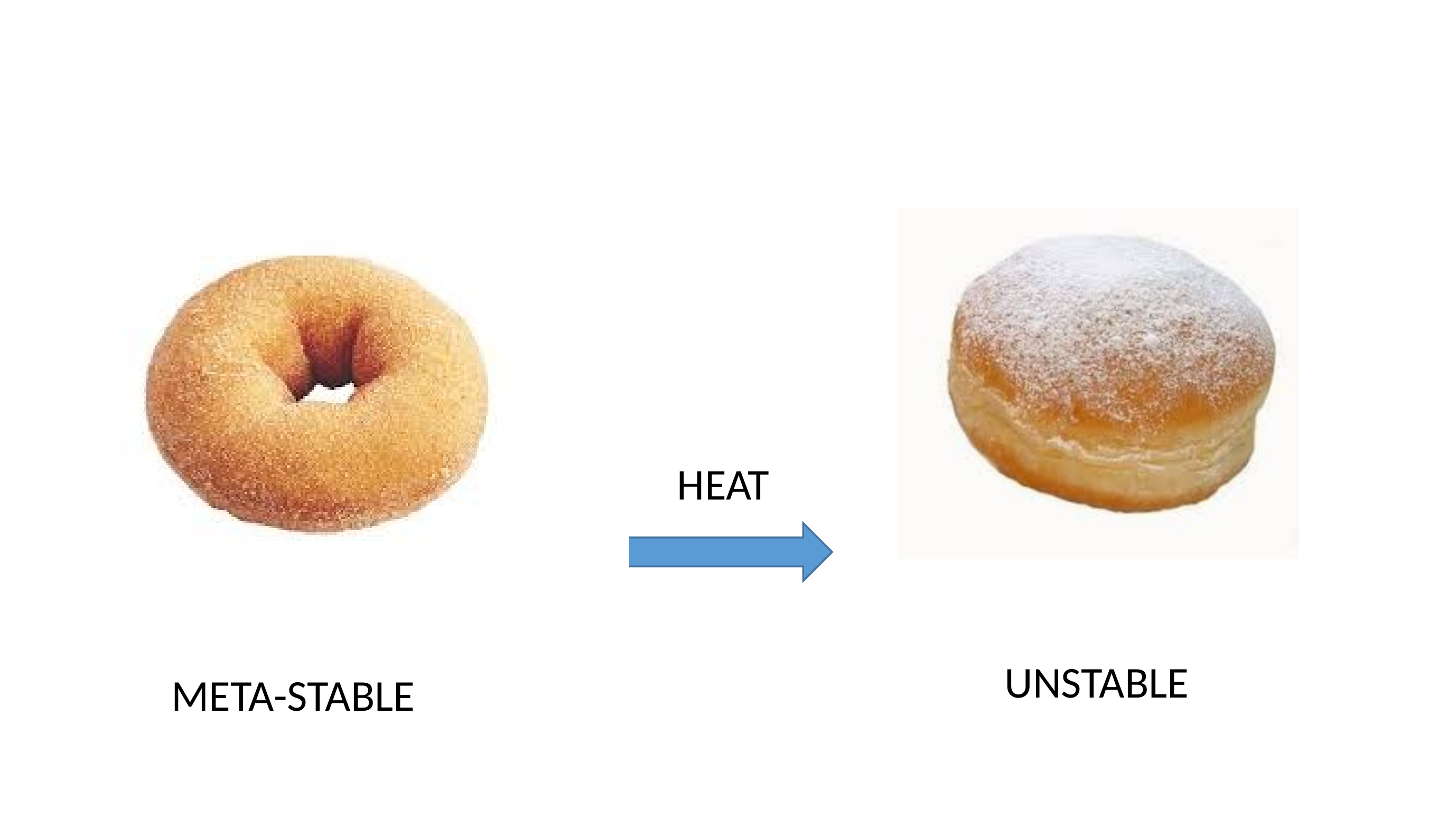} 
    \caption{\small{\emph{The left picture represents a spherical D5 brane at finite temperature. The thickness of the doughnut is set by the amount of heat that is added to the spherical 5-brane. If it gets heated up so much that the doughnut becomes a Berliner, the system decays directly through brane-flux decay. }} }
    \label{pic:berliner}
\end{figure}

This suggested resolution has then been established in much more details in  \cite{Armas:2018rsy, Armas:2019asf, Blaback:2019ucp,  Nguyen:2019syc, Nguyen:2021srl}. Even more, extra arguments for meta-stability were found by showing that the anti-D3 or better said, the spherical 5-brane setup, can be heated up with finite energy without destabilising the system. This is what one expects from gapped states: a finite amount of energy could be added without destabilising the system.  The interplay between the singularity (regularity) theorems of \cite{Cohen-Maldonado:2015ssa, Cohen-Maldonado:2016cjh} and the addition of temperature has lead to a remarkably simple picture which is sketched informally in Figure \ref{pic:berliner}. 
The spherical 5-brane can be represented as a thin ring and when heated up it becomes a fatter ring, that looks like a doughnut. The hole inside the doughnut is a necessary requirement to evade the singularities since the singularity theorems directly probe the topology of the brane setup. If the doughnut is heated up too much such that the finite temperature horizon grows to overtake the hole inside turning the doughnut into a Berliner, then the singularities appear again and create an instability. This was  confirmed in an approximation beyond the probe approximation, known as the \emph{blackfold approach}, see \cite{Armas:2018rsy, Armas:2019asf,  Nguyen:2019syc, Nguyen:2021srl}.\footnote{Another potential mechanism to resolve the singular fluxes was suggested in \cite{Hartnett:2015oda}, but shown to lead to an unstable state in \cite{Armas:2018rsy}.} Hence the blackfold approach is a good test for the meta-stability of the anti-brane state since it explicitly shows how it can be heated up until a maximal temperature. In the particular case of anti-D6 brane toy models these ideas can be verified completely explicit using fully backreacted anti-brane solutions \cite{Blaback:2019ucp}. 

\paragraph{No singular uplifts in compact geometries?}
A potential resolution to the singular bulk problem argued in \cite{Gao:2020xqh} is presented in \cite{Carta:2021lqg}. The suggested resolution however relies on various approximations, and if it takes place, it seems the corrections to the bulk geometry are significant and it is unclear whether the KKLT effective field theory is safe from these corrections.

\paragraph{No collapsing throat geometries?}
 It has been argued in a recent paper that the worry of a collapsing throat is not necessary and that the instability is absent \cite{Lust:2022xoq}. General remarks about how warped down moduli in a throat alter the EFT can be found in \cite{Blumenhagen:2019qcg, Seo:2023ssl}.

\paragraph{No flattening effects for uplifts?}
The computation done in \cite{Moritz:2017xto} that revealed the so-named flattening effects was shrouded in some confusion due to a singular contribution of divergent fluxes sourced by the gaugino condensate (and some sign issues \cite{Gautason:2018gln}). How to deal with such a divergent flux was explained in \cite{Hamada:2018qef, Hamada:2021ryq} (see also \cite{Kallosh:2019oxv, Bena:2019mte, Kachru:2019dvo}) and then the computation of the anti-brane uplift was revised in \cite{Gautason:2019jwq, Hamada:2019ack, Carta:2019rhx}; whereas references \cite{Hamada:2019ack, Carta:2019rhx} came to the conclusion that the flattening effect was gone, reference \cite{Gautason:2019jwq} did find a significant difference between the 10d and the 4d approach, making it unclear the 4d EFT can be trusted. The difference between the references arises due to a difference in the way the 10d Einstein equation was computed. Reference \cite{Gautason:2019jwq} takes the 10d action and varies with respect to the 10d metric to get the semi-classical Einstein equation:
\begin{equation}\label{semiEinstein}
    G_{\mu\nu} = \langle T_{\mu\nu}\rangle \,,
\end{equation}
where afterwards the quantum vevs of the fields is filled in to obtain $\langle T_{\mu\nu}\rangle$. The vevs of the fields is information obtained from the 4d EFT since gaugino condensation is, by definition, an IR effect. On the other hand, references \cite{Hamada:2019ack, Carta:2019rhx} notice that consistency with the 4d SUGRA equations is only obtained if \emph{in the 10d action} the gaugino vevs dependence on the volume is filled in, after which they take the variation of the `10d action' to arrive at the semi-classical Einstein equation \eqref{semiEinstein}. Since the gaugino vev is metric dependent changing the order of the procedure does change the results. Whereas \cite{Gautason:2019jwq} remains 10d covariant, references \cite{Hamada:2019ack, Carta:2019rhx} break this. Claiming that consistency with the 4d EFT is an argument for the second procedure of \cite{Hamada:2019ack, Carta:2019rhx} to be correct is in some sense circular reasoning since the whole idea was to see what can be learned from a 10d approach. On the other hand, it is not strange that 10d covariance needs to be sacrificed if one insists on writing an inherent large distance effect in small distance variables. In other words: the whole 10d method is potentially a moot point since gaugino condensation does not exist at scales as small as the KK scale, so there is no reason one tries to understand how it backreacts in 10d language. 

\paragraph{Casimir energies and hyperbolic spaces:} Instead of using supersymmetric geometries, with their associated supergravity effective field theories where dS is typically build from a combination of quantum effects (fox example gaugino condensation) with classical supersymmetry breaking sources (like anti-branes), one can simply break SUSY at the KK scale. Then naively all bets are off since there is no supersymmetry to protect the corrections to the effective field theory after compactification. This means one has to be very careful. Using the fact that compact hyperbolic geometries can have complicated circles inside them, non-trivial Casimir energies arise from loop effects. Together with the classical positive energy from the negative internal curvature, it seems de Sitter vacua arise naturally \cite{DeLuca:2021pej}. In the next section we will discuss Casimir energies in simple torus compactifications and how it can give rise to AdS vacua with small cosmological constants.

\begin{figure}[htp]
    \centering
    \includegraphics[width=14cm]{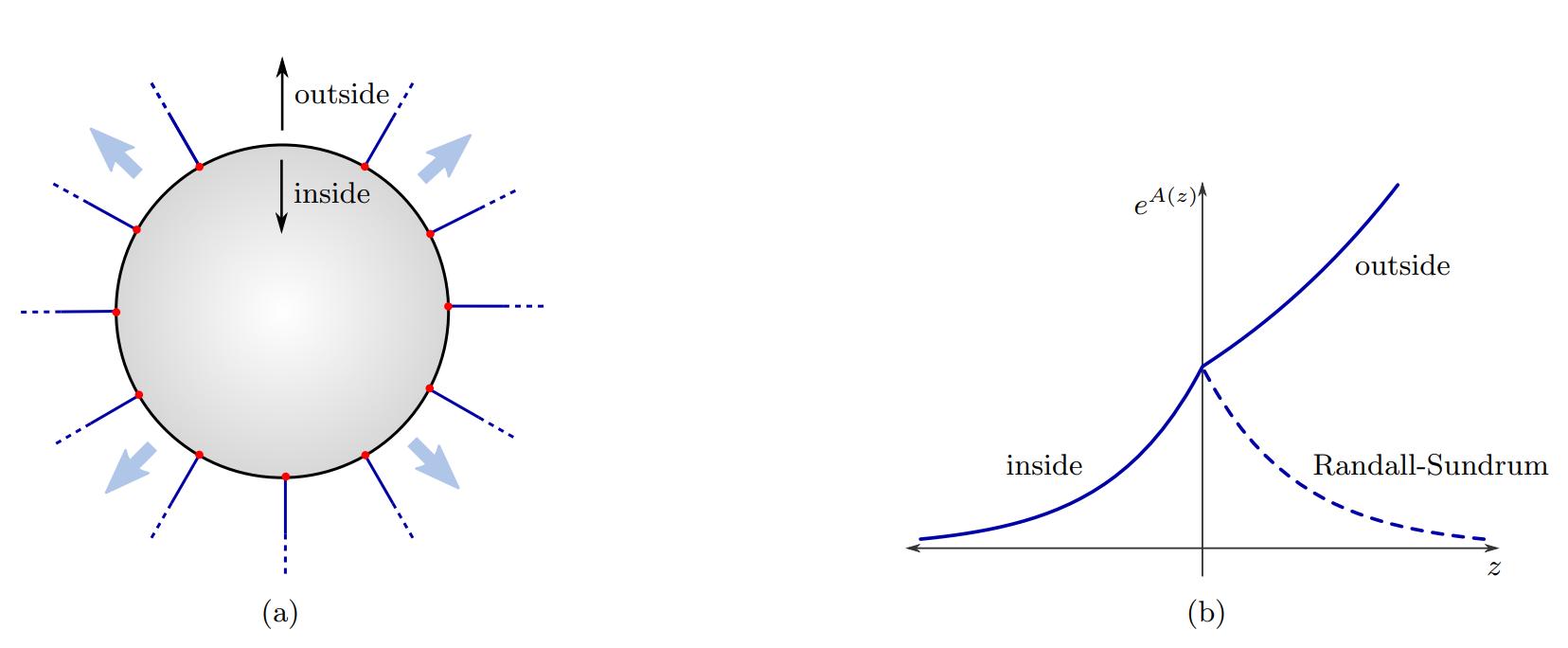}
    \caption{\small{\emph{The left picture (a) represents the 5-dimensional description of the 4D brane bubble word. The inside is the true stable vacuum and the outside is the meta-stable AdS vacuum. The bubble wall separates the two. The right picture (b) plots the 5D warp factor and compares it with the Randall-Sundrum scenario, which makes clear that the dark bubble mechanism is vastly different.}} }
    \label{pic:bubble}
\end{figure}

\paragraph{Dark bubble scenario:} A very fresh out-of-the-box construction of de Sitter vacua was suggested in \cite{Banerjee:2018qey} (see also \cite{Banerjee:2019fzz}, \cite{Banerjee:2022myh}). The idea is to take the Swampland tensions seriously and assume this implies that standard compactifications tend to get in trouble when you insist on de Sitter in the lower-dimension because it is difficult to keep the extra dimensions stable while obtaining positive vacuum energy. Instead Banerjee and collaborators contemplate \emph{brane world scenarios} which are vastly different from the standard Randall--Sundrum \cite{Randall:1999vf} or Karch--Randall \cite{Karch:2000ct} kind. The idea is in fact based on a Swampland conjecture known as the \emph{non-SUSY AdS conjecture} \cite{Ooguri:2016pdq} and states that any non-supersymmetric anti-de Sitter vacuum supported by fluxes must be unstable. In absence of perturbative instabilities such decays tend to be of the Brown-Teitelboim type \cite{Brown:1988kg}. This means that a spherical domain wall of sub-critical tension forms and inside this bubble there is a vacuum with lower energy. The insight of \cite{Banerjee:2018qey} is that matter can be localised on such a wall for instance in the form of end-points of strings stretching from the boundary. Such matter would obey 4d Einstein equations. A pictorial representation of this scenario is presented in Figure \ref{pic:bubble}. 
Even more striking is that the wall itself is expanding inside the bulk spacetime and this expansion is exactly like a de Sitter expansion. Hence such brane world scenarios are unavoidably de Sitter at late times (but also at early times! \cite{Danielsson:2022lsl}). Attempts to embed this scenario in actual string theory examples are \cite{Basile:2020mpt, Danielsson:2022lsl}. This scenario has been criticised in \cite{Mirbabayi:2022eqn}, but then it was explained in \cite{Banerjee:2022myh} how the criticism is flawed. Interestingly the model is so concrete that even basic questions in quantum cosmology can be addressed. One such a question concerns the quantum state of the universe. There are many options but two canonical choices are the Vilenkin state \cite{Vilenkin:1982de} or the Hartle--Hawking one \cite{Hartle:1983ai}. The dark bubble scenario of \cite{Banerjee:2018qey} uniquely selects the Vilenkin wave function \cite{Danielsson:2021tyb}.

\paragraph{Liberal singularities:} Allowing singularities of the orientifold kind that differ in their sub-leading behavior seems a very simple manner to get de Sitter solutions at the classical level which completely go around the no-go results constructed in this review \cite{Cordova:2018dbb, Cordova:2019cvf}. The reason is that our no-go statements feature integrated equations with certain assumptions of how the singularities integrate. These assumptions were dropped in \cite{Cordova:2018dbb, Cordova:2019cvf}, but criticised to be unphysical in \cite{Cribiori:2019clo} or claimed to lead to instabilities as pointed out in \cite{Bena:2020qpa}.

\paragraph{Other approaches:} Other interesting recent attempts to construct de Sitter which we do not discuss due to lack of spacetime are for instance \cite{Farakos:2020idt, Berglund:2022qsb, Dasgupta:2019rwt, Dasgupta:2019vjn, Dasgupta:2019gcd, Cicoli:2021dhg, Shukla:2022srx, AbdusSalam:2022krp, Bento:2021nbb,  Bena:2022cwb ,Bena:2022tro, Alexander:2023qym, Leontaris:2023obe}.

\subsection{Ads with scale separation from supergravity/strings?}

Holographic backgrounds known within String Theory are of the kind AdS$_n \times X_m$ which have a dual description in terms of a CFT$_{n-1}$. An important question that arises here is whether it is possible to decouple the mass scale associated to the internal space (the mass of KK modes) from the energy scale associated to the cosmological constant? In other words, can we ignore the $X_m$? Or even other words \cite{Alday:2019qrf, Polchinski:2009ch}; \emph{how many space dimensions emerge from a CFT in $d$ dimensions?}. Only one dimension (which is the usual popular explanation) or $10-d$ dimensions (which is what happens in all known cases)? 
While this question seems to be confined to the realms of the holographic correspondence and not bear any consequence for phenomenological model building this is incorrect. In several scenarios (like the KKLT construction for instance) where a landscape of $10^{\text{some number}}$ dS vacua is claimed to exist, always necessitates AdS vacua for which
\begin{align}
    \frac{L^2_{KK}}{L^2_{\text{AdS}}} \ll 1\,.
\end{align}
The reason for this should be evident from the previous section: most dS vacua are obtained from uplifting AdS vacua and these AdS vacua are subsequently obtained in some hybrid 10d string theory-4d EFT formalism. Such a hybrid formalism can only make sense when the machinery of 4d EFT is applicable, and that requires a separation of scales.  

We now analyse whether or not AdS vacua can be scale separated using the higher-dimensional equations of motion, in the simple setting of 11d supergravity. After that we will analyse the same question in 10d supergravity using our beloved $\rho, \tau$ scalings, which we argued in section \ref{sec:10d} are equivalent to solving certain integrated 10d equations of motion.

\paragraph{The supergravity perspective in 11d.}
Here we follow arguments that first appeared in \cite{Gautason:2015tig}. The starting point is to use 11d supergravity whose action is
\begin{align}
    S = \int d^{11}x \sqrt{-g} \left[R_{11}-\frac{1}{2} \frac{1}{4!} F_4^2\right] + \dots\,,
\end{align}
where the dots include the Chern--Simons terms that will be unimportant in the following. We consider compactifications with fluxes down to 4d, and if we assume there is no warping for simplicity we can write the traced Einstein equations as follows:
\begin{align}
    R_4 &= -\frac{4}{3} \frac{F_4^2}{4!}-\frac{8}{3}\frac{F_7^2}{7!}\,,\label{eq:R4}\\
    R_7 &= \frac{5}{3} \frac{F_4^2}{4!}+\frac{7}{3}\frac{F_7^2}{7!}\,.\label{eq:R7}
\end{align}
We keep $F_7$ so that all fluxes can be written as ``magnetic''. By that we mean that all indices of $F_4$ and $F_7$ are inside the internal space. This way of writing simplifies things since contractions like $F_4^2$ and $F_7^2$ are only among spacelike indices and so the squares are strictly positive. The above two equations then show that necessarily $R_4<0$ and $R_7 >0$. The fact that $R_4$ is negative is of course a consequence of the Maldacena--Nu\~{n}ez no-go theorem \cite{Maldacena:2000mw} and the fact that $R_7$ is positive is part of what can be called the ``Douglas-Kallosh'' no-go theorem \cite{Douglas:2010rt}. 
\begin{theorem}
Verify equations \eqref{eq:R4} and \eqref{eq:R7}.
\end{theorem}

For the sake of understanding scale separation it is instructive to consider the ratio of integrated curvatures\footnote{This kind of reasoning can be found in older works like \cite{Tsimpis:2012tu}. }
\begin{align}\label{eq:curvratio}
    \left|\frac{\int R_7}{\int R_4}\right| = \frac{5\int  \frac{F_4^2}{4!}+7\int \frac{F_7^2}{7!}}{4\int  \frac{F_4^2}{4!}+8\int \frac{F_7^2}{7!}} \leq \frac{5}{4}\,.
\end{align}
Notice that all integrals are done over 11d. We can define the two curvature radii as follows
\begin{align}
    L_R^{-2} & = \frac{1}{\text{vol}_7} \int_{\mathcal M_7} R_7\,,\\
    L_{\text{AdS}}^{-2} &= \frac{1}{\text{vol}_4} \int_{\mathcal M_4} R_4\,.
\end{align}
In defining $L_{\text{AdS}}$ one needs to take some care as the volume of $\text{AdS}_4$ is clearly infinite so some regularization is needed. But using these newly defined length scales we have that \eqref{eq:curvratio} can be rewritten as
\begin{align}
    \frac{L_{\text{AdS}}^2}{L_R^2} \leq \frac{5}{4}\,.
\end{align}
Similar bounds can be derived in 10d whenever the ingredients are fluxes and positive tension branes \cite{Gautason:2015tig}. This means that, under these conditions, it is not possible to  make the AdS curvature length bigger than the curvature length of the  internal manifold. One can wonder to what extend this means there is no scale separation between $L_{KK}$ and $L_{\text{AdS}}$?
In some cases it is possible to relate $L_R^2$ to $L_{KK}^2$, however it is not clear whether this holds in general \cite{DeLuca:2021mcj,DeLuca:2021ojx, DeLuca:2022wfq,Cribiori:2021djm, Collins:2022nux}. Note that for negatively curved geometries it is rather simple to achieve such a separation, see \cite{Andriot:2018tmb} or the appendix of \cite{Cribiori:2021djm}, but the Einstein equations show that negatively curved internal geometries are not possible within our assumptions (classical supergravity with no negative tension sources). 

\begin{theorem}
Consider the 4d compact space given by the product of two 2-spheres. Call the radii of these two spheres $R_1$ and $R_2$. Can one, at fixed total volume, tune the curvature scale to be very large or very small?
\end{theorem}

So for all manifolds for which $L_{KK}$ can not be tuned small at fixed $L_R$ we find a  generalisation of the Maldacena--Nu\~{n}ez theorem \cite{Maldacena:2000mw,deWit:1986mwo}: 
\bigskip

\emph{There are no de Sitter nor Minkowski solutions with smooth internal static compact manifolds including fluxes and positive tension sources. Furthermore, for internal geometries with the property that  $L_{KK}$ cannot be tuned small at fixed $L_R$ there will not be AdS vacua that are scale separated. This means we can also exclude AdS vacua with a small cosmological constant, where small means small with respect to the KK scale.} 
\bigskip

There are possible ways out, like the introduction of orientifold planes \cite{DeWolfe:2005uu} (this introduces negative tension sources), or use of quantum corrections \cite{Kachru:2003aw, Balasubramanian:2005zx}. Let us analyse how orientifolds can be a way out and for that we need to be in 10 dimensions. Once we have done it we will be remarkably lead to conclude that perhaps Einstein geometries with positive curvature exist for which one can take $L_{KK}$ small at fixed $L_R$ exist.

\paragraph{The effective potential perspective from 10d}
Let us try to formulate the problem- of scale separation within the framework employed in these lectures: take the following metric 10d metric in string frame
\begin{align}
    ds^2_{10} = \tau_0^{2 } \tau^{-2} ds^2_D +\rho\, ds^2_{10-D}\,.
\end{align}
Here $\tau_0$ is just the vev of $\tau$. Note that we need to require that the $D$-dimensional metric is in Einstein frame, which is equivalent to the condition \eqref{tau}

After dimensional reduction we get the following action in $D$-dimensions
\begin{align}
    S_D = \int d^D x \sqrt{-g_D} \left[\tau_0^{D-2} R_D-\tau_0^D V + \dots\right]\,.
\end{align}
With this action we can readily identify the $D$-dimensional Planck mass as $M_P = \tau_0$. We wish to compare the length scale associated with the potential at the minimum (that is the cosmological constant in $D$-dimensions) to the mass scale of the lightest modes associated to the internal space. While in principle there can be several modes whose mass scale needs to be checked we will take as a proxy the mass scale of KK-modes. We can extract the KK-scale to be $L^2_{KK} \sim \rho_0$ by simple estimation of the volume of the internal manifold. Note that when fluxes and other ingredients are introduced in compactifications this may distort the estimation of mass scale of KK-modes but it is not possible to have a general estimate of this. Then, we find from the $D$-dimensional Einstein equations that
\begin{align}
    R_D = \frac{D}{D-2} M_P^2\, V\,,
\end{align}
which implies that $L_{\text{AdS}}^{-2} = M_P^2\, V$. Therefore the ratio of the scales entering in the problem is
\begin{align}
    \frac{L^2_{KK}}{L^2_{\text{AdS}}} \sim \rho_0\, \tau_0^2\, V\,.
\end{align}
The big looming question is whether this can be made arbitrarily small. 

To try and get some insight we look at the setup studied in the previous subsection; compactifications of type IIA on a Calabi--Yau threefold (that is $D=4$ and $R_6 = 0$) with $F_0$, $H_3$, and $F_4$ fluxes as well as O6-planes. There are two relevant Bianchi identities to be considered in this case
\begin{align}
    dF_2 &= F_0 H_3 +\delta(\text{O6})\,,\\
    dF_4 &= F_2 \wedge H_3 + \delta(\text{O4})\,.
\end{align}
Notice that existence of a solution for the first one bounds the amount of $F_0$ and $H_3$ fluxes that can be added to the compactification.\footnote{Recall that the number of O6-planes is constrained by topology and is not a free parameter.} On the other hand the second Bianchi identity is automatically satisfied since we have no O4/D4 sources and no $F_2$ flux\footnote{The generalisations with $F_2$ flux are such that $F_2\wedge H_3$ vanishes.}. Given that $F_4$ does not appear in any Bianchi identity the amount of $F_4$-flux that can be added to the compactification is unbounded. The scalar potential for this compactification is
\begin{align}
    V = U_H \rho^{-3}\tau^{-2}+ U_0 \tau^{-4}\rho^3 +U_4 \tau^{-4}\rho^{-1} + U_{\text{O6}} \tau^{-3}\,.
\end{align}
The minimization conditions are
\begin{align}
    \rho\, \partial_\rho V= 0\,, &\qquad \Longrightarrow \qquad -3V_H +3 V_0 -V_4 = 0\,,\\
    \tau\, \partial_\tau V= 0\,, &\qquad \Longrightarrow \qquad -2V_H -4 (V_0 +V_4)-3V_{\text{O6}} = 0\,.
\end{align}
Using these conditions and replacing them in the scalar potential one gets that at the minimum the scalar potential is
\begin{align}
    V &= V_H+V_0+V_4 + V_{\text{O6}} =\\
    &=-\frac{4}{9}V_4 < 0\,.
\end{align}
\begin{theorem}
Verify the above equations. Verify that without O6 planes there is no solution. 
\end{theorem}
This means that AdS solutions are perhaps possible if $F_4 \neq 0$. We define the flux quantum to be $$N = \int F_4\,,$$
(ignoring factors of $\pi$)
and consider the limit $N \rightarrow \infty$. However, even if $N$ is not bounded by tadpole cancellation, it is still necessary to find a solution of the equations of motion. We therefore make sure that all terms in the full scalar potential balance each other, and we will parametrize $\tau \sim N^A $ and $\rho \sim N^B$ with $A,\,B$ to be determined. The scalar potential then scales as
\begin{align}
    V \sim U_H \, N^{-2A-3B} + U_0 \,N^{-4A+3B}+ N^{2-4A-B}+U_{\text{O6}}\, N^{-3A}\,.
\end{align}
If we take $B=1/2$ and $A=3/2$ \emph{all components scale in the same fashion with $N$}. This is non-trivial since the system is over-constrained. Let us analyze the consequences of this  
\begin{itemize}
    \item The first observation is that $V \sim N^{-3A} \rightarrow 0$. With this we find that $\rho_0 \tau_0^2 V \sim N^{-1} \rightarrow 0$. Therefore this ``solution'' achieves scale separation;
    \item The string coupling constant in the vacuum scales as $e^{-2\phi_0} = \tau_0^2 \rho_0^{-3} = N^{3/2} \rightarrow \infty$. Therefore in the large $N$ limit we achieve a limit of weak coupling;
    \item Given that $\rho_0 \sim N^{\frac{1}{2}} \rightarrow \infty$ the volume becomes infinite in the large $N$ limit.
\end{itemize}

\begin{theorem}
Is a similar magic possible in a IIB set-up? Consider for instance IIB with O5, O7 planes. What do you find? [Hint: It is not obvious, see for instance \cite{Petrini:2013ika}. To date no classical IIB solutions with scale separation are known. The attempts in \cite{Petrini:2013ika,Caviezel:2009tu} turn out problematic after all \cite{Cribiori:2021djm}]. See also \cite{Emelin:2021gzx}.
\end{theorem}

\begin{theorem}
Verify that reductions of massive IIA on Ricci-flat 7d manifolds with 06 planes can also  be scale separated upon sending $F_4$ flux to infinity. (See \cite{Farakos:2020phe} for details (and \cite{VanHemelryck:2022ynr, Farakos:2023nms}). 
\end{theorem}

It seems almost too good to be true: we seem to achieve a regime with scale separation, weak string coupling and large volume. We have shown that the equations of motion with respect to $\rho$ and $\tau$ are satisfied, however the full set of equations of motion is larger. Luckily, in \cite{DeWolfe:2005uu} it was shown that a full solution does exist within a certain approximation; the derivation in \cite{DeWolfe:2005uu} (see also \cite{Derendinger:2004jn, Camara:2005dc, Font:2019uva}) relied on the formalism of $\mathcal N=1$ supergravity in $D=4$ \cite{Grimm:2004ua} instead of the 10d equations of motion. The 4d equations show that there exists both supersymmetric and non-supersymmetric AdS$_4 \times \text{CY}_3$ vacua with sometimes all moduli stabilized \cite{Lust:2008zd, Marchesano:2020uqz}. It was later shown that these solutions solve 10d equations of motion with \emph{smeared sources} \cite{Koerber:2007hd, Acharya:2006ne, Grana:2006kf}. This means that the $\delta$-sources in the Bianchi identities are replaced with finite forms, just as we did for the Minkowski no-scale vacua \cite{Blaback:2010sj}. In our case the replacement leads to the new Bianchi identity
    \begin{align}
        dF_2 = F_0\, H_3 + Q\, \text{Re} (\Omega_3)\,.
    \end{align}
where now the delta source is replaced by a finite regular form $\text{Re} (\Omega_3)$, which is usually interpreted as consisting of O6 planes intersecting in four different directions, see Appendix C of \cite{Caviezel:2008ik}:
\begin{center}
\begin{tabular}{cccc|cccccc}
     $x_0$ & $x_1$  & $x_2$ & $x_3 $& $x_4$ &$x_5$&$x_6$ &$x_7$&$x_8$& $x_9$ \\
     \hline
     $\times$ & $\times$ & $\times$& $\times$& $\times$ & $\times$ & $\times$ & $-$ & $-$ & $-$\\
     $\times$ & $\times$ & $\times$& $\times$& $\times$ & $-$  &  $-$ & $\times$ &$\times$ & $-$\\
     $\times$ & $\times$ & $\times$& $\times$& $-$ & $\times$ & $-$ & $-$ & $\times$ & $\times$\\
     $\times$ & $\times$ & $\times$& $\times$& $-$ & $-$ & $\times$ & $\times$ & $-$ & $\times$\\
\end{tabular}
\end{center}
The crosses describe directions wrapped by an O6 plane and the minus signs are orthogonal. We see that the O6 planes intersect non-trivially inside the internal manifold.

Smearing is actually \emph{the} source of criticism for these solutions \cite{McOrist:2012yc, Banks:2006hg}. From the point of view of the equations of motion smearing is equivalent to solving the integrated equations of motion as explained in section \ref{sec:10d}, and corresponds therefore to the standard coarse-graining procedure of dimensional reduction, which constructs an EFT valid at length scales below the KK scale. 

\begin{figure}[htp]
    \centering
    \includegraphics[width=10cm]{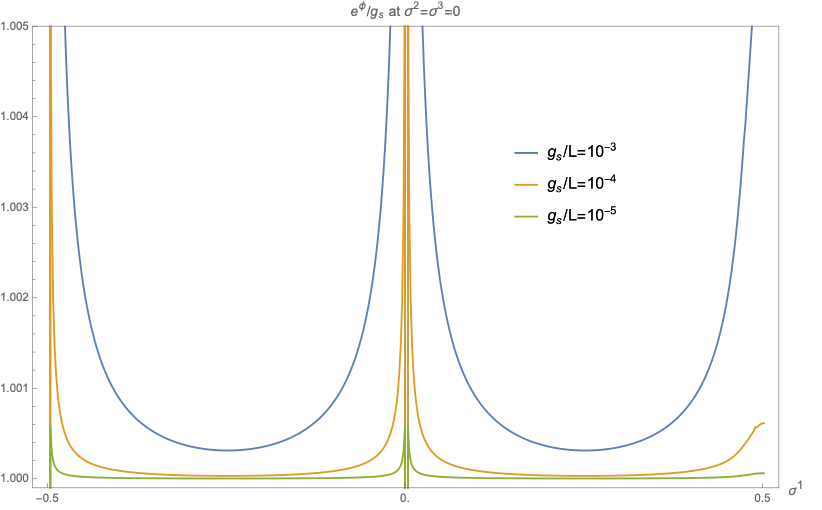}
    \caption{\small{ \emph{A plot of $e^{\phi}/g_s$ inside the extra dimensions, taken from \cite{Baines:2020dmu}. Here $g_s$ is the zero mode of $e^{\phi}$, ie the vev that one computes in the lower-dimensional theory. As $g_s/L$ becomes smaller the function $e^{\phi}/g_s$ approximates unity quickly away from the O6-plane singularities. ($L=L_{KK}$ is the size of the extra dimensions).}} }
    \label{pic:smearing}
\end{figure}

A detailed (pedagogical) example of how to move between localized and smeared sources appeared for instance in \cite{Baines:2020dmu} (and the appendix of \cite{Junghans:2020acz}) where the background $\mathbb R^{1,6} \times \mathbf T^3/\mathbb Z_2$ was considered in type IIA. There is $H_3$-flux on the $\mathbf T^3/\mathbb Z_2$ factor as well as $F_0$, and some O6-planes wrapping the Minkowski space factor. The upshot of that analysis is that the smearing approximation becomes better and better as $g_s$ becomes smaller and the volume becomes larger. For example the $\phi$ profile is depicted in figure \ref{pic:smearing} and we notice that the regions where there is a deviation from the smeared solution actually shrink to zero. This seems to suggest that smearing might not be a problem in the context of the solutions of \cite{DeWolfe:2005uu}, however this does not settle the question of its validity altogether. 
One worrisome point is the fact that there is no localized picture for the intersection of O6-planes. If we added the backreaction on the geometry of the orientifold planes the Calabi--Yau threefold becomes a manifold with $SU(3)\times SU(3)$ structure \cite{Saracco:2012wc,Marchesano:2020qvg,Junghans:2020acz} for which backreaction is understood at subleading order in a $1/N$ expansion. See also \cite{Emelin:2022cac} for a similar analysis applied to AdS$_3$ vacua. 

To summarize, we learned that O-planes are useful for obtaining scale separated vacua, but there is only an understanding of their backreaction in the smeared limit, and so there remains some doubt regarding the consistency of scale separated vacua obtained that way. 

\paragraph{Back to 11d and a mathematical conjecture.}
In the beginning of the previous paragraph we deduced that simple Freund-Rubin vacua could be scale separated on the condition we could shrink the KK scale at fixed curvature scale and we mentioned no such geometries are known. To put this in a bit more detail: we do not know of positively curved Einstein geometries for which the first eigenvalue of the (generalised) Laplace operator can be tuned arbitrarily high. Even more, recently Collins, Jafferis, Vafa, Xu and Yau conjectured no such geometries can exist \cite{Collins:2022nux}, at least for the particular case of Einstein geometries obeying
\begin{equation}
    R_{ab} =\frac{1}{L_R^2} g_{ab}\,.
\end{equation}
One could be tempted to believe that orbifolding a sphere would do the job. Consider for instance $S^n/\mathbb{Z}_k$ at large $k$. The volume goes down at fixed curvature, but for this case the volume is not a good measure of the KK scale since the only thing the orbifolding action does is killing degeneracies in the eigenvalue spectrum, but it does not remove the eigenvalues. It is remarkable how the simple computations we have done so far brings us to the forefront of mathematics as well. 

Interestingly, there is a string theory suggestion (no proof) that perhaps this mathematical conjecture is false and even Freund--Rubin vacua can be scale separated. Let us only briefly summarize this as this goes a bit beyond the level of the lecture notes. The scale separated AdS vacuum we discussed before in the limit of smeared sources relied on IIA string theory on a Calabi-Yau with O6-planes, Romans mass $F_0$, $H_3$ flux and unbounded $F_4$ flux. Interestingly from double T-duality \cite{Caviezel:2008ik} one can find new classes of solutions in IIA string theory \emph{without} Romans mass at the expense of working with generalised Calabi Yau spaces \cite{Cribiori:2021djm}. The corresponding AdS solution can then be lifted to 11-dimensional supergravity (Romans mass would be an obstacle to such a lift) and in this lift the O6-sources become pure geometry and all the rest becomes only $F_7$ flux \cite{Cribiori:2021djm}. This suggests that there exist 7-dimensional Einstein manifolds for which the KK scale can be shrunk to zero at fixed curvature scale in violation of the mathematical conjecture of \cite{Collins:2022nux}. Of course, the catch is in the word \emph{suggests} since no full 10d description of the IIA solution of \cite{Cribiori:2021djm} is given due to the O6-planes being smeared.

\paragraph{Scale separation from quantum corrections?}

The discussion above about 11 supergravity only considered classical sources in the compactification; curved extra dimensions and $p$-form fluxes. Yet, in some cases it is claimed to be possible to find compactifications with scale separation using quantum corrections, KKLT \cite{Kachru:2003aw} and LVS \cite{Balasubramanian:2005zx} being prime examples. We will follow the example of \cite{DeLuca:2022wfq} where the authors are able to build a non-supersymmetric AdS$_4$ solution of M-theory with arbitrary scale separation: the ingredient they use is Casimir energies in the internal space. The example is a compactification of M-theory on a 7-dimensional torus $\mathbf T^7$ which for simplicity is taken to be isotropic, that is all circles are chosen to have the same size. The 11d metric has the form
\begin{align}
    ds^2_{11} = \tau^{-2} ds^2_{\text{AdS}} + \rho \,ds^2_{\mathbf{T}^7}\,.
\end{align}
The internal metric is chosen such that vol$(\mathbf{T^7}) = (2 \pi)^7$. In order to compute the Casimir energy coming from fields in the internal space one should compute a one-loop diagram for the vacuum energy with the fields running in it. While in principle all fields need to be taken into account the massive states are exponentially suppressed and they can be safely neglected. Typically, one expects that due to supersymmetry the one-loop contributions of bosons and fermions cancel each other out, however in this example we will take anti-periodic boundary conditions for the fermions in the circles of the internal torus. This breaks supersymmetry by lifting the zero modes of the fermions and therefore only leaves a contribution to the vacuum energy in 4d coming only from the bosons. While there are techniques to perform such computations (see for example the book \cite{Bordag:2009zz}) in this setup it suffices to parametrize the Casimir vacuum energy as the following correction to the M-theory action
\begin{align}
    S^{\text{Casimir}} = 2 |\sigma_c| \int_{\mathcal M_{11}} d^{11}x \sqrt{-g_{11}} \, \rho^{-\frac{11}{2}} \,.
\end{align}
Here $\sigma_c$ is a constant that can be in principle computed explicitly. In addition to the Casimir energy we will add some 7-form flux threading the internal torus. The flux has to be appropriately quantized, that is
\begin{align}
    \frac{1}{(2\pi \ell_{11})^6} \int_{\mathbf{T}^7} F_7 = N_7\,,
\end{align}
where $N_7 \in \mathbb Z$ is an integer and $\ell_{11}$ is the 11d Planck length. We therefore have two contributions to the energy-momentum tensor, that is the flux and the Casimir energy. It is possible to solve the 11d Einstein equations and find the following solution
\begin{align}
    &\tau^{-2} \sim N_7^{\frac{22}{3}} |\sigma_c|^{-\frac{14}{3}}\,,\\
    &\rho \sim N_7^{\frac{4}{3}} |\sigma_c|^{-\frac{2}{3}}\,,
\end{align}
where we omitted some $\mathcal O(1)$ coefficients as we are interested in the scaling in $N_7$. Remembering that $\sigma_c$ is $\mathcal O(1)$ we find that in the large $N_7$ limit
\begin{align}
    \frac{m_{\text{KK}}^2}{|\Lambda|} \sim N_7^2\,,
\end{align}
thus achieving parametric separation in the limit of large $N_7$. This solution potentially has instabilities, likely coming from metric deformations of the torus or other effects. Another source of likely instabilities is M2-brane bubble nucleation in AdS$_4$, but no analysis of the stability of the solution is available yet.

Another, less simple, manner to employ quantum corrections to construct scale separated AdS vacua is the KKLT scenario \cite{Kachru:2003aw}. Simply from equation \eqref{AdSKKLT} one can see that scale separation is obtained in the limit of arbitrarily small $W_0$ (albeit logarithmic in small $W_0$). Such small $W_0$ are considered to be problematic in light of the tadpole conjecture since they require many fluxes to cancel against each other and this might overshoot the tadpole bound. Yet, recent breakthroughs suggest it could be achieved \cite{Demirtas:2019sip, Demirtas:2021nlu, Demirtas:2021ote}. We remain agnostic about this highly-technical matter and hopefully these lecture notes inspire the reader to study this issue in more depth. 

\subsection{Holography and scale separation} One may try to employ holographic techniques to settle the issue of whether or not scale separation is possible by constructing or proving the absence of CFTs dual to scale separated AdS vacua. The approach of translating Swampland constraints to holography has turned out fruitful, see for example \cite{Giombi:2017mxl, Baume:2020dqd,Perlmutter:2020buo, Harlow:2018jwu, Harlow:2015lma, Harlow:2020bee, Aharony:2021mpc, Conlon:2020wmc,Baume:2023msm}, but for the question of scale separation we do not know for sure since the topic is in its infancy, yet some interesting observations have appeared which we briefly summarise here. 

Recall that holography dictates that CFT operators dual to scalars in 4d are single trace scalar operators with conformal dimension
\begin{align}
    \Delta = \frac{3}{2}+\frac{1}{2}\sqrt{9+4 \,m^2 L^2_{\text{AdS}}}\,.\label{weight}
\end{align}
Here $m$ is the mass of the 4d scalar. Clearly an operator dimension is therefore a useful lead to a dual sign of scale separation once the operators dual to KK modes are identified since then we directly access the ratio $L_{\text{AdS}}/L_{KK}=L_{\text{AdS}}m_{KK}$. 

From this we conclude that putative CFT duals to AdS vacua used in phenomenology have these peculiar features:
\begin{itemize}
\item Standard properties of holographic CFTs such as a large gap to the higher spin spectrum \cite{Heemskerk:2009pn}.
    \item Lack of any relevant operators (that is, operators dual to scalars with $m^2<0$) and lack of any marginal operators ($m^2=0$). This implies that the putative dual conformal field theory would be the end point of an RG-flow, known as dead-end CFTs. The lack of tachyons (above the Breitenlohner--Freedman bound \footnote{In contrast to Minkowski space and dS space, tachyons are allowed if they are not too tachyonic , which boils down to the square root appearing in \eqref{weight} to remain real. This is called the Breitenlohner--Freedman bound.}) is demanded for AdS vacua that allow an uplift to dS vacua, since dS vacua do not allow for tachyons. 
    \item Very few low-lying relevant operators, a feature never observed in conformal field theories.
\end{itemize}
The extreme version of this, i.e. a theory with all modes stabilised at a high scale such that we end up with pure gravity, is believed to be in the Swampland. This is one of the reasons that \cite{Gautason:2018gln} conjectured there should always be at least one  low-lying operator dual to a bulk scalar with a mass of the order of the AdS scale. This seems consistent with current moduli-stabilisation scenarios. We will see below  that some Swampland conjectures take it much further and conjecture so many light operators dual to KK modes that scale separation is excluded. 

The idea of ``bootstrapping'' the landscape is not too old and started with \cite{Polchinski:2009ch} but gained more attention recently in several works, such as \cite{Alday:2019qrf, Conlon:2021cjk, Conlon:2021cjk, Conlon:2020wmc, Conlon:2018vov,  Collins:2022nux, Apers:2022zjx, Apers:2022tfm, Apers:2022vfp}. The notion of bootstrapping here is used in a general sense and does not necessarily invoke the crossing symmetry constraints. In fact,  crossing symmetry constraints seem always obeyed for any low energy effective action in AdS \cite{Heemskerk:2009pn}. Hence, to find out whether scale separated AdS vacua exist using CFT constraints one has to go beyond the standard crossing-symmetry constraints. 

We briefly review some of the most recent results in this field:
\begin{itemize}
    \item Reference \cite{Alday:2019qrf} remarks that loop diagrams in the bulk, are sensitive to scale separation since they are effected by all KK modes. Such diagrams are dual to certain non-planar CFT correlators and the main idea is then that there exists a way to define the number of large bulk dimensions purely from CFT data. Such large dimensions clearly count the number of AdS dimensions but would also count internal dimensions when they are not scale separated. So this paper does address the question of how many large bulk dimensions a holographic CFT describes, it has not been used yet to settle the consistency of scale separated AdS vacua. 
    \item In reference \cite{Lust:2022lfc} a holographic QFT is identified that ought to flow to the KKLT conformal field theory. It is possible to guess the central charge in that ultraviolet QFT, and using that $c_{\text{UV}} > c_{\text{IR}} = l^2_{\text{AdS}}$ it is possible to put a bound on $c_{\text{UV}}$ by a small topological number contradicting  $c_{\text{UV}} > l^2_{\text{AdS}}$. If correct, this invalidates the KKLT construction. 
    \item In \cite{Montero:2022ghl} it was argued that extreme scale separation, meaning that one can forget all together the extra dimensions, is inconsistent with holography in the presence of enough supersymmetry. The reason being that the R-symmetry would become a global symmetry in the bulk, which is impossible since quantum gravity does not allow global symmetries that are not gauged. 
    \item In \cite{Collins:2022nux} after checking a large number of holographic conformal field theories, a seemingly universal bound on the dimension of the first non-trivial spin 2 operator was found. This, in turn, puts bounds on the internal space which turns out to have a minimal diameter in AdS units. Reference \cite{Collins:2022nux} then conjectured to be true for all conformal field theories, which would make scale separation impossible. 
    \item In a series of recent works \cite{Conlon:2021cjk, Apers:2022zjx,Apers:2022tfm, Apers:2022vfp, Plauschinn:2022ztd}, extending \cite{Aharony:2008wz}, it was observed that the conformal dimensions obtained from supersymmetric DGKT solutions are integers. This is absolutely non-obvious. What is less special is that masses tend to be quantized in $1/L^2$ units, but that the square root in \eqref{weight} is highly special and requires explanation. An attempt to explain this is presented in \cite{Apers:2022vfp}.  Notice however that in the case where supersymmetry is broken the conformal dimensions sometimes cease to be integers \cite{Quirant:2022fpn} unless one considers the scale separated non-SUSY solutions \cite{Apers:2022zjx}. 
\end{itemize}

\subsection{Swampland arguments in favor of/against  scale separation?}

The main swampland conjecture relating to the existence of AdS solutions with scale separation comes from \cite{Lust:2019zwm} and is called the \emph{AdS distance conjecture (ADC)} and reads\footnote{A closely related conjecture that arrives at similar conclusions invokes the mass of the gravitino, see \cite{Castellano:2021yye, Cribiori:2021gbf}.}:
\bigskip

\emph{Consider quantum
gravity on $d$-dimensional AdS space with cosmological constant $\Lambda$. There exists an infinite tower of states with mass scale
$m$ which, as $\Lambda \rightarrow 0$, behaves (in Planck units) as 
\begin{equation}
 m \sim |\Lambda|^{\alpha}   
\end{equation} 
where $\alpha$ is a positive order-one number. For SUSY AdS vacua $\alpha=1/2$.}\footnote{A very interesting refinement of this conjecture has been suggested in \cite{Buratti:2020kda}, and commented upon in \cite{Apers:2022zjx}. } 
\bigskip

This last statement about supersymmetry is known as the \emph{strong AdS distance conjecture}. 
 Let us call the operator of vacuum energy in dimension $d$ to be $M_p^{d-2}\Lambda$ so that $\Lambda$ has mass dimension two, in other words $\Lambda=-1/L^2_{AdS}$. Then the strong ADC applied to a KK tower implies
\begin{align}
    L_{KK} = c \,L_{\text{AdS}}\,,
\end{align}
where $c$ is a $\mathcal O(1)$ constant and forbids scale separation for SUSY vacua. If correct, then the IIA DGKT vacua are inconsistent, and KKLT with exponentially parametric tuning of $W_0$ as well. The strong ADC is therefore quite a strong conjecture, as the name suggests. The more general ADC seems in line with all examples ever constructed within string theory, to our knowledge. 

The conjecture is based on an extension of the distance conjecture \cite{Ooguri:2006in} for scalar fields to distances on metric space. It is not clear why the distance conjecture can be applied beyond scalar fields,  but it is sensible. Yet, there seems no particular argument for the strong ADC beside the observation that well controlled examples in string theory obey it. Such examples typically feature more than 4 conserved supercharges.  Very intriguingly, \cite{Cribiori:2023ihv, Cribiori:2022trc} pointed out that  the lack of scale separation for AdS vacua with more than 4 conserved supercharges is a direct consequence of the \emph{magnetic Weak Gravity Conjecture \cite{Arkani-Hamed:2006emk}}, which, unlike the strong ADC, is a very well established conjecture. It states that in a quantum gravity theory which has a U(1) gauge field at low energies with coupling $g$ then the cut-off $\Lambda_C$ of this EFT  is bounded by
\begin{equation}
\Lambda_C < g M_p^{(d-2)/2}\,.
\end{equation}
What \cite{Cribiori:2023ihv, Cribiori:2022trc} relied upon is to use that the vacuum energy for SUSY AdS vacua with extended SUSY is fixed in terms of the gauge coupling and then demanded that the AdS length scale is above the cut-off length scale, leading to an effective absence of scale separation. Although in this case the tower does not necessarily need to be a KK tower. It most likely is in the examples we understand well, but that does not matter. The question at state is whether the AdS vacuum can be seen as a vacuum within an EFT, and for that all towers must decouple, KK or something else. The above relation with other Swampland conjectures is typical to Swampland research; the conjectures seem to form a very tied web, and conjectures that seem at first not related, often turn out to be. 

We already mentioned that the IIA vacua seem a counter-example to the strong ADC, but then again, these vacua have been criticized on the basis of using the approximation of smeared orientifolds. 

Yet, there is in fact a Swampland argument \emph{in favor} of the consistency of the IIA vacua, rather than against \cite{Shiu:2022oti} (see also \cite{Farakos:2023nms}). The basic argument is rather simple. Instead of computing the distance in metric space as $\Lambda\rightarrow 0$, we can compute the distance in scalar field space, which is were most of the evidence for the distance conjecture comes from. The reason we can do this, is that there exists an open-string scalar field that continuously varies along the IIA vacua with different size of the energy. In other words, the different IIA vacua can be seen as different minima of a single potential energy function, once one includes this scalar field. It then turns out that the distance conjecture is exactly obeyed! One could argue that this is evidence for the ADC from the ordinary distance conjecture and that the strong ADC somehow has no separate argument going for it. 

\subsection{The AdS moduli conjecture}

The two main applications of flux compactifications discussed in these lectures were 1) dS vacua and 2) AdS vacua with scale separation. We already alluded to the fact that both topics are not disconnected and to end these lecture notes we want to discuss that in a bit more detail. 

The most obvious relation between the two has been explained before; scale separation is a necessary condition for using the tools of 4d EFT to describe the fluctuations around the vacuum $AdS_4\times \mathcal{M}_6$. In fact, it is often used as a prior to construct the vacuum if the construction relies on EFT methods. Then self-consistency requires the vacuum to be stabilised at a volume for which scale separation is indeed realised and so the prior was justified. This is crucial for dS vacua since almost all claimed de Sitter compactifications start out with this assumption, where the majority of these first goes about and constructs an AdS vacuum (often SUSY) and then uplifts. 

\begin{figure}[htp]
    \centering
    \includegraphics[width=11cm]{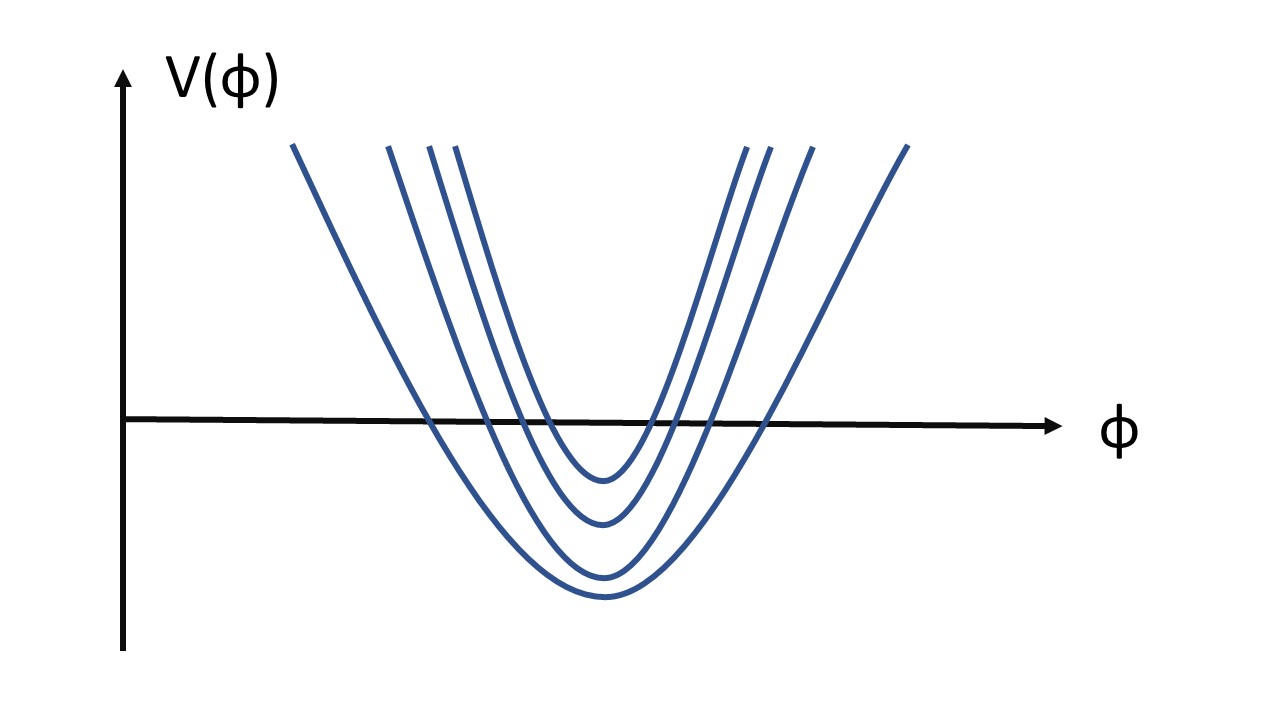}
    \caption{\small{\emph{A series of AdS potentials with ever increasing $m^2L^2_{AdS}$ combination.}} }
    \label{pic:modulicon}
\end{figure}

Uplifts are often done using warped down anti-branes but can be generalised, see for instance \cite{Saltman:2004sn}. Yet, there are more conditions beyond scale separation to guarantee a successful uplift. An obvious one is a SUSY-breaking energy that is sufficiently small. Imagine this to be possible then there is still a condition on the AdS vacuum: it needs to be sufficiently rigid and should not be destabilised by the uplift. Interestingly this condition can be put into a necessary condition on the mass spectrum, as first pointed out in \cite{Moritz:2017xto}.  The condition is most easily visualised in a picture, see \ref{pic:modulicon}.
One would like an AdS vacuum whose vacuum energy is a small negative number and at the same time the minimum lies in a sharp enough potential. This means that a small SUSY-breaking effect, which will backreact little on the assumed approximations, will be able to generate a positive vacuum energy, without causing a runaway. In practise this means that the dimensionless products
\begin{equation}\label{Adsmoduli}
m^2_{\phi}\,L^2_{AdS} \gg 1\,,    
\end{equation}
should be large enough. Here $m^2_{\phi}$ is the mass squared of the lightest scalar field. The factor $L^2_{AdS}$ in the product informs us how close the vacuum is to Minkowski space and the other factor $m^2_{\phi}$ how sharply peaked it is. In the picture \ref{pic:modulicon} we have drawn various potentials with ever increasing $m^2_{\phi} L_{AdS}^2$. 

Interestingly this product is the one that appears in the formula \eqref{weight} for the conformal weight of the dual operator. And so vacua which can be easily uplifted are such that the lowest weight is very high. But note that this is not the case in KKLT for instance \cite{Blumenhagen:2019qcg} and this is why uplifting is tedious. Inspired by the absence of known CFTs obeying \eqref{Adsmoduli} and the difficulty in achieving dS vacua it was conjectured that the condition \eqref{Adsmoduli} is a Swampland condition in the sense that EFTs obeying  \eqref{Adsmoduli} are in the Swampland \cite{Gautason:2018gln}. In other words the condition says there is always a scalar for which 
\begin{equation}\label{Adsmoduli2}
m^2_{\phi}L^2_{AdS} \sim \mathcal{O}(1)\,,    
\end{equation}
This conjecture was later named the \emph{AdS moduli conjecture} in \cite{Blumenhagen:2019vgj}. Note that this conjecture could be considered weaker than the strong ADC since it only requires that one scalar is light in AdS units (order 1 say) instead of a whole tower. 

Examples of stringy constructions that would violate this conjecture are the so-called Kallosh--Linde racetrack fine tuning scenarios \cite{Kallosh:2004yh}, which have no clear controllable dS embedding so far, consistent with the AdS moduli conjecture. 

\newpage

\section*{Acknowledgments}
We are grateful to Vincent Van Hemelryck for comments on an earlier draft and we like to thank the organisers and participants of the \emph{CERN Winter School on Supergravity, Strings and Gauge Theory 2022} and the \emph{Tehran School on the Swampland Program 2022} for inspiration to write these notes. The work of GZ is supported by the program Rita Levi Montalcini for young researchers (D.M. n. 928, December 23rd 2020).

\newpage

\appendix
\section{Forms and Hodge star}\label{app:forms}
For a general p-form we switch between component and form notation as follows:
\begin{equation}
A_p =\frac{1}{p!}A_{M_1\ldots M_p}dx^{M_1}\wedge\ldots\wedge dx^{M_p}\,.
\end{equation} 
The exterior derivative works as usual:
\begin{align}
F_{p+1}=dA_p &= \frac{1}{p!}\partial_N A_{M_1\ldots M_p}dx^{N}\wedge dx^{M_1}\wedge\ldots\wedge dx^{M_p}\\
\rightarrow\qquad & F_{M_1\ldots M_{p+1}}= (p+1)\partial_{[M_1}A_{M_2\ldots M_{p+1}]}\,,
\end{align}
where anti-symmetrization is taken with weight one, for example: $T_{[ab]}=1/2\left(T_{ab}-T_{ba}\right)$. 

In $D$ dimensions the epsilon-\emph{symbol} $\varepsilon_{M_1\ldots M_D}$ is defined as 
\begin{equation}
\varepsilon_{1\ldots D} = 1\,,
\end{equation}
and completely anti-symmetric in all indices: $\varepsilon_{[M_1\ldots M_D]} =\varepsilon_{M_1\ldots M_D}$ . To get to a volume \emph{tensor} on a space with metric $g_{
\mu\nu}$ we define:
\begin{equation}
    \epsilon_{M_1 M_2\ldots M_D} =\sqrt{|g|}\varepsilon_{M_1M_2\ldots M_D}\,.
\end{equation}
Contractions of the epsilon tensor (and symbol) obey the following relations:
\begin{equation}
\epsilon_{M_1\ldots M_q M_{q+1}\ldots M_D} \epsilon^{M_1\ldots M_q N_{q+1}\ldots N_D} =(-1)^t q!(D-q)!\delta^{[N_{q+1}}_{[M_{q+1}} \ldots \delta^{N_D]}_{M_D]}\,.      
\end{equation}
where $t$ stands for the number of timelike dimensions of the $D$-dimensional space. To define the Hodge star on all forms, it is enough to define on the following set made from a coordinate base and then extend it by linearity:
\begin{equation}
    \star_D (d x^{M_1}\wedge \ldots\wedge d x^{M_p} ) =\frac{1}{(D-p)!} \epsilon_{N_1\ldots N_{D-p}}^{\qquad M_{1}\ldots M_p}d x^{N_1}\wedge \ldots\wedge d x^{N_{D-p}}\,.
\end{equation}
For example, we have $\star \Phi \equiv \Phi\,\epsilon\equiv \sqrt{|g|}\Phi dx^0\wedge\ldots\wedge dx^D$ where  $\Phi$ denotes some zero-form (scalar). In other words, the volume tensor with all indices down is defined as:
\begin{equation}
 \epsilon= \star 1.    
\end{equation}
The $\star$ operation obeys the following properties:
\begin{align}
&(\star A_p)\wedge B_p = (\star B_p)\wedge A_p =\frac{1}{p!}A_{\mu_1\ldots\mu_p} B^{\mu_1\ldots\mu_p} \star 1\,, \\
&\star\star A_p = (-1)^{t+p(D-p)}A_p\,.
\end{align}

\section{Orientifold planes in the worldsheet}

\label{sec:oplaneapp}

One ubiquitous ingredient in type II string compactifications is orientifold planes. Their introduction is  useful for two main reason: they play an important role in tadpole cancellation providing a source of negative charge without completely breaking supersymmetry, and they cut the supersymmetry preserved by the background in half thus allowing to have $\mathcal N=1$ solutions in 4d of type II string theories. In this appendix we  provide a brief review of orientifolds, starting with their worldsheet description and then moving to a spacetime perspective. For a more complete treatment which also includes original references we point to \cite{Polchinski:1996na}.

The usual way orientifold planes are introduced is via worldsheet: they are indeed pretty exotic objects and the natural path to their introduction is via the study of un-oriented strings. When trying to introduce un-oriented strings in the context of type II strings one encounters several consistency constraints which lead to the formulation of type I string theory. Type I string theory secretly already contains an orientifold plane, specifically an O9-plane, and the remaining orientifold planes are found via toroidal compactification and application of T-duality. We will start by discussing un-oriented strings and then study T-duality of type I string theory to discover all the other orientifold planes.

\subsection{Un-oriented closed strings}

In the description of un-oriented strings we will focus solely on the bosonic worldsheet oscillations. In the case of superstrings there are also fermionic worldsheet oscillations but it is sufficient to focus on the bosonic ones to capture the salient features of un-oriented strings and orientifolds.

Recall that the fields describing the oscillation of strings are described as scalar fields $X^\mu(t,\sigma)$ on the worldsheet, where the index $\mu$ runs over the number of spacetime dimensions. Here $t$ is the time coordinate on the worldsheet and $\sigma$ is the space coordinate on the worldsheet which is periodic in the case of closed strings, that is $\sigma \sim \sigma + 2 \pi$. Equations of motion on the worldsheet decompose all oscillations of fields into a linear superposition of left and right moving waves, that is
\begin{align} X^\mu(t,\sigma) = X^\mu_L(t + \sigma) + X_R^\mu(t-\sigma)\,,
\end{align}
where the left movers $X^\mu_L(t+\sigma)$ are a function of only $t+\sigma$ and the right movers are a function of $X_R^\mu(t-\sigma)$ are a function of only $t-\sigma$. Having introduced this notation we can start talking about the worldsheet parity operation $\Omega$: its action is to simply map $X^\mu(t,\sigma)$ to $X^\mu(t,\pi - \sigma)$. This operation ends up swapping left and right movers. This gives a classical definition of the worldsheet parity, and it can be easily extended to an operator $\Omega$ on the Hilbert space of string theory. Many states in the Hilbert space will not be invariant under $\Omega$, however, given that clearly $\Omega^2 = 1$,\footnote{Some care has to be taken in the definition of $\Omega$ for the fermionic fields on the worldsheet.} we can easily build a projection operator that retains only states are invariant under $\Omega$. Calling this projector $\Pi_{\Omega}$ its definition is obviously
\begin{align}\Pi_{\Omega} = \frac{1}{2} \left(\mathbb I + \Omega \right)\,,
\end{align}
where $\mathbb I$ is the identity operator. By applying $\Pi_\Omega$ on the Hilbert state of string theory one retains only states that are worldsheet parity invariant thus giving un-oriented strings. At the level of the spectrum of closed strings the projection cuts out some string fluctuations from the spectrum: for instance at the massless level of the bosonic string we have a graviton, an anti-symmetric two-form and a dilaton. For the un-oriented string the anti-symmetric two-form is projected out leaving therefore only the graviton and the dilaton. Unfortunately the tachyon that plagues the bosonic string is not projected out and therefore the projection $\Pi_\Omega$ does not stabilise the bosonic string. 
Having discussed briefly what happens in the bosonic string we can move to the case of superstrings: as usual here there are some additional consistency constraints, first and foremost the projection can be done only for type IIB string theory.\footnote{The intuition is the following: the operation $\Omega$ swaps left and right movers on the string worldsheet, and therefore requires a worldsheet theory that is left-right symmetric. That happens for type IIB given that the GSO projection is the same for left and right movers, but fails for type IIA where the projection is different in the two sectors.} Moreover the projection requires the presence of an open string sector for consistency, which leads to the formulation of type I string theory eventually. Before moving onto how orientifold planes are defined let us talk briefly about the case of un-oriented open strings.

\subsection{Un-oriented open strings}

When it comes to the definition of the operator $\Omega$ on open strings there is not much of a difference: the action is that of a parity transformation mapping the spatial coordinate $\sigma$ to $\pi - \sigma$. But to fully define the action on any open string state it is necessary to define the action of $\Omega$ on Chan--Paton factors. Recall that Chan--Paton factors are a set of non-dynamical degrees of freedom attached to each endpoint of an open string. To fully specify an open string state in addition to the oscillator content it is necessary to choose the Chan--Paton factor, and their information can be packaged in an $N \times N$ matrix $\lambda$ in the case where there are $N$ distinct Chan--Paton factors at the endpoints.\footnote{The matrix $\lambda$ is nothing but the tensor product of the two Chan--Paton labels at each endpoint of a string.} The system is invariant under $U(N)$ transformations that act on $\lambda$ as $U \lambda U^\dag$  which is a reflection of the $U(N)$ gauge symmetry: indeed the dynamics of open strings is described at the massless level by a $U(N)$ Yang--Mills theory. To fully define $\Omega$ it is necessary to specify its action on $\lambda$: indeed $\Omega$ swaps the two endpoints of an open string thus acting on $\lambda$ via transposition. However it is possible to define a more general action as
\begin{align} \Omega: \lambda \ \mapsto \ M \lambda^T M^{-1}\,,
\end{align}
for some matrix $M$. Requiring that $\Omega^2 = 1$  forces $M$ to be either symmetric or anti-symmetric. In the former case the system will end up having a $SO(2N)$ gauge symmetry and in the latter case it will have a $USp(2N)$ gauge symmetry.\footnote{Here we implicitly assumed that the number of Chan--Paton factors is even.} For the case of type I it is necessary to take $M$ to be symmetric to have a consistent theory therefore giving a $SO(32)$ theory.

\subsection{From un-oriented strings to orientifolds}

The common way to discover orientifold planes is via T-duality of type I string theory. We have discussed how to take type IIB string theory and do a projection to un-oriented strings giving type I string theory. However it is also known that type IIB string theory when defined on a manifold of the form $\mathcal M_9 \times S^1$ is equivalent to type IIA on $\mathcal M_9 \times S^1$ via T-duality. Therefore this suggest that taking type I string theory on a circle and applying T-duality might give a way to define un-oriented strings in type IIA string theory. Let us follow this route then: we need to understand how to combine the action of $\Omega$ and a T-duality transformation on the worldsheet fields $X^\mu(t,\sigma)$. Recall that performing a T-duality along the coordinate $i$ results in the transformation
\begin{align} X^i(t,\sigma) = X^i_L(t + \sigma) + X_R^i(t-\sigma) \  \mapsto \  \hat X^i(t,\sigma) =  X_L^i(t + \sigma) -  X_R^i(t-\sigma)\,.
\end{align}
Moreover the effect of $\Omega$ is to swap left and right movers. Therefore in terms of the dual coordinate $\hat X^i(t,\sigma)$ the worldsheet parity acts as
\begin{align} \hat X^i(t,\sigma) \rightarrow -\hat X^i(t,\pi - \sigma)\,.
\end{align}
This is a combination of a pure worldsheet parity and a spacetime parity transformation. Successive applications of T-duality along different directions yield a similar result: in the dual frame a worldsheet parity will also be combined with a spacetime parity along every dual coordinate. Notice that we no longer have a pure un-oriented version of string theory: given that we are performing a spacetime reflection any string state will have a mirror version with opposite orientation and related by a spacetime transformation. Only at the fixed point of the spacetime parity we do recover a pure un-oriented string theory: this locus is called an \emph{orientifold plane}. The general action of an orientifold will be $\Omega \mathcal R$ where $\mathcal R$ is a spacetime reflection, that is it is an operation that gives the identity when applied twice. Orientifolds will act as a mirror for strings relating states on one side to states on the other side. We can briefly add some properties of orientifold planes that are specific to superstrings:

\begin{itemize}
\item We only focused on the orientifold action on bosonic worldsheet fields. When taking into account the action on fermions in order to make sure that the operation squares to one it is necessary to change it to $\Omega \mathcal R (-1)^{F_L}$ where $F_L$ is the left-moving space-time fermion number;
\item The dimension of orientifold planes is fixed by which type II string theory we are considering. The fixed locus we just encountered in type IIA has eight space dimensions, making it a O8-plane. By successive application of T-duality one finds that type IIA string theory has O$p$-planes with $p$ even and type IIB has O$p$-planes with $p$ odd. By using the language of O-planes it is possible to understand type I string theory as type IIB string theory with an O$9$-plane;
\item Notice that when introducing an O$9$-plane in type IIB string theory we went to type I string theory, which has half the supersymmetry. Applying T-duality does not change the supersymmetry of the solution, implying that orientifold planes break half of the supersymmetry.
\end{itemize}

\subsection{Orientifold planes and the worldsheet topology expansion}

Given a string diagram on a worldsheet $\Sigma$ its contribution to a given correlation function will be weighted by a factor $g_s^{- \chi(\Sigma)}$ where $g_s = e^{\phi}$ is the string coupling constant and $\chi(\Sigma)$ is the Euler characteristic of $\Sigma$. Therefore, similarly to how in quantum field theory it is possible to expand correlation functions in terms of the number of loops, it is possible to expand any string correlation function in terms of the Euler number of the worldsheet provided one considers a regime where $g_s \ll 1$. In the case of a string theory model with closed strings only it is interesting to note that
\begin{align}\chi(\Sigma) = 2-2g(\Sigma)\,,
\end{align}
where $g(\Sigma)$ is called the genus of $\Sigma$ and can be identified with the number of holes in $\Sigma$. Therefore in this scenario it is possible to identify the number of loops with the number of holes in the worldsheet. The main advantage of string theory in this situation is that there is a unique worldsheet for a given genus giving therefore a much reduced number of diagrams to compute compared to the case of ordinary quantum field theory. The worldsheet with no loops is nothing but the usual Riemann sphere and any worldsheet with genus $g$ with $g>0$ is nothing but a doughnut with $g$ holes.
While this gives a concise description of how to organise any string amplitude it does not cover the case we are interested in, that is the case of orientifold planes (and it does not include any scenario with open strings either). In order to cover these cases it is necessary to expand the set of worldsheet topologies that are included in the computation of string correlation functions. When only closed strings are included (without orientifolds) the worldsheets that are included are simply closed manifolds of complex dimension one (oftentimes called Riemann surfaces), and we just discussed how to classify them for the sake of the genus expansion of string amplitudes. To include orientifolds and open strings it is necessary to introduce crosscaps and boundaries on the worldsheet. Let us cover these two separately:
\begin{itemize}
\item A crosscap is inserted on a two dimensional surface by excising a small disk and closing back the surface by identifying the antipodal points of the boundary of the disk. This is a local operation but it makes a surface un-orientable. For instance a M\"obius strip can be obtained by taking a disk and inserting a crosscap in it. Similarly a Klein bottle is a Riemann sphere in which two crosscaps have been added. Given that the worldsheet ceases to be orientable in the presence of any number of crosscaps it means that strings will no longer be oriented;
\item The introduction of boundaries on the worldsheet signals the presence of open strings given that those boundaries are identified with open string endpoints.

\end{itemize}
Luckily enough it is still sufficiently easy to classify worldsheet topologies in the presence of boundaries and crosscaps and compute their relative weight in string amplitudes. The overall factor of each diagram still is $g_s^{-\chi(\Sigma)}$, however the formula for the Euler characteristic is modified to
\begin{align}\chi(\Sigma) = 2-2g - n_b-n_c\,,
\end{align}
where here $n_b$ is the number of boundaries and $n_c$ is the number of crosscaps. For example when computing the one-loop vacuum energy of type II string theories one simply has to consider the torus as a diagram, the only diagram with zero Euler characteristic. When going to type I string theory which is un-orientable and has open strings one needs to consider all possible diagrams with zero Euler characteristic, that is the Klein bottle (which has $g=0$, $n_b= 0$, and $n_c = 2$), the cylinder (which has $g=0$, $n_b = 2$, and $n_c=0$), and the M\"obius strip (which has $g=0$, $n_b = 1$, and $n_c = 1$). All these four diagrams added up give the one-loop vacuum energy of the system.

\subsection{Orientifold properties: tension and charges}

It is natural to wonder what kind of effect does the introduction of an orientifold plane has on the geometry of space-time. Given that orientifolds (and branes) are dynamical objects they will have their own energy-momentum tensor thus implying that they source gravitational field. Moreover it is important to understand how they couple to the remaining fields in the low-energy action describing type II string theories. Quite remarkably it is possible to write a quite compact action describing the coupling of O-planes to the various bosonic fields of type II string theories. The resulting action for a O$p$-plane wrapping a manifold $\Sigma_{p+1}$ is the sum of a DBI-action and a Wess--Zumino action
\begin{align} S = S_{\text{DBI}} + S_{\text{WZ}}\,,
\end{align}
where we have that
\begin{align} S_{\text{DBI}} &= - T_p\, \mu_p \int_{\Sigma_{p+1}} d^{p+1}x e^{-\phi} \sqrt{-\text{det}(g)}\,,\\
 S_{\text{WZ}} &= - Q_p \mu_p \int_{\Sigma_{p+1}} C \wedge \sqrt{\frac{\mathcal L(R_T/4)}{\mathcal L(R_N/4)}}\,.
\end{align}
These formulas have a lot of information and we will spend some time unpacking them. Let us start from the DBI-action: we can clearly see that O-planes couple to the metric and the dilaton and therefore become sources of both fields. The coupling to the metric is measured by the volume of the cycle wrapped by the O-plane similarly to the case of other extended objects in string theory. The quantity $\mu_p$ is the tension of a D$p$-brane and it is given by
\begin{align} \mu_p = \frac{2 \pi }{(2 \pi \sqrt{\alpha'})^{p+1}}\,.
\end{align}
The quantity $T_p$ is the ratio between the tension of an O-plane and a D-brane and it is given by
\begin{align}T_p = -2^{p-5}\,.
\end{align}
Quite remarkably this is \emph{negative}. This is not an issue because orientifold planes are not dynamical, but it is this property that makes O-planes so valuable in flux compactifications as they can cancel the positive energy density sourced by branes and fluxes in a supersymmetric fashion. We will shortly discuss how tensions are computed in string theory, but before that let us discuss the Wess--Zumino term of the O-plane action. The quantity $Q_p$ is again the ratio of the charge of an O$p$-plane and a D$p$-brane, and its value matches exactly $T_p$. This is nothing but a manifestation of the fact that orientifold planes are supersymmetric objects. The term $C$ inside the action is the formal sum of all Ramond--Ramond potentials, implying that an orientifold plane will act as a source of these fields. In the formula $R_T$ and $R_N$ denote the curvatures of the tangent and normal bundles respectively, and the term $\mathcal L$ is the Hirzebruch $\mathcal L$-polynomial, which can be expressed in terms of Pontryagin classes as
\begin{align} \sqrt{\mathcal L(R/4)} = 1+ \frac{(4 \pi^2 \alpha')^2}{96} p_1(R) - \frac{(4\pi^2 \alpha')^4}{10240} p_1^2(R) + \frac{7 (4 \pi \alpha')^4}{23040} p_2(R)+ \dots\,.
\end{align}
The main observation concerning the Wess--Zumino action is that again the charge sourced by O-planes is negative, thus allowing to cancel all charges sourced by D-branes and fluxes without having to introduce anti-branes that would lead to a breakdown of supersymmetry.

As we highlighted the most remarkable property of O-planes is the fact that their tensions and charges have the opposite sign of the one of D-branes. It is quite odd to have a negative tension object in a theory with gravity, and in the following we will briefly outline how the computation of tension and charges are done in string theory. In keeping with the fact that string theory has no free dimensionless parameters we will see that the fundamental theory will completely fix the values of tensions and charges. We will discuss the computation method for charges and tensions of both D-branes and O-planes because of their similarity. 

\begin{figure}[htp]
    \centering
    \includegraphics[width=14cm]{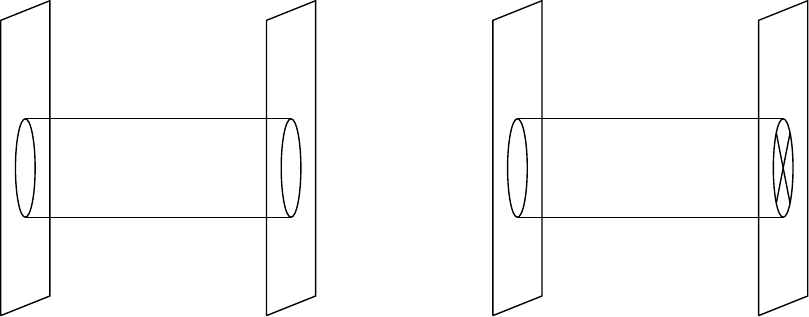}
    \caption{\small{\emph{The diagrams computing tension and charge of D-branes and O-planes. On the left, the exchange of strings between two D-branes. On the right, the exchange between a D-brane and an O-plane (with the O-plane on the right side). The difference between the two diagrams is the presence of the crosscap on top of the O-plane, drawn here with a cross inside the oval.}}}
    \label{pic:oplane}
\end{figure}

In order to compute tensions and charges one considers an exchange of closed strings between two objects which can be two D-branes or one D-brane and an O-plane: in the first case we will compute the charge and tension of D-branes and in the second case the charge and tension of O-planes. The diagrams are sketched in figure \ref{pic:oplane}. In the case of an exchange between two D-branes the diagram has the topology of an annulus, that is it has genus zero, two boundaries, and no crosscap. For the case of the exchange between a D-brane and an O-plane one simply replaces one boundary with a crosscap, thus giving a M\"obius strip.\footnote{Looking at figure \ref{pic:oplane} it is clear that we are considering the case in which one closed string is emitted from one object and lands on the other one, which clearly computes the net force between them. However there is a different perspective that can be useful to consider: if we consider a stretched string between the two objects one and have its endpoints move in a loop we recover the same exact diagram! This is an instance of open-closed duality in which a closed string phenomenon can be interpreted dually in terms of open strings physics.} From the formula of the diagrams it is possible to extract that the various objects exchange various fields between each other (gravitons, dilaton, Ramond--Ramond fields), thus allowing to extract the couplings constant directly. The result of the diagram as a matter of fact is that the net force between the objects involved actually vanishes which is nothing but a manifestation of supersymmetry. We will not go into the details of the computation but simply point to \cite{Polchinski:1996na} for those interested in seeing how the diagrams are computed. 

\subsection{Geometrization of O6 planes}

So far we have discussed some properties of orientifold planes as seen from the string worldsheet, which implies that all the discussion is valid at weak coupling. Accessing the fate of orientifold planes at strong coupling is a difficult question given that a fully non-perturbative definition of String Theory is still not completely available, however employing some dualities it is possible to explore some strong coupling regimes and peek into how O-planes behave in such regimes. While it is not important for the lectures we cite one famous example of this: it is well known that non-perturbative effects break the O7-plane of type IIB String Theory into a couple of 7-branes \cite{Sen:1996vd}, see \cite{Denef:2008wq} for a review. The other important example we would like to discuss is the one of the O6-plane: in the main text we discuss several instances of solutions that employ O6-planes and therefore it may be useful to understand their behaviour at strong coupling, that is their uplift to M-theory. The result is well known: the uplift of an O6-plane to M-theory is a smooth space with a Atiyah--Hitchin metric \cite{Atiyah:1985dv}. We will not need the detailed metric, but we would like to stress that the uplift to M-theory of O6-planes is purely geometrical and smooth, and the same happens with D6-branes. Therefore any question that one may have about type IIA compactifications that admit an uplift to M-theory can be recast into a purely geometrical question without the addition of localized sources. Some care has to be taken though: first, the uplift is valid only at very large string coupling, so a solution of type IIA at weak string coupling does not admit a well behaved uplift to M-theory and therefore any question is better posed within the framework of type IIA String Theory. Second, even in a situation of strong coupling, there may be some hurdles in the uplift, most notably the presence of a non-zero Romans mass.

{\small
\bibliography{refs}}
\bibliographystyle{utphys}
\end{document}